\documentclass[twocolumn,epjc3]{svjour3}  

\journalname{Eur. Phys. J. A}

\usepackage{lipsum}
\usepackage{xcolor}
\usepackage{hyperref}
\usepackage{amsmath,amssymb} 
\usepackage{widetext}
\usepackage{hhline}

\usepackage{cite}

\usepackage{graphicx}

\usepackage{subfigure}

\usepackage{tikz}

\hypersetup{colorlinks=true,citecolor= blue,citecolor= blue}

\begin{document}
\title{Jacobi no-core shell model for \texorpdfstring{$p$}{p}-shell hypernuclei}
\author{Hoai Le\thanksref{addr1,e1}
\and Johann Haidenbauer\thanksref{addr1,e2}
\and Ulf-G. Mei{\ss}ner\thanksref{addr2,addr1,addr3,e3}
\and Andreas Nogga\thanksref{addr1,e4}
}
\thankstext{e1}{e-mail: h.le@fz-juelich.de}
\thankstext{e2}{e-mail: j.haidenbauer@fz-juelich.de}
\thankstext{e3}{e-mail: meissner@hiskp.uni-bonn.de}
\thankstext{e4}{e-mail: a.nogga@fz-juelich.de}

\institute{IAS-4, IKP-3 and JHCP, Forschungszentrum J\"ulich, D-52428 J\"ulich, Germany \label{addr1}
           \and
           HISKP and BCTP, Universit\"at Bonn, D-53115 Bonn, Germany \label{addr2}
           \and 
           Tbilisi State University, 0186 Tbilisi, Georgia \label{addr3}
}

\date{August 26th, 2020}

\maketitle

\begin{abstract}
We extend the recently developed Jacobi no-core shell model to hypernuclei. Based on the coefficients of 
fractional parentage for ordinary nuclei, we define a basis where the hyperon is the spectator particle. We then 
formulate transition coefficients to states that single out a hyperon-nucleon pair which allow us to implement 
a hypernuclear many-baryon Hamiltonian for $p$-shell hypernuclei. As a first application, we use the 
basis states and the transition coefficients to calculate the ground states of 
$^{4}_{\Lambda}$He, $^{4}_{\Lambda}$H, $^{5}_{\Lambda}$He, $^{6}_{\Lambda}$He, 
$^{6}_{\Lambda}$Li, and $^{7}_{\Lambda}$Li and, additionally, the first excited states of 
$^{4}_{\Lambda}$He, $^{4}_{\Lambda}$H, and $^{7}_{\Lambda}$Li. 
In order to obtain converged results, we employ the similarity renormalization group (SRG) to soften 
the nucleon-nucleon and hyperon-nucleon interactions. Although the dependence on this 
evolution of the Hamiltonian is significant, we show that a strong correlation of the results 
can be used to identify preferred SRG parameters. This allows for meaningful 
predictions of hypernuclear binding and excitation energies. 
The transition coefficients will be made publicly available as HDF5 data files. 
\end{abstract}


\section{Introduction}
After more than 65 years of research on hypernuclei, our knowledge of the 
interaction of hyperons with nucleons or with other hyperons 
still remains on a modest level. This situation
is rather unsatisfactory given the important role hyperons 
play for various aspects of nuclear physics as well as
for astrophysics \cite{Tolos:2020aln,Gal:2016boi,Gandolfi:2015jma,Chatterjee:2015pua,Weissenborn:2011ut}. For example, as extensively discussed in recent years, the hyperon interaction could 
have a significant impact on the properties of neutron 
stars \cite{Gandolfi:2015jma,Chatterjee:2015pua,Weissenborn:2011ut}. 
The reason for the large uncertainty is the tremendous 
difficulty to perform scattering experiments involving hyperons 
and the fact that no two-baryon bound state has been 
found so far, except for the well known deuteron. 
An important source of information has been the
spectroscopy of hypernuclei \cite{Hashimoto:2006aw}.
New experiments are planned at facilities like J-PARC, 
FAIR, MAMI and JLab
\cite{Feliciello:2015dua,Schonning:2020ndl,Garibaldi:2018lxi,Ohnishi:2019cif,Rappold:2020yia,Jude:2019qqd}, 
some to study the scattering of hyperons on nucleons, but 
mostly measurements of bound states of ordinary nuclei with hyperons.
Such new and very probably more precise data will not only 
be phenomenologically 
interesting, but also enable us to explore the underlying 
interactions in more detail. The latter is now possible 
because even fairly complex systems can be treated 
theoretically on a microscopic level, 
thanks to improved algorithms and 
increasing computational resources. Indeed, nowadays, one 
can solve the Schr\"odinger equation for hypernuclei  
up to the $p$-shell based on realistic and rather elaborate
baryon-baryon interactions \cite{Wirth:2017bpw,Wirth:2014apa,Le:2019gjp}. Thus, it has
become feasible to study detailed features of the baryonic forces, 
like the spin-dependence of hypernuclear interactions, which 
are inaccessible in direct scattering experiments. 
With these theoretical advances, the new data on hypernuclei will
definitely provide valuable input to pin down the underlying interactions. Eventually, the hypernuclear data 
could be directly utilized in fits of interaction parameters. 

However, a direct use of hypernuclear data 
requires solving the hypernuclear many-body 
problem many \\ times and, therefore, calls for 
a very efficient calculation scheme. 
Several methods have been employed in the past 
to study hypernuclei. For local interactions, configuration 
space methods, e.g. hyperspherical harmonics, 
Green's function Monte Carlo, expansion in Gaussians or stochastic variational method (SVM), have been successfully used 
to predict properties of light hypernuclei
\cite{Nemura:2002fu,Hiyama:2010zzb,Contessi:2018qnz,Contessi:2019csf,Lonardoni:2014bwa}.
For very light systems, that goal can be likewise achieved by 
solving the Faddeev- or Yakubovsky equations in 
momentum space \cite{Miyagawa:1993rd,Miyagawa:1995sf,Nogga:2001ef,Nogga:2013pwa,Haidenbauer:2019boi,Le:2019gjp}. 
Those methods allow one also to deal with non-local two-body
interactions, but it is difficult to extend the approaches to larger systems.
Alternatively, shell model calculations have been a  
quite successful tool to understand properties of hypernuclei, 
in particular the energy level splittings   \cite{Gal:1978jt,Millener:1985yf,Millener:2012zz,Gal:2013ooa}. 
However, that approach requires specific effective interactions that 
are not easily related to free-space baryon-baryon 
interactions. The same disadvantage also holds for density 
functional approaches, which have 
been applied to rather complex hypernuclei  \cite{Mei:2015pca,Lu:2014wta}. Recently, 
nuclear lattice  effective field theory (NLEFT) has been extended to hypernuclei 
using the impurity lattice Monte Carlo technique \cite{Frame:2020mvv}. Although 
this first study has been performed with somewhat simplified (spin-independent) 
interactions, that method promises the application of free-space 
interactions up to medium-heavy hypernuclei. 

One specifically interesting approach to tackle bound baryon systems 
is the no-core shell model (NCSM) \cite{BARRETT2013131}. An essential tool is here the
representation in terms of a harmonic oscillator (HO) basis. 
There are several variants of the approach. In most applications 
so far, a single-particle Slater-determinant basis has been chosen. 
This realization has been very successfully employed for studying ordinary nuclei and even hypernuclei  
\cite{Gazda:2015qyt,Wirth:2014apa,Wirth:2016iwn,Wirth:2017lso},  especially, when the
so-called importance truncation is implemented \cite{Wirth:2014apa,Wirth:2016iwn,Wirth:2017lso}. 
Highly accurate results for binding energies,
excitation energies and even radii have been obtained. Generally, 
the problem becomes very high dimensional,  
not least because the center-of-mass (CM) motion cannot be 
separated off and because angular momentum and isospin 
conservation cannot be exploited to limit the basis size. 

Such a complication can be avoided by using a Jacobi relative 
coordinate basis. This, however, requires 
a very tedious antisymmetrization for the nucleonic states \cite{Navratil:1999pw,Liebig:2015kwa}.
Nevertheless, the method can be  advantageous when many calculations are required for variations of 
the underlying interactions, e.g. in fitting procedures,
since the antisymmetrization and other preparatory steps can be accomplished independently of the interactions. 
The final step of the calculation itself can then be much more 
efficiently performed than in the standard NCSM so that it becomes 
feasible to solve the problem hundreds or even thousands of times 
or with limited computational resources. 
The work of Gazda et al. \cite{Gazda:2015qyt,Wirth:2014apa} has already been 
employing this Jacobi NCSM (J-NCSM) for $s$-shell hypernuclei. 
It is the main aim of the present work to extend the J-NCSM approach 
to $p$-shell hypernuclei. The new approach 
is then used to study in more detail the $^{4}_{\Lambda}$He, $^{5}_{\Lambda}$He, $^{6}_{\Lambda}$Li and $^{7}_{\Lambda}$Li
systems based on the next-to-leading order (NLO) hyperon-nucleon 
(YN) interaction derived within chiral effective field theory (EFT)
 \cite{Polinder:2006zh,Haidenbauer:2013oca,Haidenbauer:2019boi}. 
For interactions from chiral EFT, it 
is possible to obtain reliable uncertainty estimates of the results \cite{Haidenbauer:2019boi,Le:2019gjp}, 
utilizing different orders of the chiral expansion and/or by
exploiting the regulator (cutoff) dependence of these interactions 
(where the latter method provides only a lower limit for the error). 
For ordinary nuclei, such estimates are 
now regularly performed \cite{Binder:2018pgl,Epelbaum:2019zqc}. 
 
As usual, the NCSM 
requires a further softening of the nucleon-nucleon (NN) and YN interactions. To this aim, we apply the 
similarity renormalization group (SRG) to the NN and YN potentials \cite{Wegner:1994ann,Bogner:2007hn}. This method has the advantage that 
an effective interaction can be systematically derived from the starting NN and YN interactions, which  can then be equally well
employed in momentum space and HO space. 
The SRG evolution gives rise also to so-called {\it induced} three-body and many-body forces. 
In the present study, we will not take into account such induced 
many-body forces (for the application of 
the SRG induced YNN forces see \cite{Wirth:2016iwn,Wirth:2017lso,Wirth:2019cpp}). 
Therefore, a part of this work is devoted to study the SRG  
dependence of the binding energies, excitation energies and $\Lambda$-separation energies. 

In Section~\ref{sec:basis}, we start with a definition of our basis states based on the totally antisymmetrized 
nucleonic states defined in \cite{Liebig:2015kwa}. Practical calculations can only be performed when 
the transition matrix elements to states that single out NN or YN pairs are known. The calculation of these 
matrix elements is explained in detail in Section~\ref{sec:transition}. This already concludes the description 
of the Jacobi NCSM. As mentioned above, for explicit calculations, 
we, however, also need soft interactions. In Section~\ref{sec:srg},
we therefore discuss the basic features of chiral interactions and their SRG evolvement  including the impact
on the binding energy for $^{3}_{\Lambda}$H when the SRG-induced 
three-baryon force (3BF)  is neglected. For this study, we will make use of  solutions based on
the Faddeev equations.  The application of the Jacobi NCSM then follows in Section~\ref{sec:results}.
We first present  a detailed benchmark for $^{4}_{\Lambda}$H/$^{4}_{\Lambda}$He to Yakubovsky results and  then continue
towards $A=5$ to $7$ hypernuclei. Our conclusions are finally given in Section~\ref{sec:concl}. 
Some technicalities are relegated to the appendices.

\section{NCSM basis in Jacobi coordinates}
\label{sec:basis}
The  translationally invariant many-body Hamiltonian of a system consisting of  $(A-1)$ nucleons and a
single-strangeness  hyperon $Y$ ($Y=\Lambda$ or $\Sigma$)
in  Jacobi  relative coordinates can be written as follows
\begin{eqnarray} \label{eq:hamiltonian}
H &  = & H^{S=0}  + H^{S=-1} \cr
&= & \sum_{i < j=1}^{A-1} \Big( \frac{2p^{2}_{ij}}{M(t_Y)} + V^{NN}_{ij} \Big)\nonumber \\[2pt]
 & & + \sum_{i=1}^{A-1} 
\Big( \frac{m_N + m(t_Y)}{M(t_Y)} \frac{p^2_{iY}}{2\mu_{NY}}  +  V^{YN}_{iY}\cr
& & \qquad \qquad +\frac{1}{A-1} \big(m(t_Y) - m_{\Lambda}\big) \Big). 
\end{eqnarray}
Here, $m_N$, $m(t_Y)$ and $\mu_{NY}$ are  nucleon-, hyperon-, and their reduced masses, respectively, which we define by 
$m_N =2 m_n m_p/(m_n + m_p)$, $m(t_Y =0) = m_{\Lambda}$, and $m(t_Y =1) = (m_{\Sigma^+} + m_{\Sigma^{-}} + m_{\Sigma^{0}})/3$.
For simplicity,  we  assume isospin symmetry. A generalization to unequal   masses 
within the isospin multiplet of nucleons and of $\Sigma$'s  is  straightforward but will not be considered here. 
The total rest mass of the system,   $M(t_Y)=(A-1)m_N + m(t_Y)$,  
depends  explicitly  on the hyperon isospin $t_Y$  because an explicit $\Lambda$-$\Sigma$ conversion is allowed. 
The term  $m(t_Y) - m_{\Lambda}$ then 
  accounts for the  difference in the rest masses of the two hyperons. The relative Jacobi momenta of NN and YN pairs,
\begin{equation}
    p_{ij} = \frac{1}{2}(k_i - k_j),
\end{equation}
and
\begin{equation}
    p_{iY } = \frac{m(t_Y)}{m_N + m(t_Y)} k_{i}  - \frac{m_N}{m_N + m(t_Y)} k_Y 
\end{equation}  
are linear combinations of the momenta $k_i$ and $k_Y$ of 
the i-th nucleon and the hyperon, respectively.    $V^{NN}_{ij} $ and  $V^{YN}_{iY}$ are the corresponding 
  NN and YN potentials.
  
Since hyperons ($\Lambda$, $\Sigma$) and nucleons are 
distinguishable,    hypernuclear basis functions,  denoted as \\  $|\alpha^{* (Y)}\rangle$,  can be    formed
by  coupling  the hyperon  HO  states $|Y\rangle$,  which describe the relative motion of   a single
hyperon $Y$ with respect to the CM of the 
 $(A-1)$N core,    to the fully antisymmetrized  states  of  the core $|\alpha_{(A-1)N}\rangle$ 
 \begin{eqnarray} \label{eq:hbasis}
& &  \big |\alpha^{* (Y)}(\mathcal{N}J T)\big \rangle =   |\alpha_{(A-1)N}\rangle \otimes  |Y\rangle \nonumber \\[3pt]
 & & =  | \mathcal{N}{J}{T}, \alpha_{(A-1)N}\, n_{Y}I_Y t_Y ; \nonumber \\[3pt]
&    & \qquad (J_{A-1} (l_Y s_Y)I_Y){J}, 
(T_{A-1} t_Y) {T} \rangle
 \equiv | \begin{tikzpicture}[baseline={([yshift=-.5ex]current bounding box.center)},scale=0.6]
                \filldraw[color=black, ultra thick, fill=gray ]  (0.,0.) circle(0.22cm);
                \filldraw[red]  (0.6,0.) circle (0.7mm) ;
                \draw[baseline,thick, -]  (0.22,0.) -- (0.53,0.);
                \end{tikzpicture}  \rangle,
\end{eqnarray}
where $\alpha_{(A-1)N}$ stands for  a complete  set of all necessary  quantum numbers characterizing the fully
antisymmetrized states of  an 
$(A-1)$N system: the total HO energy quantum number $\mathcal{N}_{{A-1}}$, total angular momentum ${J}_{{A-1}} $, 
isospin ${T}_{{A-1}}$,  and  the state indices   $\zeta_{A-1}$  (that distinguish different $|\alpha_{(A-1)N}\rangle$
states with the same
set of  $\mathcal{N}_{A-1}, J_{A-1}$  and $T_{A-1}$).  These  antisymmetrized  states for $A \ge 4$
systems are computed iteratively  starting from the naturally  antisymmetrized basis 
 for two nucleons, for more 
detail we refer to Ref.~\cite{Liebig:2015kwa}.
 Here, the superscript $(*Y)$  represents  the
separation of the  hyperon $Y$ from the $(A-1)$N core.   The hyperon states
 $|Y\rangle$  are  described by a  similar set of   quantum numbers:
the HO energy quanta $n_Y$, the orbital angular  momentum $l_Y$ and spin   $s_Y$ which  are coupled to 
the  relative angular momentum $I_Y$, and the  isospin $t_Y$ as well.  The last line in Eq.~(\ref{eq:hbasis})
defines the ordering in which  the quantum numbers of  the two subclusters are combined to form the 
total angular momentum and  total isospin of the system,  $J$ and  $T$,   respectively, whose values are
given by the physical state of interest. Also, for  practical realization, the total HO quantum numbers
$\mathcal{N}$  of the basis states   are  constrained by the maximum number of the single-particle oscillators
$\mathcal{N}_{max}$ (also referred to as the model space size),  i.e.   $\mathcal{N} =
\mathcal{N}_{A-1} +2n_Y +l_Y \leq  \mathcal{N}_{max}$.  
The state index $\zeta$  that distinguishes different  basis states $|\alpha^{* (Y)}\rangle$ with the same
$\mathcal{N}, J $  and  $T$  is omitted for simplifying  the notation. Finally,  on the right-hand
side of Eq.~(\ref{eq:hbasis}), the  graphical representation of  the basis is shown.  The small  red circle denotes 
a hyperon spectator while  the big black circle represents  the  system of $(A-1)$N.

\section{Separation of   NN and  YN pairs}
With the basis defined in Eq.~(\ref{eq:hbasis}), the matrix elements of the Hamiltonian in Eq.~(\ref{eq:hamiltonian})  now read
 \begin{eqnarray} \label{eq:matrixelement}
\langle \alpha^{*(Y)} | H | \alpha^{\prime *(Y)} \rangle & = & \, \langle \alpha^{*(Y)} | H^{S=0} |
 \alpha^{\prime *(Y)} \rangle \nonumber \\[3pt]
& &  +\,  \langle \alpha^{*(Y)} | H^{S=-1} | \alpha^{ \prime *(Y)} \rangle.
\end{eqnarray}
The  basis states $|\alpha^{*(Y)}\rangle$ are however not suitable for evaluating   the   $H^{S=0}$ and 
$H^{S=-1}$ matrix elements as they do not depend explicitly on the relative coordinates of the involved NN or YN pairs.
To facilitate the  evaluation of Eq.~(\ref{eq:matrixelement}),
we expand  $|\alpha^{*(Y)}\rangle$ in  
two additional bases of intermediate states  $ |\big(\alpha^{*(2)}\big)^{*(Y)}\rangle $  and
$|\alpha^{*(YN)} \rangle$ that  explicitly single out the active  NN or a YN pairs, respectively. Also
 the  superscripts   represent  subsystems that are separated  out. 
Clearly, the former states $ |\big(\alpha^{*(2)}\big)^{*(Y)}\rangle $ are needed for evaluating the first
part  in  Eq.~(\ref{eq:matrixelement})  involving $H^{S=0}$,  while the latter  ones are   necessary  for the
evaluation of  the second part that involves $H^{S=-1}$.

The  first  set  of auxiliary states  $ |\big(\alpha^{*(2)}\big)^{*(Y)}\rangle $   can be directly  constructed
by  coupling  the  hyperon states $|Y\rangle$, depending on  Jacobi coordinates  of $Y$ relative to the
CM of  (A-1)N,   to the  $(A-1)$N states  that consist of 
   antisymmetrized subclusters of $(A-3)$N  and 2N. In the notation of Ref.~\cite{Liebig:2015kwa}, this reads 
%
\begin{eqnarray}\label{eq:NNsingleout}
& & \big |\big(\alpha^{*(2)}\big)^{*(Y)}\big \rangle =  | \alpha^{*(2)}_{(A-1)} \rangle \otimes |Y \rangle \nonumber \\[4pt]
& & \quad =  \big | {\mathcal{\tilde{N}}} {J} {T}, \alpha^{*(2)}_{(A-1)}\, \tilde{n}_{Y} \tilde{I}_Y \tilde{t}_Y ; \nonumber \\
&  & \qquad (J^{*(2)}_{A-1} (\tilde{l}_Y  {s}_Y)\tilde{I}_Y) {J}, 
(T^{*(2)}_{A-1} \tilde{t}_Y) {T} \big \rangle  \equiv \big| \,\begin{tikzpicture}[baseline={([yshift=-0.2ex]current bounding box.center)},scale=0.65]
                \filldraw[color=black, ultra thick, fill=gray ]  (0.,0.) circle(0.19cm);
                \filldraw[red]  (-0.4,-0.45)  circle (0.7mm) ;
                \filldraw[black] (-0.74,0.24)   circle(0.55mm); 
                \filldraw[black] (-0.74,-0.24)  circle(0.55mm); 
                \draw[baseline,thick, -]  (-0.74,-0.24) -- (-0.74,0.24);
                \draw[baseline,thick, -]  (-0.74,0.) -- (-0.2,0.0);
                \draw[baseline,thick, -]  (-0.4,-0.38) -- (-0.4,0.0);
                \end{tikzpicture} \big\rangle.
\end{eqnarray}
Here,  $ \alpha^{*(2)}_{(A-1)} $  stands for the total HO energy quantum number $\mathcal{N}_{\alpha^{*(2)}_{(A-1)}},$
the total angular momentum  $J^{*(2)}_{A-1}$, isospin $T^{*(2)}_{A-1}$ and  state index $\zeta^{*(2)}_{A-1}$, as
introduced in \cite{Liebig:2015kwa}. Naturally, the
total HO energy quantum number $\tilde{\mathcal{N}}$ in Eq.~(\ref{eq:NNsingleout}) is 
also restricted by $\tilde{\mathcal{N}}
\leq \mathcal{N}_{max}$. 
 With the graphical representations of  $|\big(\alpha^{*(2)}\big)^{*(Y)} \rangle$ and
 $|\alpha^{*(Y)}\rangle$, one can quickly relate the expansion coefficients  
$ \big \langle \alpha^{*(Y)} | \big(\alpha^{*(2)}\big)^{*(Y)} \big\rangle $
 to  the transition coefficients of the ${(A-1)}$N system  $
  \langle \begin{tikzpicture}[baseline={([yshift=-.5ex]current bounding box.center)},scale=0.5]
                \filldraw[color=black, ultra thick, fill=gray ]  (0.,0.) circle(0.22cm);
                \end{tikzpicture}
|\begin{tikzpicture}[baseline={([yshift=-0.5ex]current bounding box.center)},scale=0.5]
                \filldraw[color=black, ultra thick, fill=gray ]  (0.,0.) circle(0.19cm);
                \filldraw[black] (-0.74,0.24)   circle(0.55mm); 
                \filldraw[black] (-0.74,-0.24)  circle(0.55mm); 
                \draw[baseline,thick, -]  (-0.74,-0.24) -- (-0.74,0.24);
                \draw[baseline,thick, -]  (-0.74,0.) -- (-0.2,0.0);
                \end{tikzpicture} \big\rangle_{A-1},$
              \begin{eqnarray} \label{eq:overlapNN}
 \langle \alpha^{*(Y)} | \big(\alpha^{*(2)}\big)^{*(Y)}\rangle  & = & 
\langle \begin{tikzpicture}[baseline={([yshift=-.5ex]current bounding box.center)},scale=0.6]
                \filldraw[color=black, ultra thick, fill=gray ]  (0.,0.) circle(0.22cm);
                \filldraw[red]  (0.6,0.) circle (0.7mm) ;
                \draw[baseline,thick, -]  (0.22,0.) -- (0.53,0.);
                \end{tikzpicture} 
|\begin{tikzpicture}[baseline={([yshift=-0.2ex]current bounding box.center)},scale=0.6]
                \filldraw[color=black, ultra thick, fill=gray ]  (0.,0.) circle(0.19cm);
                \filldraw[red]  (-0.4,-0.45)  circle (0.7mm) ;
                \filldraw[black] (-0.74,0.24)   circle(0.55mm); 
                \filldraw[black] (-0.74,-0.24)  circle(0.55mm); 
                \draw[baseline,thick, -]  (-0.74,-0.24) -- (-0.74,0.24);
                \draw[baseline,thick, -]  (-0.74,0.) -- (-0.2,0.0);
                \draw[baseline,thick, -]  (-0.4,-0.38) -- (-0.4,0.0);
                \end{tikzpicture} \rangle \nonumber \\[2pt]
  & = & \delta_{spectator}
 \langle \begin{tikzpicture}[baseline={([yshift=-.5ex]current bounding box.center)},scale=0.6]
                \filldraw[color=black, ultra thick, fill=gray ]  (0.,0.) circle(0.22cm);
                \end{tikzpicture}
|\begin{tikzpicture}[baseline={([yshift=-0.5ex]current bounding box.center)},scale=0.6]
                \filldraw[color=black, ultra thick, fill=gray ]  (0.,0.) circle(0.19cm);
                \filldraw[black] (-0.74,0.24)   circle(0.55mm); 
                \filldraw[black] (-0.74,-0.24)  circle(0.55mm); 
                \draw[baseline,thick, -]  (-0.74,-0.24) -- (-0.74,0.24);
                \draw[baseline,thick, -]  (-0.74,0.) -- (-0.2,0.0);
                \end{tikzpicture} \big\rangle_{A-1},
\end{eqnarray}
for which an explicit expressions has been derived in \cite{Liebig:2015kwa,LiebigPhD:2013}. 
The Kronecker symbol $\delta_{spectator}$ is to ensure
 the conservation of the quantum numbers of the hyperon  and the $(A-1)$N  system, 
\begin{align*}
\begin{split}    
&\delta_{spectator} = \delta_{\mathcal{N} \mathcal{\tilde{N}}}\delta_{Y} \delta_{core},\\[3pt]
& \delta_{Y }   =  \delta_{n_Y \tilde{n}_Y}  \delta_{l_Y \tilde{l}_Y} \delta_{I_Y \tilde{I}_Y}  \delta_{t_Y \tilde{t}_Y},\\
&\delta_{core}  =  \delta_{\mathcal{N}_{A-1}  \mathcal{N}^{*(2)}_{A-1} } \delta_{{J_{A-1}}  {J^{*(2)}_{A-1}} } 
\delta_{{T_{A-1}}  {T^{*(2)}_{A-1}} }.  
\end{split}
\end{align*}

Hence, the 
matrix elements of the nucleonic Hamiltonian  $\langle \alpha^{*(Y)} | H^{S=0} | \alpha^{\prime *(Y)} \rangle$ now become
\begin{eqnarray} \label{eq:evalHNN}
& & \langle \alpha^{*(Y)}  |   H^{S=0} |  \alpha^{\prime *(Y)}  \rangle \nonumber \\  
&  & \quad = \langle \begin{tikzpicture}[baseline={([yshift=-.5ex]current bounding box.center)},scale=0.6]
                \filldraw[color=black, ultra thick, fill=gray ]  (0.,0.) circle(0.22cm);
                \filldraw[red]  (0.6,0.) circle (0.7mm) ;
                \draw[baseline,thick, -]  (0.22,0.) -- (0.53,0.);
                \end{tikzpicture} 
|\begin{tikzpicture}[baseline={([yshift=-0.2ex]current bounding box.center)},scale=0.6]
                \filldraw[color=black, ultra thick, fill=gray ]  (0.,0.) circle(0.19cm);
                \filldraw[red]  (-0.4,-0.45)  circle (0.7mm) ;
                \filldraw[black] (-0.74,0.24)   circle(0.55mm); 
                \filldraw[black] (-0.74,-0.24)  circle(0.55mm); 
                \draw[baseline,thick, -]  (-0.74,-0.24) -- (-0.74,0.24);
                \draw[baseline,thick, -]  (-0.74,0.) -- (-0.2,0.0);
                \draw[baseline,thick, -]  (-0.4,-0.38) -- (-0.4,0.0);
                \end{tikzpicture} \rangle
\langle \begin{tikzpicture}[baseline={([yshift=-0.2ex]current bounding box.center)},scale=0.6]
                \filldraw[color=black, ultra thick, fill=gray ]  (0.,0.) circle(0.19cm);
                \filldraw[red]  (0.4,-0.45)  circle (0.7mm) ;
                \filldraw[black] (0.74,0.24)   circle(0.55mm); 
                \filldraw[black] (0.74,-0.24)  circle(0.55mm); 
                \draw[baseline,thick, -]  (0.74,-0.24) -- (0.74,0.24);
                \draw[baseline,thick, -]  (0.74,0.) -- (0.2,0.0);
                \draw[baseline,thick, -]  (0.4,-0.38) -- (0.4,0.0);
                \end{tikzpicture} 
| H^{S=0} | 
\begin{tikzpicture}[baseline={([yshift=-0.2ex]current bounding box.center)},scale=0.6]
                \filldraw[color=black, ultra thick, fill=gray ]  (0.,0.) circle(0.19cm);
                \filldraw[red]  (-0.4,-0.45)  circle (0.7mm) ;
                \filldraw[black] (-0.74,0.24)   circle(0.55mm); 
                \filldraw[black] (-0.74,-0.24)  circle(0.55mm); 
                \draw[baseline,thick, -]  (-0.74,-0.24) -- (-0.74,0.24);
                \draw[baseline,thick, -]  (-0.74,0.) -- (-0.2,0.0);
                \draw[baseline,thick, -]  (-0.4,-0.38) -- (-0.4,0.0);
                \end{tikzpicture} \rangle
\langle \begin{tikzpicture}[baseline={([yshift=-0.2ex]current bounding box.center)},scale=0.6]
                \filldraw[color=black, ultra thick, fill=gray ]  (0.,0.) circle(0.19cm);
                \filldraw[red]  (0.4,-0.45)  circle (0.7mm) ;
                \filldraw[black] (0.74,0.24)   circle(0.55mm); 
                \filldraw[black] (0.74,-0.24)  circle(0.55mm); 
                \draw[baseline,thick, -]  (0.74,-0.24) -- (0.74,0.24);
                \draw[baseline,thick, -]  (0.74,0.) -- (0.2,0.0);
                \draw[baseline,thick, -]  (0.4,-0.38) -- (0.4,0.0);
                \end{tikzpicture} |
  \begin{tikzpicture}[baseline={([yshift=-.5ex]current bounding box.center)},scale=0.6]
                \filldraw[color=black, ultra thick, fill=gray ]  (0.,0.) circle(0.22cm);
                \filldraw[red]  (-0.6,0.) circle (0.7mm) ;
                \draw[baseline,thick, -]  (-0.53,0.) -- (-0.22,0.);
                \end{tikzpicture} \rangle \nonumber \\[3pt]
& & \quad = \delta_{spectator} 
 \langle \begin{tikzpicture}[baseline={([yshift=-.5ex]current bounding box.center)},scale=0.6]
                \filldraw[color=black, ultra thick, fill=gray ]  (0.,0.) circle(0.22cm);
                \end{tikzpicture} 
|\begin{tikzpicture}[baseline={([yshift=-0.5ex]current bounding box.center)},scale=0.6]
                \filldraw[color=black, ultra thick, fill=gray ]  (0.,0.) circle(0.19cm);
                \filldraw[black] (-0.74,0.24)   circle(0.55mm); 
                \filldraw[black] (-0.74,-0.24)  circle(0.55mm); 
                \draw[baseline,thick, -]  (-0.74,-0.24) -- (-0.74,0.24);
                \draw[baseline,thick, -]  (-0.74,0.) -- (-0.2,0.0);
                \end{tikzpicture} \rangle
\langle \begin{tikzpicture}[baseline={([yshift=-0.5ex]current bounding box.center)},scale=0.6]
                \filldraw[color=black, ultra thick, fill=gray ]  (0.,0.) circle(0.19cm);
                \filldraw[black] (0.74,0.24)   circle(0.55mm); 
                \filldraw[black] (0.74,-0.24)  circle(0.55mm); 
                \draw[baseline,thick, -]  (0.74,-0.24) -- (0.74,0.24);
                \draw[baseline,thick, -]  (0.74,0.) -- (0.2,0.0);
                \end{tikzpicture} 
| H^{S=0} | 
\begin{tikzpicture}[baseline={([yshift=-0.5ex]current bounding box.center)},scale=0.6]
                \filldraw[color=black, ultra thick, fill=gray ]  (0.,0.) circle(0.19cm);
                \filldraw[black] (-0.74,0.24)   circle(0.55mm); 
                \filldraw[black] (-0.74,-0.24)  circle(0.55mm); 
                \draw[baseline,thick, -]  (-0.74,-0.24) -- (-0.74,0.24);
                \draw[baseline,thick, -]  (-0.74,0.) -- (-0.2,0.0);
                \end{tikzpicture} \rangle
\langle \begin{tikzpicture}[baseline={([yshift=-0.5ex]current bounding box.center)},scale=0.6]
                \filldraw[color=black, ultra thick, fill=gray ]  (0.,0.) circle(0.19cm);
                \filldraw[black] (0.74,0.24)   circle(0.55mm); 
                \filldraw[black] (0.74,-0.24)  circle(0.55mm); 
                \draw[baseline,thick, -]  (0.74,-0.24) -- (0.74,0.24);
                \draw[baseline,thick, -]  (0.74,0.) -- (0.2,0.0);
                \end{tikzpicture} |
  \begin{tikzpicture}[baseline={([yshift=-.5ex]current bounding box.center)},scale=0.6]
                \filldraw[color=black, ultra thick, fill=gray ]  (0.,0.) circle(0.22cm);
                \end{tikzpicture} \rangle, 
\end{eqnarray}
with summations over intermediate states $ 
|\begin{tikzpicture}[baseline={([yshift=-0.2ex]current bounding box.center)},scale=0.5]
                \filldraw[color=black, ultra thick, fill=gray ]  (0.,0.) circle(0.19cm);
                \filldraw[red]  (-0.4,-0.45)  circle (0.7mm) ;
                \filldraw[black] (-0.74,0.24)   circle(0.55mm); 
                \filldraw[black] (-0.74,-0.24)  circle(0.55mm); 
                \draw[baseline,thick, -]  (-0.74,-0.24) -- (-0.74,0.24);
                \draw[baseline,thick, -]  (-0.74,0.) -- (-0.2,0.0);
                \draw[baseline,thick, -]  (-0.4,-0.38) -- (-0.4,0.0);
                \end{tikzpicture} \rangle$
being implied.  The remaining  unknown term  in Eq.~(\ref{eq:evalHNN})  is 
simply the matrix elements of  $H^{S=0}$ in the basis of \break $(A-1)$ nucleons.
 
Similarly, in order to construct   the intermediate \\  states $|\alpha^{*(YN)} \rangle$,    one combines the
states describing a YN pair,  $|YN \rangle$, with the antisymmetrized basis of an  $(A-2)$N system,
$|\alpha_{(A-2)}\rangle$
\begin{eqnarray}
\label{eq:YNsingleout}
& & |\alpha^{*(YN)} \rangle = |\alpha_{YN} \rangle \otimes |\alpha_{A-2}\rangle \nonumber \\[2pt]
& & \quad = | \mathcal{N} \mathcal{J} \mathcal{T}, \alpha_{YN}\, n_{\lambda} \lambda \,\alpha_{A-2};  ((l_{YN}(s_Y s_N)S_{YN}) \nonumber \\[2pt]
& & \qquad J_{YN} (\lambda J_{A-2})I_{\lambda})\mathcal{J}, ( (t_Y t_N) T_{YN}T_{A-2}) \mathcal{T} \rangle \nonumber \\
& & \quad \equiv \big| \,\begin{tikzpicture}[baseline={([yshift=-0.5ex]current bounding box.center)},scale=0.6]
                \filldraw[color=black, ultra thick, fill=gray ]  (0.,0.) circle(0.20cm);
                \filldraw[red] (-0.7,0.24)   circle(0.7mm); 
                \filldraw[black] (-0.7,-0.24)  circle(0.55mm); 
                \draw[baseline,thick, -]  (-0.7,-0.24) -- (-0.7,0.17);
                \draw[baseline,thick, -]  (-0.7,0.) -- (-0.2,0.0);
                \end{tikzpicture} \big\rangle.
\end{eqnarray}
Again, $|\alpha_{YN}\rangle$ and $|\alpha_{A-2}\rangle$ represent the complete sets  of 
quantum numbers  characterizing the states of the two-body hyperon-nucleon and the $(A-2)$N 
subsystems. Note that,  in contrast to two-nucleon states, there is no antisymmetry requirement
for  $|\alpha_{YN}\rangle$. The
relative motion of the $(A-2)$N   cluster  with respect to the separated out 
YN pair is specified   by the HO energy number $n_{\lambda}$ and  the  orbital 
angular momentum ${\lambda}$. For evaluating  the overlap $\langle \alpha^{*(Y)} | \alpha^{*(YN)}\rangle$, we   need to exploit another set of
auxiliary states $|\big(\alpha^{*(1)}\big)^{*(Y)}\rangle$ in which  a hyperon and  a nucleon are explicitly
singled out 
\begin{eqnarray} \label{eq:signleoutNandY}
\big| \big (\alpha^{*(1)}\big )^{*(Y)} \big \rangle &= &|\alpha^{*(1)}_{A-1}\rangle \otimes |Y\rangle \nonumber \\[3pt]
 & = &| \tilde{\mathcal{N}} {{J}}  {{T}}, \alpha^{*(1)}_{(A-1)} \, n_Y I_Y \tilde{t}_Y; \nonumber \\[3pt]
 & &   (J^{*(1)}_{A-1} (l_Y s_Y) I_Y)  {J}, (T^{*(1)}_{A-1} \tilde{t}_Y) {T} \rangle \nonumber \\[3pt]
  &\equiv& \big| \begin{tikzpicture}[baseline={([yshift=-.2ex]current bounding box.center)},scale=0.6]
                \filldraw[color=black, ultra thick, fill=gray ]  (0.,0.) circle(0.22cm);
                \filldraw[black]  (0.7,0.) circle (0.5mm) ;
                \draw[baseline,thick, -]  (0.22,0.) -- (0.65,0.);
                \filldraw[red]  (0.4,-0.4) circle (0.7mm) ;
                \draw[baseline,thick, -]  (0.4,-0.33) -- (0.4,0.);
                \end{tikzpicture} \big\rangle.
\end{eqnarray}
With the help of Eq.~(\ref{eq:signleoutNandY}), the transition coefficients $\langle \alpha^{*(Y)} | \alpha^{*(YN)}\rangle$
can  be  computed in  two steps   as follows
\begin{eqnarray} \label{eq:overlapYN}
\langle \alpha^{*(Y)} | \alpha^{*(YN)}\rangle & = &
 \langle \begin{tikzpicture}[baseline={([yshift=-.5ex]current bounding box.center)},scale=0.6]
                \filldraw[color=black, ultra thick, fill=gray ]  (0.,0.) circle(0.22cm);
                \filldraw[red]  (0.6,0.) circle (0.7mm) ;
                \draw[baseline,thick, -]  (0.22,0.) -- (0.53,0.);
                \end{tikzpicture} |  
 \begin{tikzpicture}[baseline={([yshift=-.2ex]current bounding box.center)},scale=0.6]
                \filldraw[color=black, ultra thick, fill=gray ]  (0.,0.) circle(0.22cm);
                \filldraw[black]  (0.7,0.) circle (0.5mm) ;
                \draw[baseline,thick, -]  (0.22,0.) -- (0.65,0.);
                \filldraw[red]  (0.4,-0.4) circle (0.7mm) ;
                \draw[baseline,thick, -]  (0.4,-0.33) -- (0.4,0.);
                \end{tikzpicture} \rangle 
\langle
\begin{tikzpicture}[baseline={([yshift=-.2ex]current bounding box.center)},scale=0.6]
                \filldraw[color=black, ultra thick, fill=gray ]  (0.,0.) circle(0.22cm);
                \filldraw[black]  (0.7,0.) circle (0.5mm) ;
                \draw[baseline,thick, -]  (0.22,0.) -- (0.65,0.);
                \filldraw[red]  (0.4,-0.4) circle (0.7mm) ;
                \draw[baseline,thick, -]  (0.4,-0.33) -- (0.4,0.);
                \end{tikzpicture}   | 
\begin{tikzpicture}[baseline={([yshift=-0.5ex]current bounding box.center)},scale=0.6]
                \filldraw[color=black, ultra thick, fill=gray ]  (0.,0.) circle(0.20cm);
                \filldraw[red] (-0.7,0.24)   circle(0.7mm); 
                \filldraw[black] (-0.7,-0.24)  circle(0.55mm); 
                \draw[baseline,thick, -]  (-0.7,-0.24) -- (-0.7,0.17);
                \draw[baseline,thick, -]  (-0.7,0.) -- (-0.2,0.0);
                \end{tikzpicture} \rangle \nonumber \\
                & = & \delta_{spectator} 
 \langle \begin{tikzpicture}[baseline={([yshift=-.5ex]current bounding box.center)},scale=0.6]
                \filldraw[color=black, ultra thick, fill=gray ]  (0.,0.) circle(0.22cm);
                \end{tikzpicture} |
 \begin{tikzpicture}[baseline={([yshift=-.5ex]current bounding box.center)},scale=0.6]
                \filldraw[color=black, ultra thick, fill=gray ]  (0.,0.) circle(0.22cm);
                \filldraw[black]  (0.7,0.) circle (0.5mm) ;
                \draw[baseline,thick, -]  (0.22,0.) -- (0.65,0.);
                \end{tikzpicture} \rangle_{A-1} 
          \langle \begin{tikzpicture}[baseline={([yshift=-.2ex]current bounding box.center)},scale=0.6]
                \filldraw[color=black, ultra thick, fill=gray ]  (0.,0.) circle(0.22cm);
                \filldraw[black]  (0.7,0.) circle (0.5mm) ;
                \draw[baseline,thick, -]  (0.22,0.) -- (0.65,0.);
                \filldraw[red]  (0.4,-0.4) circle (0.7mm) ;
                \draw[baseline,thick, -]  (0.4,-0.33) -- (0.4,0.);
                \end{tikzpicture}  |
\begin{tikzpicture}[baseline={([yshift=-0.5ex]current bounding box.center)},scale=0.6]
                \filldraw[color=black, ultra thick, fill=gray ]  (0.,0.) circle(0.20cm);
                \filldraw[red] (-0.7,0.24)   circle(0.7mm); 
                \filldraw[black] (-0.7,-0.24)  circle(0.55mm); 
                \draw[baseline,thick, -]  (-0.7,-0.24) -- (-0.7,0.17);
                \draw[baseline,thick, -]  (-0.7,0.) -- (-0.2,0.0);
                \end{tikzpicture} \rangle.
\end{eqnarray}
Here also an explicit summation over the auxiliary states  $| \big (\alpha^{*(1)}\big )^{*(Y)}  \rangle = |  
 \begin{tikzpicture}[baseline={([yshift=-.2ex]current bounding box.center)},scale=0.6]
                \filldraw[color=black, ultra thick, fill=gray ]  (0.,0.) circle(0.22cm);
                \filldraw[black]  (0.7,0.) circle (0.5mm) ;
                \draw[baseline,thick, -]  (0.22,0.) -- (0.65,0.);
                \filldraw[red]  (0.4,-0.4) circle (0.7mm) ;
                \draw[baseline,thick, -]  (0.4,-0.33) -- (0.4,0.);
                \end{tikzpicture} \rangle  $ is assumed.  
Clearly, the first overlap $  \langle \begin{tikzpicture}[baseline={([yshift=-.5ex]current bounding box.center)},scale=0.5]
                \filldraw[color=black, ultra thick, fill=gray ]  (0.,0.) circle(0.22cm);
                \filldraw[red]  (0.6,0.) circle (0.7mm) ;
                \draw[baseline,thick, -]  (0.22,0.) -- (0.53,0.);
                \end{tikzpicture} |  
 \begin{tikzpicture}[baseline={([yshift=-.2ex]current bounding box.center)},scale=0.5]
                \filldraw[color=black, ultra thick, fill=gray ]  (0.,0.) circle(0.22cm);
                \filldraw[black]  (0.7,0.) circle (0.5mm) ;
                \draw[baseline,thick, -]  (0.22,0.) -- (0.65,0.);
                \filldraw[red]  (0.4,-0.4) circle (0.7mm) ;
                \draw[baseline,thick, -]  (0.4,-0.33) -- (0.4,0.);
                \end{tikzpicture} \rangle $
is  essentially given by the coefficients of fractional parentage (cfp)
$ 
 \langle \begin{tikzpicture}[baseline={([yshift=-.5ex]current bounding box.center)},scale=0.5]
                \filldraw[color=black, ultra thick, fill=gray ]  (0.,0.) circle(0.22cm);
                \end{tikzpicture} |
 \begin{tikzpicture}[baseline={([yshift=-.5ex]current bounding box.center)},scale=0.5]
                \filldraw[color=black, ultra thick, fill=gray ]  (0.,0.) circle(0.22cm);
                \filldraw[black]  (0.7,0.) circle (0.5mm) ;
                \draw[baseline,thick, -]  (0.22,0.) -- (0.65,0.);
 \end{tikzpicture}\rangle_{A-1}$
of an $(A-1)$N system,  which  basically determine the  antisymmetrized basis of  $(A-1)$ nucleons in
terms of  the $| \alpha^{*(1)}_{(A-1)}\rangle$ states  \cite{Liebig:2015kwa}, and is therefore well known. 
Hence, only the second transition $ 
               \langle \begin{tikzpicture}[baseline={([yshift=-.2ex]current bounding box.center)},scale=0.5]
                \filldraw[color=black, ultra thick, fill=gray ]  (0.,0.) circle(0.22cm);
                \filldraw[black]  (0.7,0.) circle (0.5mm) ;
                \draw[baseline,thick, -]  (0.22,0.) -- (0.65,0.);
                \filldraw[red]  (0.4,-0.4) circle (0.7mm) ;
                \draw[baseline,thick, -]  (0.4,-0.33) -- (0.4,0.);
                \end{tikzpicture} |               
                \begin{tikzpicture}[baseline={([yshift=-0.5ex]current bounding box.center)},scale=0.5]
                \filldraw[color=black, ultra thick, fill=gray ]  (0.,0.) circle(0.20cm);
                \filldraw[red] (-0.7,0.24)   circle(0.7mm); 
                \filldraw[black] (-0.7,-0.24)  circle(0.55mm); 
                \draw[baseline,thick, -]  (-0.7,-0.24) -- (-0.7,0.17);
                \draw[baseline,thick, -]  (-0.7,0.) -- (-0.2,0.0);
                \end{tikzpicture} \rangle $
in Eq.~(\ref{eq:overlapYN}) needs to be taken care of.  This transition  is  
a transformation between different Jacobi coordinates and therefore given by the general coordinate
transformation formula derived in \cite{Liebig:2015kwa}. We  skip the detailed derivation  but
provide the final expression in \ref{Appsec:transition}. Finally, a summation over the intermediate states
$
|\begin{tikzpicture}[baseline={([yshift=-.2ex]current bounding box.center)},scale=0.5]
                \filldraw[color=black, ultra thick, fill=gray ]  (0.,0.) circle(0.22cm);
                \filldraw[black]  (-0.7,0.) circle (0.5mm) ;
                \draw[baseline,thick, -]  (-0.22,0.) -- (-0.7,0.);
                \filldraw[red]  (-0.4,-0.4) circle (0.7mm) ;
                \draw[baseline,thick, -]  (-0.4,-0.33) -- (-0.4,0.);
                \end{tikzpicture} \rangle
$ is carried out.  Let us again stress that  both, the transition coefficients
$\langle \alpha^{*(Y)} | \big(\alpha^{*(2)}\big)^{*(Y)}\rangle$
and $\langle \alpha^{*(Y)} | \alpha^{*(YN)} \rangle$,  are  independent of 
the HO frequency  (HO-$\omega$)  as well as of the interactions employed.  They  can  therefore be
prepared in advance  and   stored 
in the machine-independent HDF5 format so  that the parallel input and  output can be  performed most
efficiently. The corresponding files can be found at \cite{datapub}. 

Once the transition coefficients $\langle \alpha^{*(Y)} | \alpha^{*(YN)}\rangle$ are known, the single-strangeness
Hamiltonian matrix
elements $\langle \alpha^{*(Y)} | H^{S=-1}| \alpha^{\prime *(Y)}\rangle$  are  computed similarly as in  Eq.~(\ref{eq:evalHNN}):
\begin{eqnarray} \label{eq:matHYN}
& &  \langle \alpha^{*(Y)} |   H^{S=-1  }  |   \alpha^{\prime*(Y)}\rangle     \nonumber \\
& & \quad  = \langle \begin{tikzpicture}[baseline={([yshift=-.5ex]current bounding box.center)},scale=0.6]
                \filldraw[color=black, ultra thick, fill=gray ]  (0.,0.) circle(0.22cm);
                \filldraw[red]  (0.6,0.) circle (0.7mm) ;
                \draw[baseline,thick, -]  (0.22,0.) -- (0.53,0.);
                \end{tikzpicture} | 
\begin{tikzpicture}[baseline={([yshift=-0.5ex]current bounding box.center)},scale=0.6]
                \filldraw[color=black, ultra thick, fill=gray ]  (0.,0.) circle(0.20cm);
                \filldraw[red] (-0.7,0.24)   circle(0.7mm); 
                \filldraw[black] (-0.7,-0.24)  circle(0.55mm); 
                \draw[baseline,thick, -]  (-0.7,-0.24) -- (-0.7,0.17);
                \draw[baseline,thick, -]  (-0.7,0.) -- (-0.2,0.0);
                \end{tikzpicture} \rangle
\langle \begin{tikzpicture}[baseline={([yshift=-0.5ex]current bounding box.center)},scale=0.6]
                \filldraw[color=black, ultra thick, fill=gray ]  (0.,0.) circle(0.20cm);
                \filldraw[red] (0.7,0.24)   circle(0.7mm); 
                \filldraw[black] (0.7,-0.24)  circle(0.55mm); 
                \draw[baseline,thick, -]  (0.7,-0.24) -- (0.7,0.17);
                \draw[baseline,thick, -]  (0.7,0.) -- (0.2,0.0);
                \end{tikzpicture} | H^{S=-1} | 
\begin{tikzpicture}[baseline={([yshift=-0.5ex]current bounding box.center)},scale=0.6]
                \filldraw[color=black, ultra thick, fill=gray ]  (0.,0.) circle(0.20cm);
                \filldraw[red] (-0.7,0.24)   circle(0.7mm); 
                \filldraw[black] (-0.7,-0.24)  circle(0.55mm); 
                \draw[baseline,thick, -]  (-0.7,-0.24) -- (-0.7,0.17);
                \draw[baseline,thick, -]  (-0.7,0.) -- (-0.2,0.0);
                \end{tikzpicture} \rangle
\langle \begin{tikzpicture}[baseline={([yshift=-0.5ex]current bounding box.center)},scale=0.65]
                \filldraw[color=black, ultra thick, fill=gray ]  (0.,0.) circle(0.20cm);
                \filldraw[red] (0.7,0.24)   circle(0.7mm); 
                \filldraw[black] (0.7,-0.24)  circle(0.55mm); 
                \draw[baseline,thick, -]  (0.7,-0.24) -- (0.7,0.17);
                \draw[baseline,thick, -]  (0.7,0.) -- (0.2,0.0);
                \end{tikzpicture} |
  \begin{tikzpicture}[baseline={([yshift=-.5ex]current bounding box.center)},scale=0.65]
                \filldraw[color=black, ultra thick, fill=gray ]  (0.,0.) circle(0.22cm);
                \filldraw[red]  (-0.6,0.) circle (0.7mm) ;
                \draw[baseline,thick, -]  (-0.53,0.) -- (-0.22,0.);
                \end{tikzpicture} \rangle. 
\end{eqnarray}

Thus, the evaluation of the matrix elements\break  $\langle \alpha^{*(Y)} | H^{S=0} | \alpha^{\prime*(Y)} \rangle$ and
$\langle \alpha^{*(Y)} | H^{S=-1} | \alpha^{\prime*(Y)} \rangle$ can be traced back to multiplications  of 
very large  but sparse matrices. As usual, we solve   the eigenvalue problem  using the Lanczos method
so that these matrix multiplications must be  computed again and again. Therefore,  an efficient
method to evaluate such product matrices is extremely important. More details on the technical realization
are given in Ref.~\cite{LePhD:2020}.

\label{sec:transition}
\section{SRG evolution for chiral NN and YN interactions}
\label{sec:srg}
We  follow the formalism  initially applied by Wegner \cite{Wegner:1994ann} to solid state physics and later
employed by  Bogner, Furnstahl and Perry \cite{Bogner:2006pc} to nuclear  interactions,  which defines
the SRG evolution in terms of a unitary transformation depending on a flow parameter $s$
\begin{equation}\label{eq:SRGtrans}
H_{s}^{} = U_{s}^{} H_0 U^{\dagger}_s \equiv T_{rel}^{} + V_s^{}.
\end{equation}  
Here $H_0 = H_{s=0}$ is the initial (bare) Hamiltonian and $T_{rel}$ is the intrinsic relative kinetic operator
that also includes the mass  difference  term when one allows for particle conversions in the Hamiltonian. The
 parameter $s$ has the unit of energy\textsuperscript{-2} and varies continuously from zero to $\infty$.
Note that, although the flow equation is solved with respect to 
$s$, for characterizing the SRG-evolved potentials,  we will utilize 
a more intuitive variable
\begin{equation}
\lambda=\left(\frac{4 \mu^2}{s}\right)^{1/4}~,
\end{equation}  
with $\mu={m_N\, m_\Lambda}/{(m_N+m_\Lambda)}$ for YN interactions and $\mu = {m_N}/{2}$ for NN forces. A
similar definition for $\lambda$ was introduced in  \cite{Bogner:2006pc}. $\lambda$ can be 
(to some approximation) identified with the width of the band for which the SRG evolved matrix elements of the interaction are non-zero. 
By differentiating the transformation Eq.~(\ref{eq:SRGtrans}),  one obtains  the
evolution  equation for the Hamiltonian 
\begin{equation}
\label{eq:SRGflow}
   \frac{d H_s}{ds} = \frac{d V_s}{ds} = [\eta_{s}, H_s]
\end{equation}
where the generator 
\begin{equation}
    \eta_{s}^{} = \frac{dU_s^{}}{ds} U^{\dagger}_s  = -\eta^{\dagger}_{s}
\end{equation}
is an anti-hermitian operator.   Usually, $ \eta_{s}$  is  taken as a commutator of an hermitian operator
$G_s$  with the Hamiltonian, $\eta_s = [ G_s, H_s]$.
The operator  $G_s$ is often chosen such that the evolved Hamiltonian $H_s$ possesses  a desired form. 
For our purpose of decoupling the low- and high-momentum components,  the simplest, but yet very useful generator,
is the relative kinetic energy excluding the mass shift.   We take
\begin{equation}\label{eq:SRGgenerator}
  G_s = \frac{p^2}{2 \mu}
\end{equation}
with $p$  being the particles relative momentum. The flow equation Eq.~(\ref{eq:SRGflow}) now becomes an operator equation
\begin{equation}\label{eq:SRGflows2}
  \frac{d V_s}{ds} = \Big[\Big[\frac{p^2}{2\mu}, V_s\Big], H_s\Big] \ . 
\end{equation}
This is then solved   in  a partial-wave relative momentum basis
\begin{equation}\label{eq:srgbasis}
    | p \,(ls)J; \, t_1 m_{t_1}S_1\, \,t_2 m_{t_2}S_2 \rangle \equiv  |p \alpha \rangle ,
\end{equation}
where  $l$  is the orbital angular momentum that combines with the total spin $s$ to form the total
angular momentum $J$. Further, $(t_i, m_{t_i}, S_i)_{i=1,2}$ are  sets of the intrinsic quantum numbers  that
distinguish  different particle states:  isospin, isospin projection and strangeness.  The  normalization of
the basis states Eq.~(\ref{eq:srgbasis})  simply reads
\begin{equation}\label{eq:srgbasis_norm}
  \sum_{\alpha} \int dp  p^2  \,|p \alpha \rangle  \langle p \alpha |  = 1. 
\end{equation}
After  projecting Eq.~(\ref{eq:SRGflows2}) onto the basis Eq.~(\ref{eq:srgbasis}), one obtains  the flow equation in form of an integro-differential equation
\begin{eqnarray}\label{eq:srg_eq}
\frac{dV^{\alpha \alpha^{\prime}}_{s}(p p^{\prime}) }{ds} &   =  &  \Big [ T^{\alpha}_{rel}(p) \frac{p^{\prime 2}}{2\mu^{\alpha^{\prime}}  } + T^{\alpha^{\prime}}_{rel}(p^{\prime}) \frac{p^{ 2}}{2\mu^{\alpha}  } \nonumber \\[2pt]
& & -T^{\alpha}_{rel}(p) \frac{p^{2}}{2\mu^{\alpha}  } - T^{\alpha^{\prime}}_{rel}(p^{\prime}) \frac{p^{\prime 2}}{2\mu^{\alpha^{\prime}}  }     \Big ]
  V^{\alpha \alpha^{\prime}}_{s}(p p^{\prime})\nonumber \\[3pt]
  & & + \sum_{\tilde{\alpha}}\int_{0}^{\infty}  dk  k^2 \Big [   \frac{p^{2}}{2\mu^{\alpha}} +  \frac{p^{\prime 2}}{2\mu^{\alpha^{\prime}}} 
   - \frac{k^2}{\mu^{\tilde{\alpha}}}
  \Big] \nonumber \\[2pt]
  & &   \qquad\quad  \times V^{\alpha \tilde{\alpha}}_s(pk)  V^{\tilde{\alpha} \alpha^{\prime}}_s(k p^{\prime}).
\end{eqnarray}
Here,  the reduced mass $\mu$ and $T_{rel}$ depend explicitly  on the particle states $\alpha$ since physical
masses are employed for the SRG evolution.
We solve the  flow equation Eq.~(\ref{eq:srg_eq}) numerically using a non-equidistant  momentum
grid characterized by  the ultraviolet momentum cutoff  $p_{max}$ and  $N$ Gauss-Legendre  integration \\ points $p_n$
with corresponding weights $w_n  (n=1,\cdots N)$. Since the initial potentials often vary at low momenta
faster than at high momenta, it is useful to define the grid such that it is sparse at high momenta but 
denser at the low-momentum region.   

\begin{figure}[tbp]
\centering  
 \vskip 0.5cm
\includegraphics[width=7.5cm]{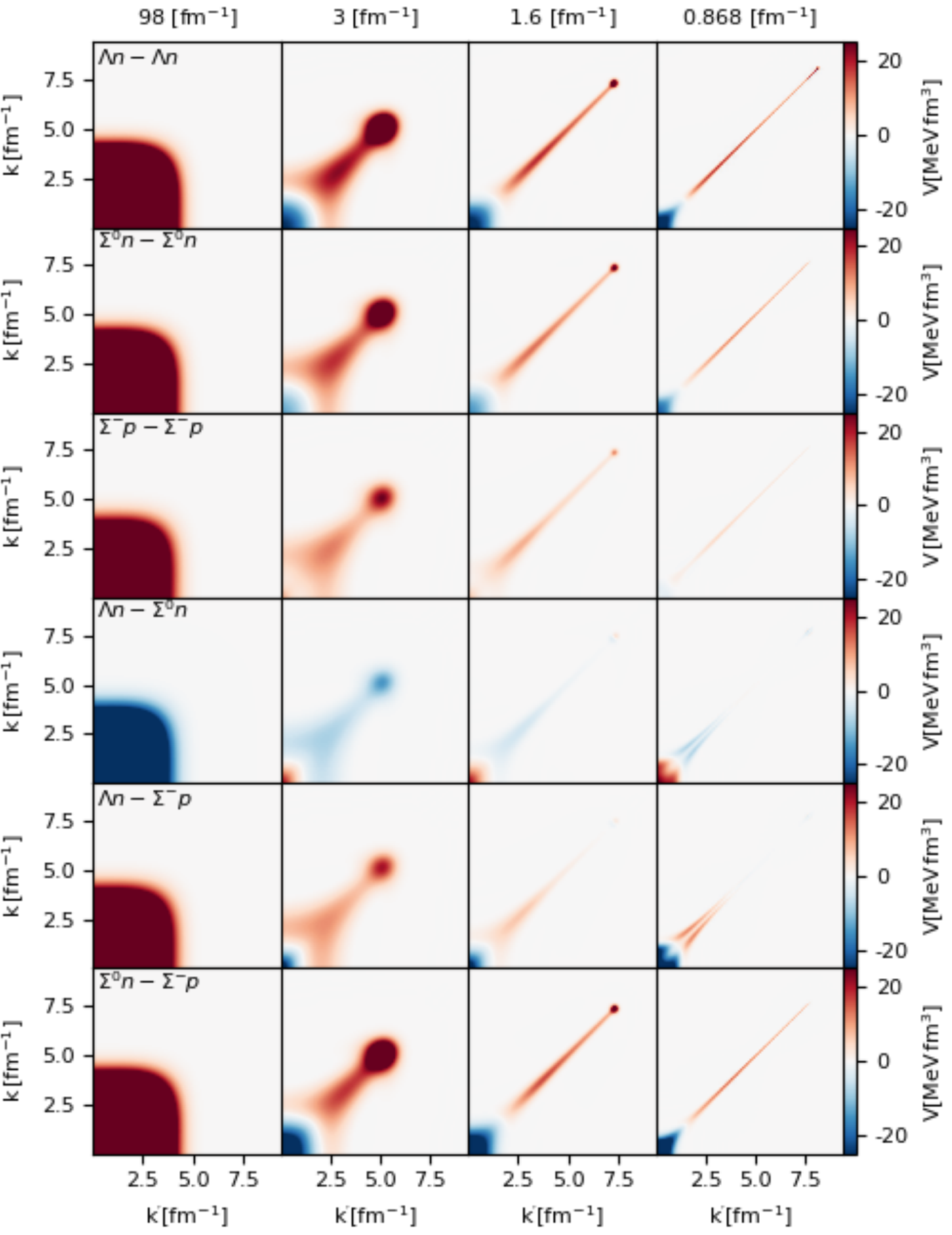}
 \caption{Contour plot of the YN potential matrix elements  for all possible particle channels with  charge $Q=0$ and  in the $^1S_0$ partial wave. The potentials are  evolved to four different values of the YN flow parameter: 
    $\lambda_{YN} =98$~fm\textsuperscript{-1}  (first column, almost non-evolved), $\lambda_{YN}= 3$~fm\textsuperscript{-1}
     (second column, slightly evolved), $\lambda_{YN} =1.6$~fm\textsuperscript{-1} (third column)  and 
     $\lambda_{YN} =0.868$~fm\textsuperscript{-1} (last column). The initial  potential is the YN  NLO interaction with a regulator of $\Lambda_{Y}= 650$~MeV.}
     \label{fig:srgYN-NLO650V} 
\end{figure}

Discretizing the flow equation leads to a set of coupled differential equations which is then solved using the
advanced multi-step Adams PECE (Predict Estimation Correct Estimation) method \cite{hindmarsh2005sundials}.
The SRG-evolution of the YN interaction NLO19 with a regulator 
of  $\Lambda_{Y}=650$~MeV is illustrated in  Fig.~\ref{fig:srgYN-NLO650V}. The contour plots are the potentials for
all the  particle channels with  zero charge  
and  in the  $^1S_0$  partial wave. The initial  potential NLO19(650) is evolved to four different
values of the YN flow parameter:  $\lambda_{YN} = 98$~fm\textsuperscript{-1} (almost non-evolved, bare
interaction),  $\lambda_{YN} = 3$~fm\textsuperscript{-1} (slightly evolved),  $\lambda_{YN} = 1.6$~fm\textsuperscript{-1}
(commonly used) and the extreme case $\lambda_{YN} = 0.868$~fm\textsuperscript{-1}. 
As expected,  the SRG evolution steadily drives the potentials toward a diagonal form
decoupling the low- and higher-momentum states.  While the bare NLO19
shows a strong repulsive  behavior  for almost all particle channels over the entire momentum range, 
the SRG-evolved potentials become  slightly attractive  at low momenta   but  remain repulsive at high momenta. 
 
We explicitly checked that NN and YN scattering observables remain unchanged 
by this unitary transformation. At this point, we neglect induced three-baryon forces 
(3BFs). In this approximation, the evolution of NN and YN forces is not linked to each other and we can choose $\lambda_{NN}$ and $\lambda_{YN}$ independently. 

As a first application, we  apply the SRG transformed interactions 
to obtain binding energies $E(^3_\Lambda {\rm H})$ and the $\Lambda$ separation energies
$B_\Lambda(^3_\Lambda {\rm H}) = E(^2 {\rm H})$  - $E(^3_\Lambda {\rm H})$   of $^3_\Lambda$H. 

Since the $^3_\Lambda$H is predominantely  a weakly bound $\Lambda$ to  a significantly stronger bound
deuteron, it is very difficult to obtain converged results for the 
binding energies using the NCSM. Therefore, for this study, we use solutions based 
on Faddeev equations (see \ref{sec:FYequations}). With this method, 
an accuracy of 1~keV for these energies is routinely achieved. 

\begin{figure}[tbp]
\begin{center}
\includegraphics[width=8.0cm]{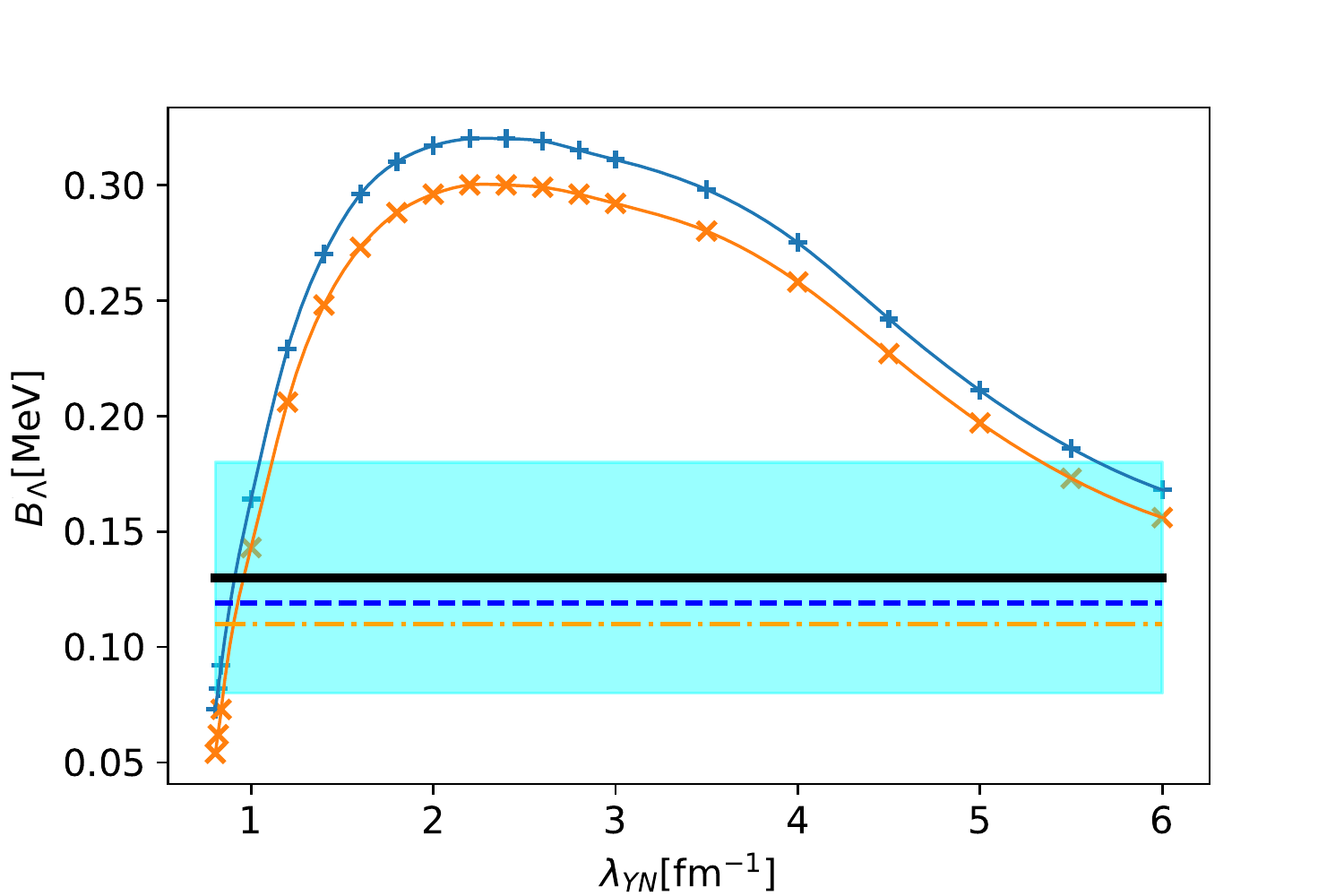}
\end{center}
\caption{\label{fig:hyptri} Dependence of $B_\Lambda(^3_\Lambda {\rm H})$ on $\lambda_{YN}$ for $\lambda_{NN} = 1.6$
(blue +) and 2.4~fm$^{-1}$ (orange x). Starting point of the NN SRG evolution is the Idaho-N$^3$LO(500)
interaction \cite{Entem:2003ft}. For YN, the NLO19(600) interaction \cite{Haidenbauer:2019boi} is used. 
The black solid horizontal line and cyan band indicates the experimental value \cite{Davis:2005mb} and its uncertainty.
The blue dashed and orange dash-dotted lines are results  for the bare YN interaction  and for $\lambda_{NN} = 1.6$ and 
2.4~fm$^{-1}$, respectively. }  
\end{figure}

In Fig.~\ref{fig:hyptri}, $B_\Lambda(^3_\Lambda {\rm H})$ is  shown for one typical choice of the NN and YN
starting interactions. It can be seen that the dependence on the flow parameter of the NN interaction is
of the order of 20~keV. But, unfortunately, it is also clear that 
the dependence on $\lambda_{YN}$ is rather significant, indicating 
a non-negligible contribution of SRG induced three-baryon interactions. 
We will discuss later in Section~\ref{sec:resultscorrelation} how 
this issue could be possibly resolved 
without explicitly taking the induced 3BFs into account. 
Note that, for $\lambda_{YN} \lesssim 1.0$~fm$^{-1}$, the separation energy is in fair 
agreement with experiment and the result of calculations  based on the bare YN interaction.

\section{Results}
\label{sec:results}
As first application of the Jacobi NCSM,  we employ the  approach to investigate 
some interesting hypernuclear systems up to the $p$-shell.  Since 
3BFs are not included in the current study, our primary  focus  will be the impact of different chiral NN
and YN interactions  as well as their SRG evolution on the separation energies. For the
NN interaction we consider the next-to-next-to-next-to-leading order
potential
from the Idaho group with  a regulator  of $\Lambda_{N} = 500$~MeV  (Idaho-N\textsuperscript{3}LO(500))
\cite{Entem:2003ft},  and  the high-order semilocal momentum-space (SMS)
potential regularized with  $\Lambda_{N} = 450$~MeV  \\ 
(SMS N\textsuperscript{4}LO+(450)) \cite{Reinert:2017usi}. 
Two chiral potentials at next-to-leading order,
namely NLO13 and NLO19 \cite{Haidenbauer:2013oca,Haidenbauer:2019boi}  with the  range of regulators
$\Lambda_{Y}=550-650$~MeV, are chosen for the YN interaction. 
In all calculations, contributions of the NN and YN
potentials in partial waves higher than $J = 6$ are left  out. The high partial waves affect the energies
only by a few keV.  For simplicity, the electromagnetic part of the 
NN interaction \cite{Wiringa:1994wb} as well as
the Coulomb point-like contribution in some YN channels are not 
included in the SRG evolution, but only added afterwards. 
We observed that evolving these interactions changes hypernuclear binding energies only by few keV.

\subsection{Extrapolation of the binding energies}

Due to the finite truncation in the single-particle Hilbert space, results from the NCSM calculations are
dependent on the HO frequency $\omega$ as well  as on the
model space size $\mathcal{N}$. Both parameters can be understood 
in terms of an ultraviolet and infrared cutoff. Based on this 
insight, theoretically founded extrapolations can be performed 
with respect to the infrared cutoff \cite{Coon:2012ab,Furnstahl:2012qg,More:2013rma,Wendt:2015nba}. 
This is especially interesting for the 
calculation of expectation values of long range operators, like 
radii, because the infrared dependence is pronounced in this case. 
Since we will be most concerned about the ultraviolet dependence, 
we follow here a simple, but practical approach. 

\begin{figure}[t!]
\centering
\includegraphics[width=7.0cm,trim={0.0cm 0.0cm 0.0cm 0.7cm}, clip]{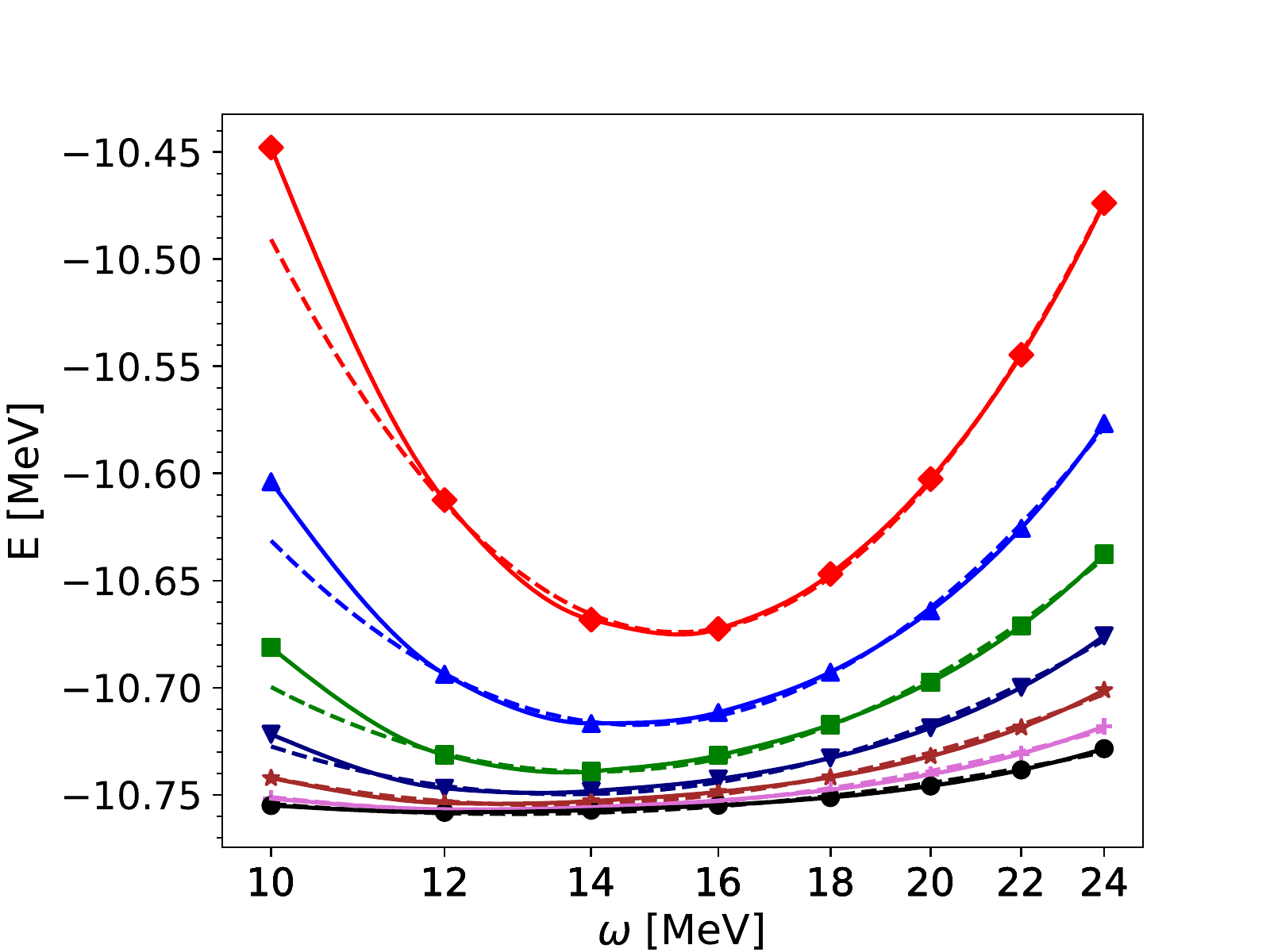}
\caption{$E{(^4_{\Lambda}\text{He},0^+)}$ as a function of 
HO $\omega$. Solid lines  with different colors and
  symbols represent numerical results for different model spaces $\mathcal{N}.$ Dashed lines are obtained using the
  ansatz  Eq.~(\ref{eq:HOansatz}).  The calculations are based on the Idaho-N\textsuperscript{3}LO(500) (NN) and NLO19(600) (YN)
  interactions, SRG-evolved to  $\lambda_{NN} =1.6$~fm\textsuperscript{-1} and $\lambda_{YN} = 2.00$~fm\textsuperscript{-1},
  respectively.}
\label{fig:omegadepent4He}
\end{figure}

In order to obtain converged binding energies, and, at the same time,
to be able to systematically estimate the numerical  uncertainties, we 
follow   a  two-step procedure as employed  in \cite{Liebig:2015kwa}. The first step is to minimize
(eliminate) the HO-$\omega$ dependence.  For each model space size $\mathcal{N}$, we first  calculate the
binding energies, $E(\omega, \mathcal{N})$,  for  a range of HO-$\omega$ and then utilize the following ansatz, 
\begin{align}\label{eq:HOansatz}
E(\omega, \mathcal{N}) = E_{\mathcal{N}}  + \kappa(\text{log}(\omega) - \text{log}(\omega_{\text{opt}}))^2,
\end{align}
to extract the lowest binding energy $E_{\mathcal{N}}$ for the considered model space $\mathcal{N}$ and the
corresponding  optimal HO frequency $\omega_{\text{opt}}$.   As an example,  we show in Fig.~\ref{fig:omegadepent4He}
the HO-$\omega$ dependence of  $E({^{4}_{\Lambda} \text{He},0^{+}})$ for model space  $\mathcal{N}$  varying
from $10$ to $22$. We notice that  the optimal frequency  $\omega_{\text{opt}}$ shifts to lower values as the
model space size $\mathcal{N}$  increases, and the $\omega$-dependence of $E(\omega, \mathcal{N})$  
flattens out as we move forward to the largest model space $\mathcal{N}_\text{max}$. 

\begin{figure}[t!]
\centering
\includegraphics[width=7.0cm]{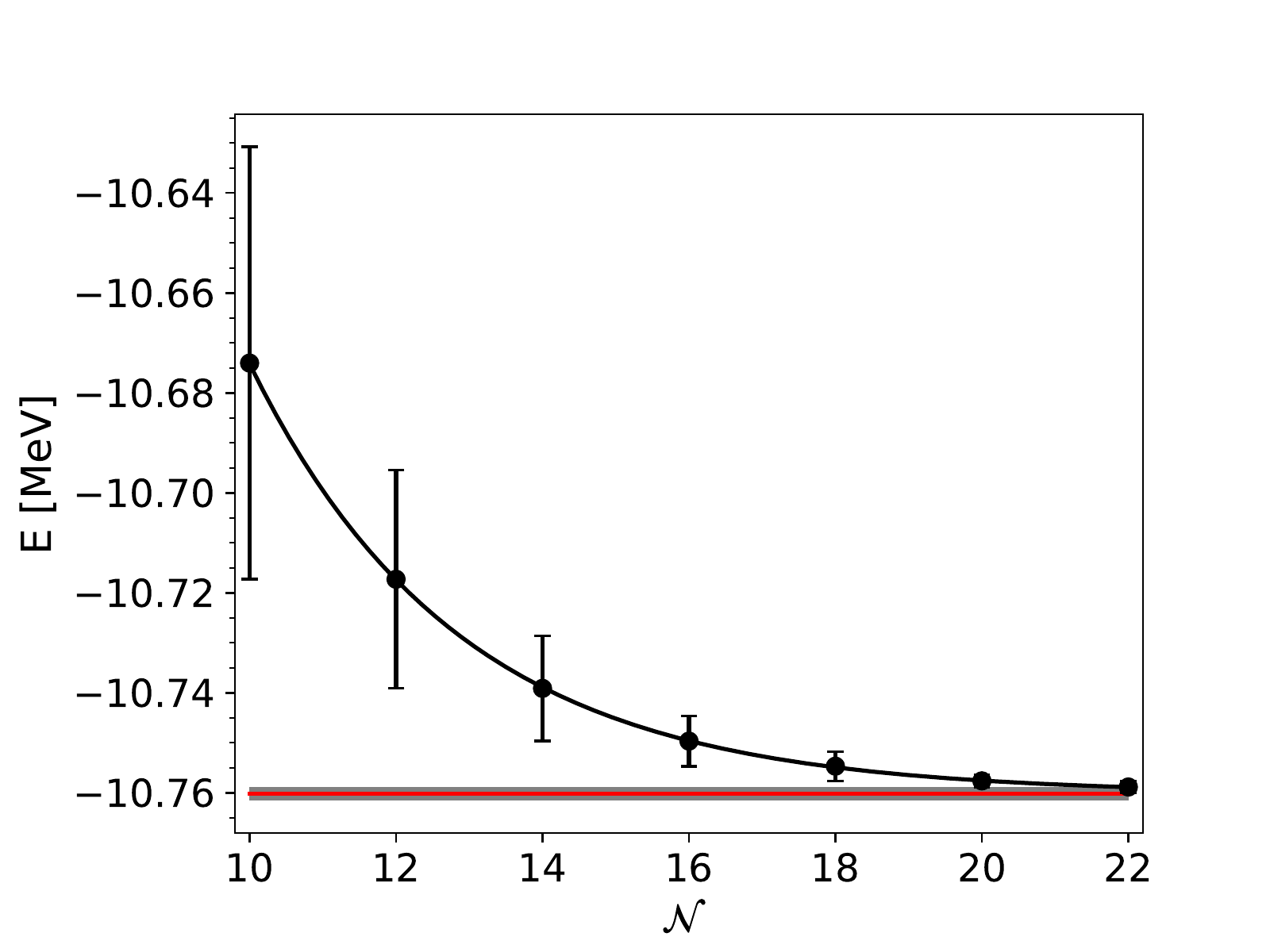}
\caption{$\mathcal{N}$-dependence of $E{(^4_{\Lambda} \text{He},0^+)}$. The symbols and uncertainties represent results extracted from Eq.~(\ref{eq:HOansatz}).
The black line is obtained using Eq.~(\ref{eq:Ndepend}). 
The (red) straight line with shaded area indicates the converged result and its uncertainty. Same description of interactions as in Fig.~\ref{fig:omegadepent4He}.}
\label{fig:Ndepend4He}
\end{figure}

In the second step, the binding energies with the minimal $\omega$-dependence, $E_{\mathcal{N}}$,  are  used
for extrapolating to  a converged result in infinite  model space  assuming an exponential   ansatz
\begin{align}\label{eq:Ndepend}
E_{\mathcal{N}} = E_{\infty} + A e^{-B\mathcal{N}}~.
\end{align}
The confidence interval for each $E_{\mathcal{N}}$ in Eq.~(\ref{eq:Ndepend}) can be determined either from
the spread of the energy in the vicinity of  $\omega_{\text{opt}}$ or from  the slope between two successive energies,
$E_{\mathcal{N}}$ and $E_{\mathcal{N}+2}$.  The latter is mostly employed in our calculations. It should however be
stressed  that the two ways of assigning confidence intervals are practically equivalent and lead to   the same
results  within the numerical uncertainties.  The  determined  intervals   will  serve  as a weight 
for each $E_{\mathcal{N}}$ in the model-space fit using  the ansatz in  Eq.~(\ref{eq:Ndepend}).  The model-space
extrapolation for $E(^4_{\Lambda}\text{He},0^+)$ is illustrated in Fig.~\ref{fig:Ndepend4He}. The final
uncertainty (shaded area) is then taken as the  difference between  the extrapolated  $E_{\infty}$  and
$E_{ {\mathcal{N}} \text{max}}$.

\begin{figure}[htbp]
\centering
\includegraphics[width=7.0cm]{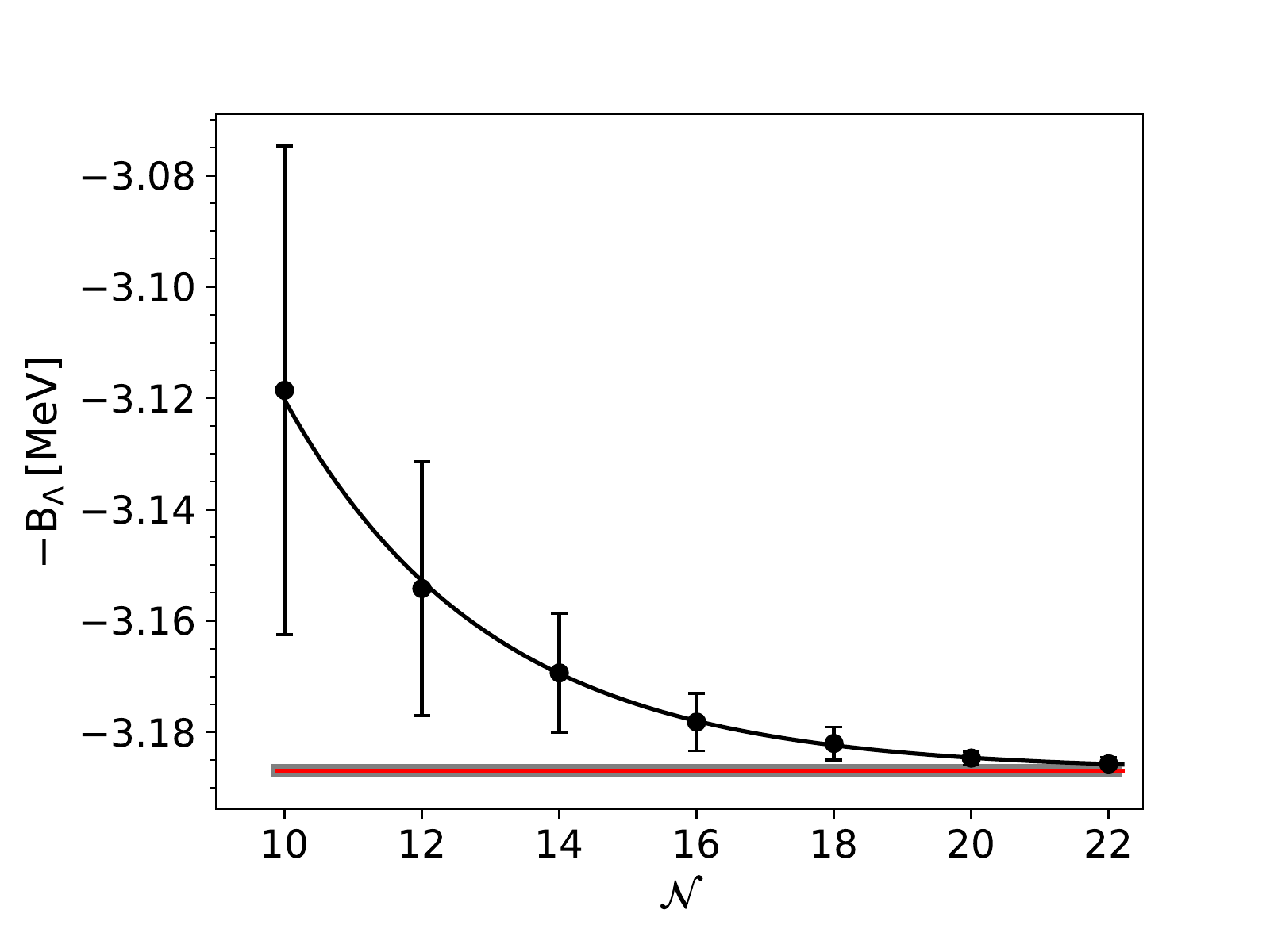}
\caption{$\mathcal{N}$-dependence of $B_\Lambda{(^4_{\Lambda} \text{He},0^+)}$. Same description as in Fig.~\ref{fig:Ndepend4He}. }
\label{fig:Ndepend4HeBlambda}
\end{figure}

\begin{figure*}[htbp] 
      \begin{center}
      \hspace{0.5cm}{
      \subfigure[]{\includegraphics[width=0.45\textwidth]{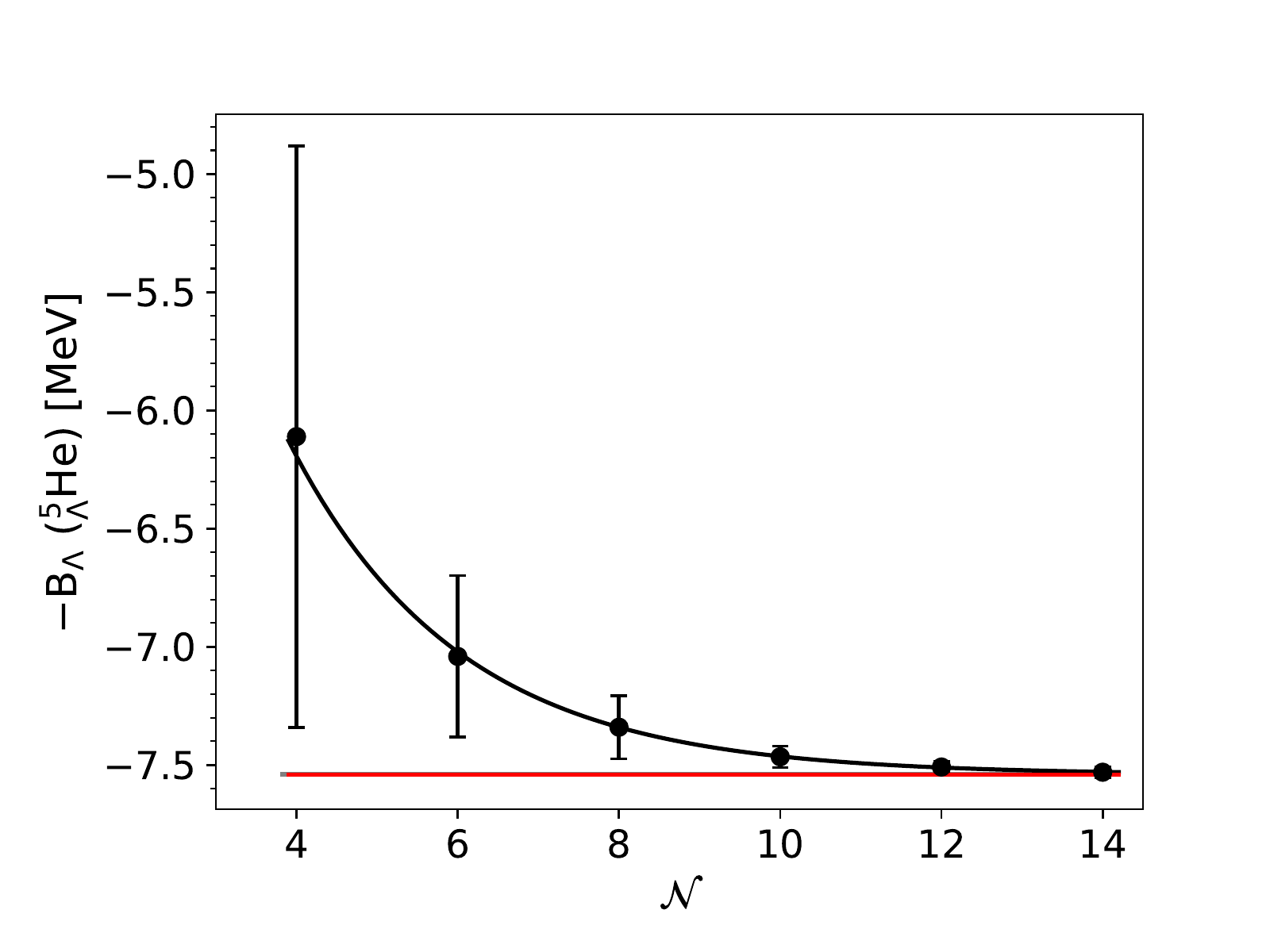}}
      \subfigure[]{\includegraphics[width=0.45\textwidth]{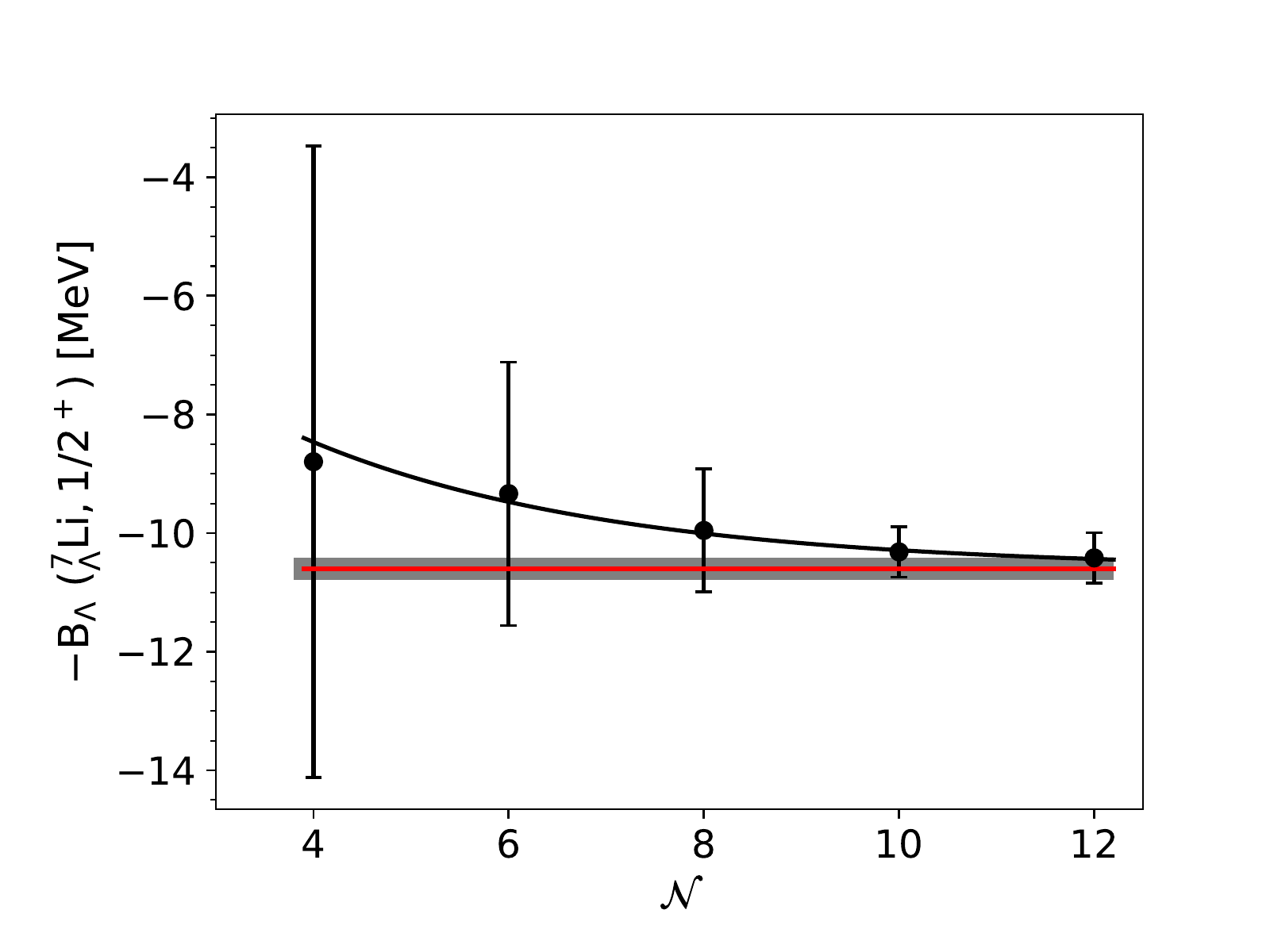}} }
      \end{center}
      \caption{$\mathcal{N}$-dependence of:  (a)  $B_{\Lambda}(^5_{\Lambda} \text{He})$,   (b)  $B_{\Lambda}(^7_{\Lambda} \text{Li}, \frac{1}{2}^+ 0)$. Same description as in Fig.~\ref{fig:Ndepend4He}. The  Idaho-N\textsuperscript{3}LO NN and   NLO19(600) YN potentials are SRG evolved to
      $\lambda_{NN} =1.6$ fm\textsuperscript{-1} and $\lambda_{YN} =2.6$ fm\textsuperscript{-1}, respectively.}
\label{fig:Ndepend5HeBlambda_7Lilambda}
         \end{figure*}

In hypernuclear physics, we are  generally  more interested in the  so-called
$\Lambda-$separation energy, $B_{\Lambda}$, which is  defined as the difference between the binding energies
of a hypernucleus  and of the corresponding parent nucleus. Hence,   
\begin{align} \label{eq:Eseparation}
B_{\Lambda}(^4_{\Lambda}\text{He}) = E(^3\text{He}) - E(^4_{\Lambda}\text{He})~.
\end{align}
Following  the definition  Eq.~(\ref{eq:Eseparation}), in principle, one 
can  subtract the separation energy for each $\omega$ and
$\mathcal{N}$,
\begin{align} \label{eq:Eseparation_omega_N}
B_{\Lambda}(^4_{\Lambda}\text{He},\omega, \mathcal{N} ) = E(^3\text{He},\omega,\mathcal{N}) -
 E(^4_{\Lambda}\text{He}, \omega, \mathcal{N})~,
\end{align}
and  then employ the described    two-step procedure to extrapolate the converged $B_{\Lambda}$. We  have however
observed that  for each model space size $\mathcal{N}$, the useful ranges of  $\omega$ and hence the
optimal frequencies $\omega_{\text{opt}}$ for the nuclear core $^3{\text{He}}$ and
 hypernucleus $^4_{\Lambda}$He are somewhat different. It is 
therefore advisable to  
eliminate the $\omega$-dependence of the binding energies of  $^3\text{He}$ and  $^4_{\Lambda}$\text{He} separately. 
After that, one subtracts $B_{\Lambda}(\mathcal{N})$  for every model space $\mathcal{N}$
\begin{align} \label{eq:EseparationN}
B_{\Lambda}(^4_{\Lambda}\text{He},\mathcal{N}) = E(^3\text{He},\mathcal{N}) - E(^4_{\Lambda}\text{He},\mathcal{N})~,
\end{align}
and  utilizes the ansatz Eq.~(\ref{eq:Ndepend}) to extract the converged result for $B_{\Lambda}(^4_{\Lambda}\text{He})$
together with its uncertainty, see  Fig.~\ref{fig:Ndepend4HeBlambda}. Clearly,
the $\Lambda$-separation energy
  $B_{\Lambda}(^4_{\Lambda}\text{He})$     exhibits a slightly faster
convergence pattern as  that of the binding energy $E(^4_{\Lambda}\text{He})$.
This tendency is also  observed for all other  investigated hypernuclei.  For completeness, 
the model-space extrapolations of $B_{\Lambda}(^5_{\Lambda}\text{He})$ and $B_{\Lambda}(^7_{\Lambda}\text{Li},1/2^+)$ are shown  in Fig.~\ref{fig:Ndepend5HeBlambda_7Lilambda}. 

It is stressed that there is no fundamental reason that the separation energies monotonically converge with increasing model space, but we observed this monotonic behavior in all systems computed so far. This motivated  the use of   Eq.~(\ref{eq:Ndepend}) for the 
extrapolation of the separation energies.  Note that  the resulting energies are consistent with results of Fadeev-Yakubovsky (FY) calculations and/or with a fit of the ${\cal N}$ dependence to a constant. The latter way of fitting is less preferable since it generally leads to larger uncertainties.  


Let us   finally  emphasize   that,  although the described procedure is computationally rather expensive, 
it allows for a systematic and, most importantly, reliable extraction of the final results of the NCSM calculations.
Within the Jacobi-basis formalism such a robust extrapolation 
is feasible and yields plausible results
for light $p$-shell hypernuclei  as one will see in the following sections. 

\subsection{Benchmark results for \texorpdfstring{$^4_{\Lambda}$}{4-Lambda-}He}
As mentioned above, to  validate  the J-NCSM we benchmark  our converged results  with the binding energies
obtained  when solving the FY equations \cite{Nogga:2001ef}. More details 
 are given in \ref{sec:FYequations}. 
 \renewcommand{\arraystretch}{1.5}
 \begin{table}
  \vskip 0.3cm
\begin{center}
   \setlength{\tabcolsep}{0.2cm}
\begin{tabular}{|c| cc| cc| }
  \cline{1-5}
$\lambda_{YN}$ & \multicolumn{2}{c|}{$0^+$}      &\multicolumn{2}{c|}{$1^+$}  \\
  $[\text{fm\textsuperscript{-1}}]$\,  & J-NCSM  & FY & \multicolumn{1}{l}{J-NCSM }   & F-Y    \\
  \cline{1-5}
1.6   &  -10.700(1)\,  & -10.70  &  -9.863(3) & -9.86 \\
3.0   &  -10.751(6)\,  & -10.77  &  -9.81(1) & -9.82\\
14.0   &  -9.27(8) \, & -9.31(3)  &   & \\
\hline
\end{tabular}
\end{center}
\caption{Ground-  and excited-state energies (in MeV) of $^4_\Lambda$He obtained from the Faddeev-Yakubovsky (FY) and
 J-NCSM  approaches. The calculations are
 based on the Idaho-N\textsuperscript{3}LO(500) NN interaction, 
 SRG-evolved to $\lambda_{NN} =1.6$~fm\textsuperscript{-1}, and the NLO19(600) YN potential, evolved 
 to three different SRG flow values, namely $\lambda_{YN} = 1.6$, $3.0$ 
 and  $14.0$~fm\textsuperscript{-1}.}
 \label{tabe:validation}
\end{table}

The binding energies for the ground state $(0^+)$
and first excited state $(1^+)$ of $^4_{\Lambda}$He are tabulated in Table~\ref{tabe:validation}. Clearly,
within the numerical accuracy of better than $20$ keV, the two approaches, J-NCSM and FY, 
agree very nicely.  

\subsection{Effects of NN chiral  interactions on \texorpdfstring{$B_{\Lambda}$}{B-Lambda}}
\begin{figure}[htbp]
\centering
\includegraphics[width=7.0cm]{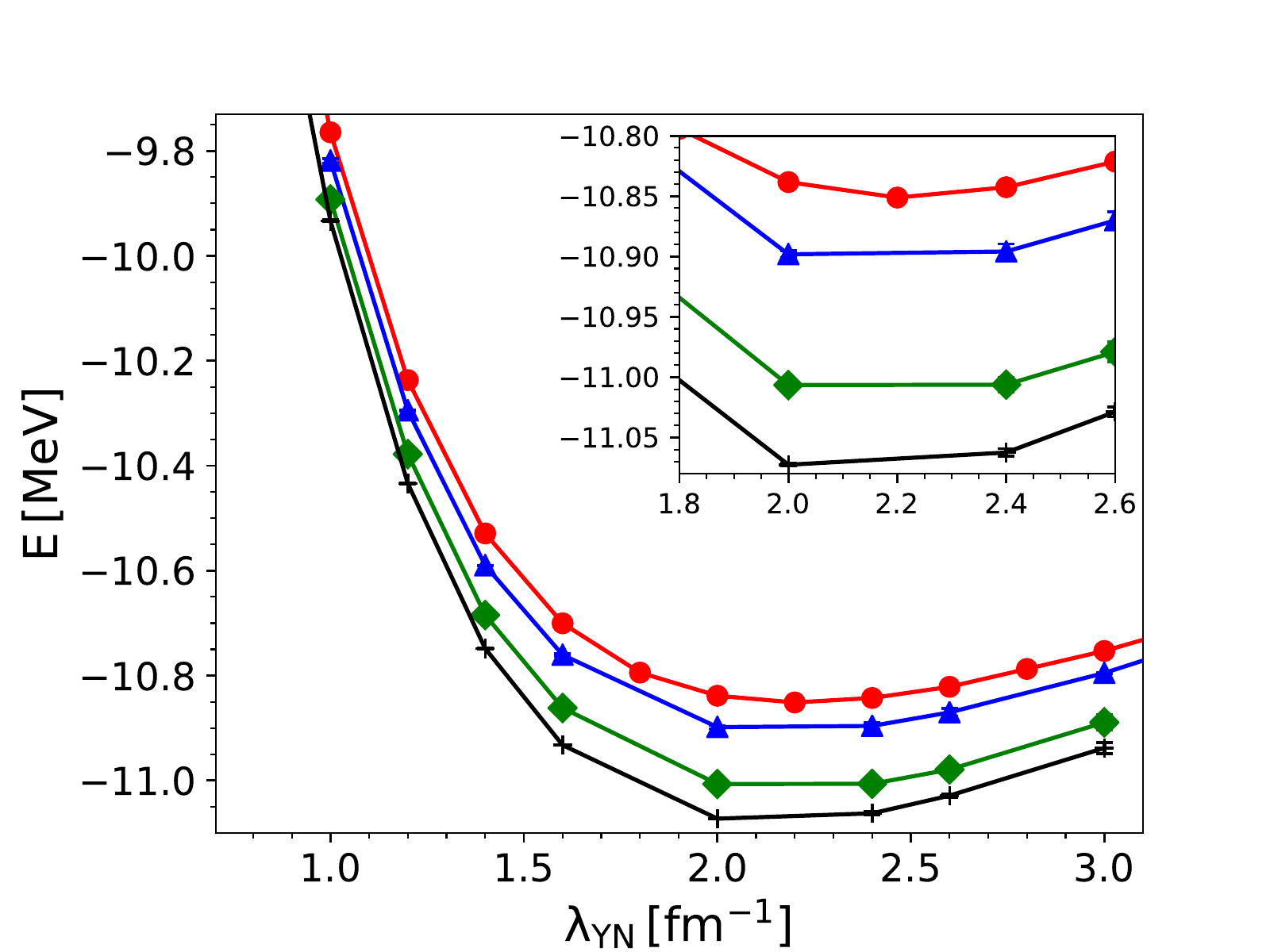}
      \caption{$E(^4_{\Lambda} \text{He},0^+)$ as a function of $\lambda_{YN}$.   The  calculations are
        based on the NLO19(600) YN potential  and four chiral NN interactions: the Idaho-N\textsuperscript{3}LO(500)
        (red circles),  two improved chiral-N\textsuperscript{4}LO (blue triangles) and
        chiral-N\textsuperscript{2}LO (green diamonds) interactions  regularized in configuration space with a
        cutoff $R=0.9$~fm \cite{Epelbaum:2014efa,Epelbaum:2014sza}, and the SMS N\textsuperscript{4}LO{+}(450)
        potential (black crosses).  All NN potentials are  evolved to a flow parameter of
        $\lambda_{NN}=1.6$~fm\textsuperscript{-1}.  The error bars show the estimated numerical uncertainties.}
\label{fig:Eb_4Helambdag_vs_lambdaYN}
\end{figure}

It is known that the nuclear binding energy  $E(^3\text{He})$ and 
consequently  $E(^4_{\Lambda}\text{He})$  are very sensitive to the 
employed NN potentials when three-nucleon (3N) and 
higher-body forces are not included. 
This is noticeable in the binding energies of 
the $^4_{\Lambda}\text{He}(0^+)$ state shown in Fig.~\ref{fig:Eb_4Helambdag_vs_lambdaYN}, obtained for various NN forces: 
the Idaho-N\textsuperscript{3}LO(500), the improved chiral N\textsuperscript{2}LO and N\textsuperscript{4}LO with
a  configuration-space regulator of $R = 0.9$~fm \cite{Epelbaum:2014efa,Epelbaum:2014sza} and the SMS
N\textsuperscript{4}LO{+}(450). 
\begin{figure}[htbp]
\centering
\includegraphics[width=7.0cm]{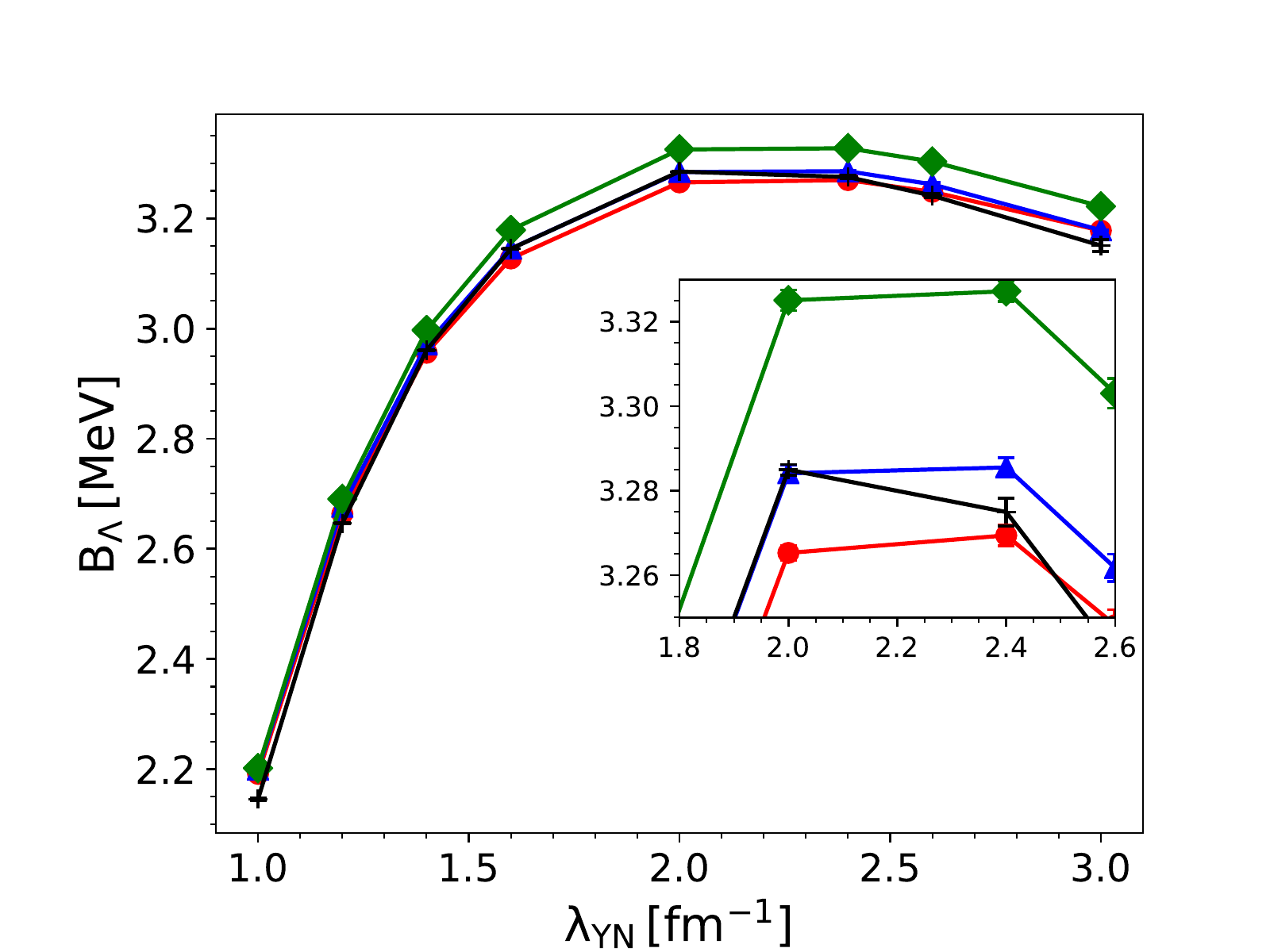}
\caption{$B_{\Lambda}(^4_{\Lambda} \text{He},0^+)$ as a function of
$\lambda_{YN}$.  Same description of the curves as in Fig.~\ref{fig:Eb_4Helambdag_vs_lambdaYN}.}
\label{fig:Blambda_4Helambdag_vs_lambdaYN}
\end{figure}
All NN forces are evolved to an SRG parameter of $\lambda_{NN} =1.6$~fm\textsuperscript{-1}. For that value
overall the binding energies of the $A=3$ to $6$ nuclei are reasonably well described. Of course, this requirement can be fulfilled within a certain range of $\lambda_{NN}$ values so that
the actual choice is to some extent arbitrary. 
The YN potential
is evolved to a wide range of  flow parameters,  $1.0 \leq \lambda_{YN}\leq 3.0$~fm\textsuperscript{-1}. 
One clearly sees  that the  binding-energy  variations  due to  different chiral NN forces  can be as
large as 270~keV.  However, being evolved to the same $\lambda_{NN}=1.6$~fm\textsuperscript{-1}, these NN
potentials have a rather similar impact  on  the  $\Lambda$ removal energy, in particular for  low  SRG-YN
flow parameters $\lambda_{YN} \leq 1.6$~fm\textsuperscript{-1} where there is practically 
no difference  in\break $B_{\Lambda}(^4_{\Lambda}$He, $0^+$), see also Fig.~\ref{fig:Blambda_4Helambdag_vs_lambdaYN}.
For higher values of $\lambda_{YN}$,   the discrepancies
among the  computed values of  $B_{\Lambda}(^4_{\Lambda}$He, $0^+$)  somewhat increase but  remain   relatively
small, about $50$~keV at most (at $\lambda_{YN} =2.0$~fm\textsuperscript{-1}).  We stress that a similar behavior
is also observed for the $\Lambda$-separation energies of
 $^4_{\Lambda}$He$(1^+)$,  $^5_{\Lambda}$\text{He} and $^7_{\Lambda}$\text{Li}$(\frac{1}{2}^+, 0)$. 

      \begin{figure*}[htbp] 
      \begin{center}
      \hspace{0.5cm}{
      \subfigure[]{\includegraphics[width=0.45\textwidth]{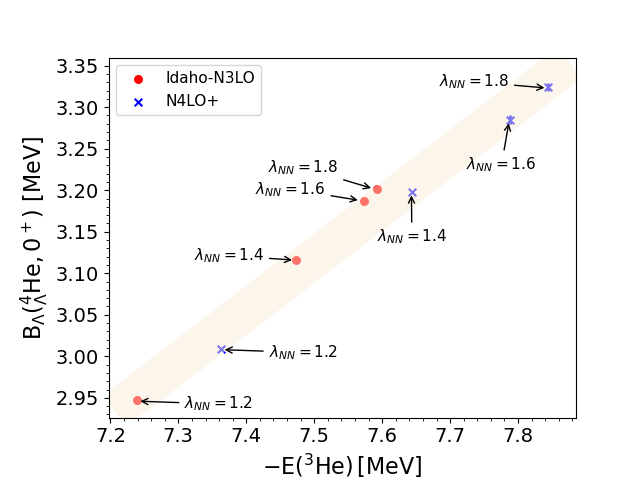}}
      \subfigure[]{ \includegraphics[width=0.45\textwidth]{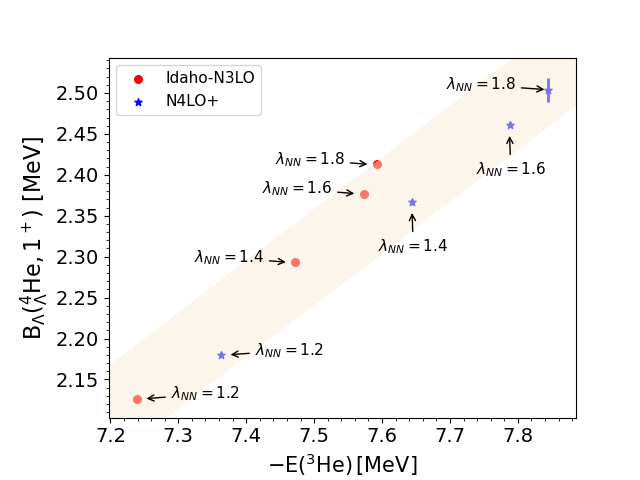}}\\ 
      \hspace{0.5cm}\subfigure[]{\includegraphics[width=0.45\textwidth]{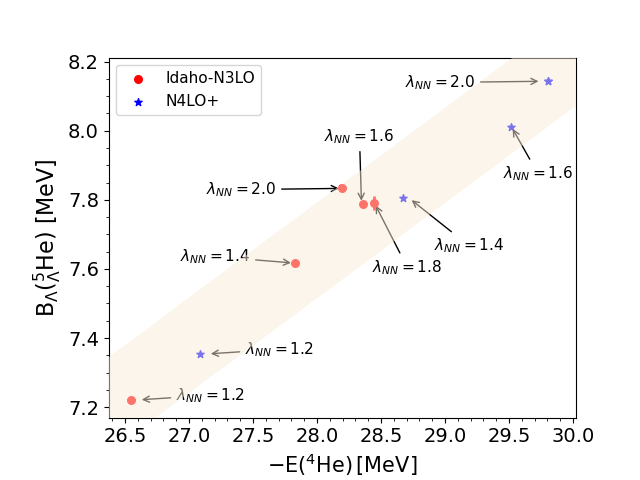}}
      \subfigure[]{ \includegraphics[width=0.45\textwidth]{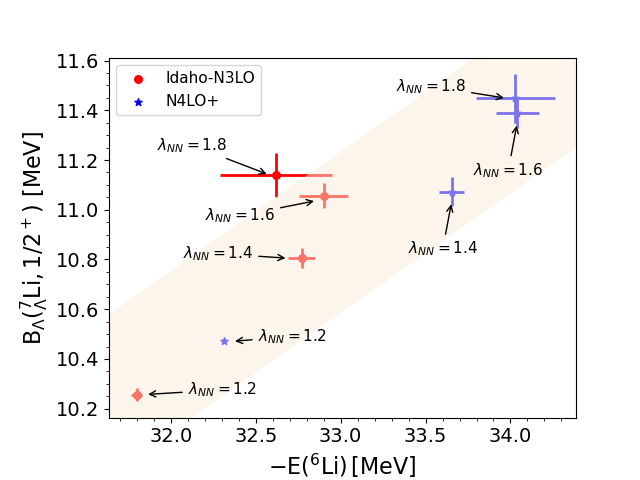}}}
      \end{center}
      \caption{${\Lambda}$-separation energies 
        versus binding energies of the nuclear core: (a)  $^4_{\Lambda} \text{He}(0^+)$ and $^3$He,
        (b)   $^4_{\Lambda} \text{He}(1^+)$ and $^3$He,
        (c)  $^5_{\Lambda} \text{He}$ and $^4$He, (d)   $^7_{\Lambda} \text{Li}(\frac{1}{2}^+, 0)$ and $^6$Li.
        The  calculations are based on the Idaho-N\textsuperscript{3}LO(500) (red circles) and the
        SMS N\textsuperscript{4}LO{+}(450) (blue asterisks) NN potentials, evolved to
        several  values of  $\lambda_{NN}$, in combination with the NLO19(600) YN
        interaction, SRG evolved to $\lambda_{YN} = 2.0$~fm\textsuperscript{-1}. 
        The error bars show the estimated numerical uncertainties.}
\label{fig:Blambda_compare_lambdaNN_andVNN}
         \end{figure*}
 
Hence, in order to further explore the effect of the NN interaction on $B_{\Lambda}$,
we   perform calculations  using the two most accurate  NN potentials, namely Idaho-N\textsuperscript{3}LO(500)
and SMS N\textsuperscript{4}LO{+}(450) evolved to several $\lambda_{NN}$ flow variables.  It is remarked  that,
although  these two  NN potentials describe the available NN scattering data almost perfectly,   they indeed
have very different matrix  elements, particularly in the high-momentum region.  It is therefore  of great interest
to study their  predictions for $B_{\Lambda}(A=4-7)$ more carefully.   To speed up the convergence of the results,      
the NLO19(600) YN potential is evolved to a flow parameter of 
$\lambda_{YN} = 2.0$~fm\textsuperscript{-1}. This specific choice of $\lambda_{YN}$  is based on the above observation
(cf. Fig.~\ref{fig:Blambda_4Helambdag_vs_lambdaYN})
that the largest  discrepancy in $B_{\Lambda}$  is generally observed  at that flow parameter.  The results for the
$A=4-7$ hypernuclei  are displayed  in 
Fig.~\ref{fig:Blambda_compare_lambdaNN_andVNN}  where  
the  $\Lambda$-separation energies are   plotted   against the binding energies of the corresponding  core nucleus.
For the chosen YN flow parameter, the hypernuclei are strongly 
overbound compared to experiment. We show the results here 
to emphasize the effect of different NN interactions. 
For a direct comparison with the experimental separation energies, see below.
The energies  obtained with the Idaho-N\textsuperscript{3}LO(500)  and  SMS N\textsuperscript{4}LO+(450) potentials
are denoted by red squares and blue crosses, respectively.  Also, the  error bars  are    added  in order to
indicate    the estimated numerical  uncertainties, which in many cases  are hardly visible. The light colored
bands indicate the variation of the 
separation energies depending on the 
binding energy of the core nucleus. 
Evidently, there is a general trend
that stronger nuclear binding energies  lead
to larger $\Lambda$-separation energies. Furthermore, the overall variations in the  $\Lambda$-separation
energies  of the two states $^4_{\Lambda}\text{He}(0^+, 1^+)$ due to the change in the $^3$He core binding  energies
are noticeable, i.e. around 400~keV (see  panels (a), (b)). 
However, the width of the band is rather small, of the order of  80~keV only.  For the  $^5_{\Lambda}$He system,  panel (c),
the variation of  $B_{\Lambda}$   stemming  from the  SRG evolution of the individual  NN interactions  is roughly
600~keV while the overall discrepancy caused by    these two  NN potentials can be twice as large.
It can be also clearly seen that the width  of the band for $^5_{\Lambda}$He is  rather large, about 220~keV.
However, given the considerable variation in  $B_{\Lambda}(^5_{\Lambda}\text{He})$, the relative width (roughly $22\%$ of the 1~MeV total variation for all NN interactions employed)
is  of the same order of  magnitude  as that  for  the two states of  $^4_{\Lambda}\text{He}$.   Similarly, the effect
of the  SRG-NN evolution on   $B_{\Lambda}(^7_{\Lambda}\text{Li}, \frac{1}{2}^+)$ for  the  SRG-YN flow parameter
of $\lambda_{YN}=2.0$~fm\textsuperscript{-1} is also pronounced. Here, one of the
individual NN potentials, i.e. the Idaho-N$^3$LO(500),  already causes a discrepancy 
in $B_{\Lambda}(^7_{\Lambda}\text{Li}, \frac{1}{2}^+)$ of about  0.8~MeV, which is almost twice the variation in
$B_{\Lambda}(^5_{\Lambda}\text{He})$. The total variation when considering both interactions is however similar for both hypernuclei, namely 1.1~MeV.  But the relative variation (i.e. the relative width of  the  colored
band   in panel (d)) is rather large,  about 400 keV ($40\%$ of the 1.1~MeV). For  larger $\lambda_{NN}$ ($\lambda_{NN} > 1.6$~fm\textsuperscript{-1}), the numerical uncertainties become visible for $^7_\Lambda$Li and its core. Since the larger $\lambda_{NN}$ significantly increase the width of the band, its width might be further reduced when more converged calculations become available also for these flow 
parameters. In any case, one can expect from the correlations shown in Fig.~\ref{fig:Blambda_compare_lambdaNN_andVNN}  that the dependence of $B_{\Lambda}$
on the nuclear interactions can be substantially reduced once the 3N forces are properly included so that
nuclear core binding energies are in fair agreement with experiment. 
Work in this direction is in progress.
  
\subsection{Effects of the NLO YN interactions on \texorpdfstring{$B_{\Lambda}$}{B-Lambda}}
\label{sec:YNeffect}
We are now in the position to study  the impact of the NLO13 and NLO19 YN interactions on the $\Lambda$-separation
energies. The two NLO potentials are  practically equivalent in terms of describing  two-body YN  observables.
Furthermore, by construction, they  reproduce the experimental binding energy of  $^3_{\Lambda}$H within its 
uncertainty (of order of 50~keV). However, as discussed in 
Ref.~\cite{Haidenbauer:2019boi}, the NLO19 interaction is 
characterized by a
different (somewhat weaker) $\Lambda$-$\Sigma$ transition strength, particularly in the $^3{S}_1$ partial-wave
channel, a feature that   is believed to be closely related to the strength of chiral YNN forces
\cite{Haidenbauer:2019boi,Wirth:2016iwn}. The latter is  expected to manifest itself  in the predictions of
observables  (e.g.~separation energies) for  $A \ge 4$ hypernuclei and in infinite nuclear matter.    
Indeed, it has been found that the NLO19 potential is more 
attractive in the medium than NLO13 \cite{Haidenbauer:2019boi}. 
In addition, in that work, the possible impact of the NLO13
and  NLO19 potentials on the $A=4$ hypernucleus has been thoroughly
investigated, using the Faddeev-Yakubovsky approach.
We provide here again
results for the spin-doublet states  of   $^4_{\Lambda}\text{He}$  for benchmarking. Furthermore,
we extend the study to the $A=5-7$ hypernuclei.  
For our purpose, it is sufficient to choose the SMS
N\textsuperscript{4}LO+(450) potential  with $\lambda_{NN} =1.6$~fm\textsuperscript{-1}.
\begin{figure*}[htbp] 
\begin{center}
\subfigure[]{\includegraphics[width=0.4\textwidth]{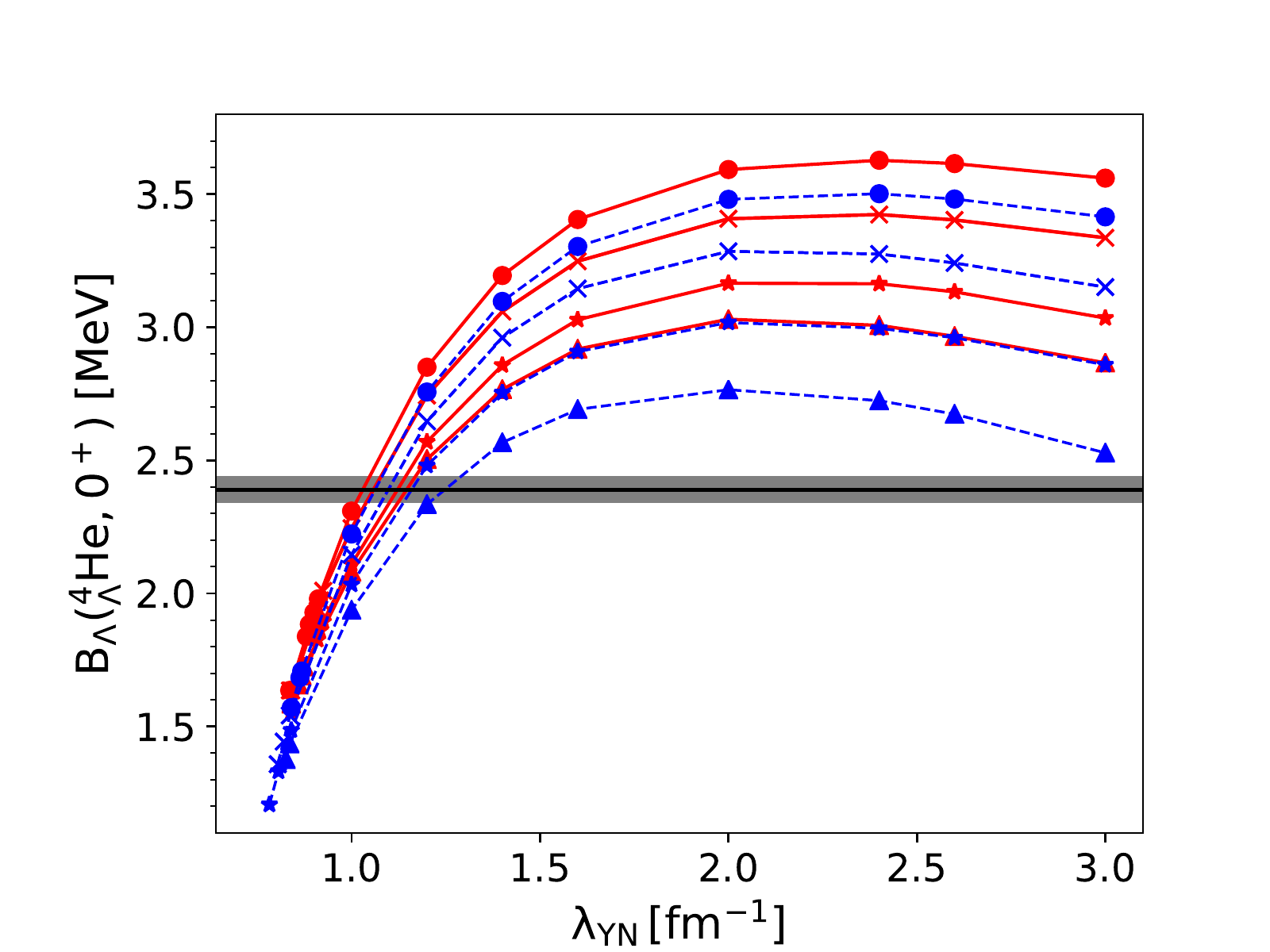}}
\subfigure[]{\includegraphics[width=0.4\textwidth]{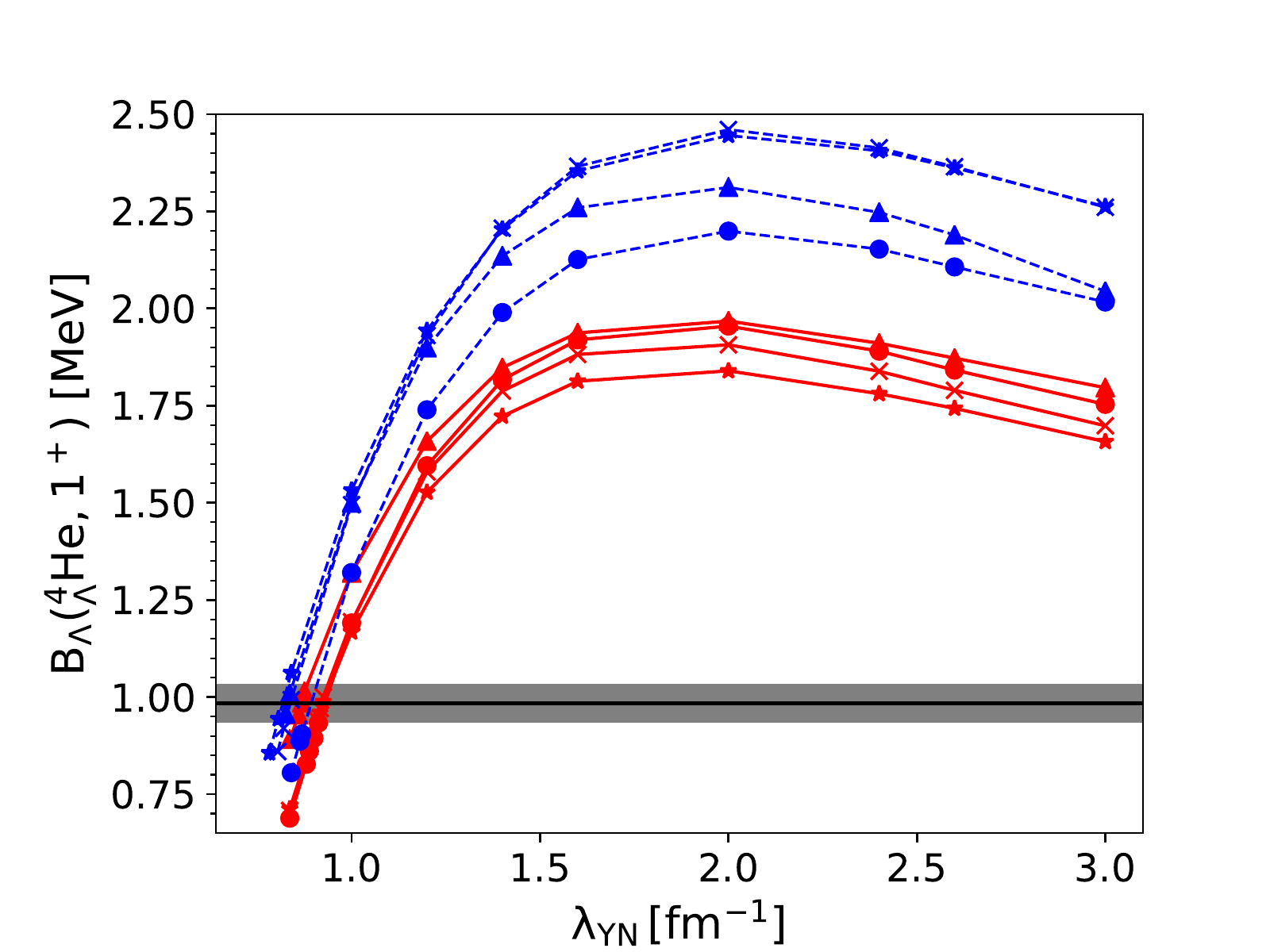}}\\
\subfigure[]{\includegraphics[width=0.4\textwidth]{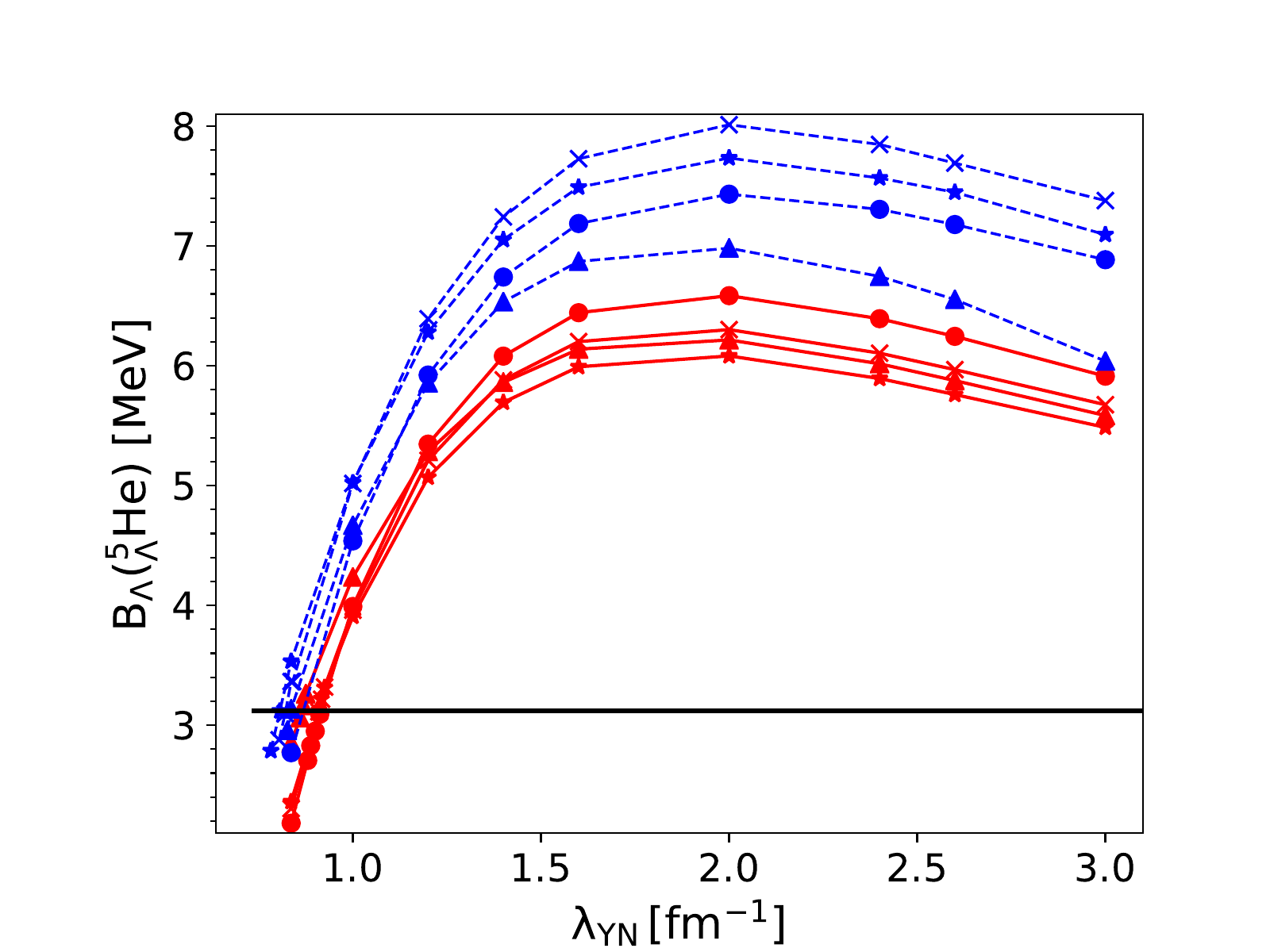}}
\subfigure[]{\includegraphics[width=0.4\textwidth]{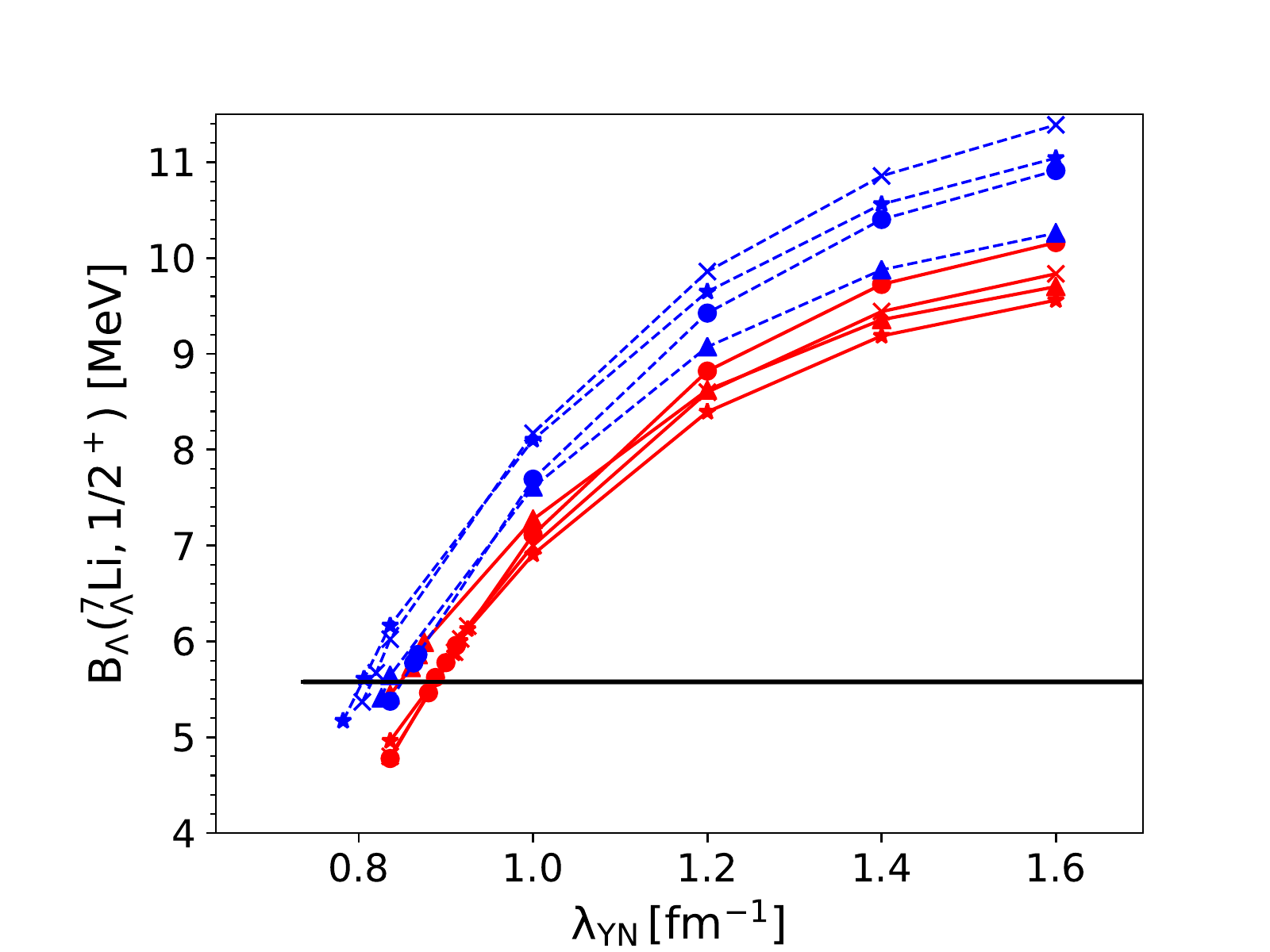}}\\
\subfigure[]{\includegraphics[width=0.4\textwidth]{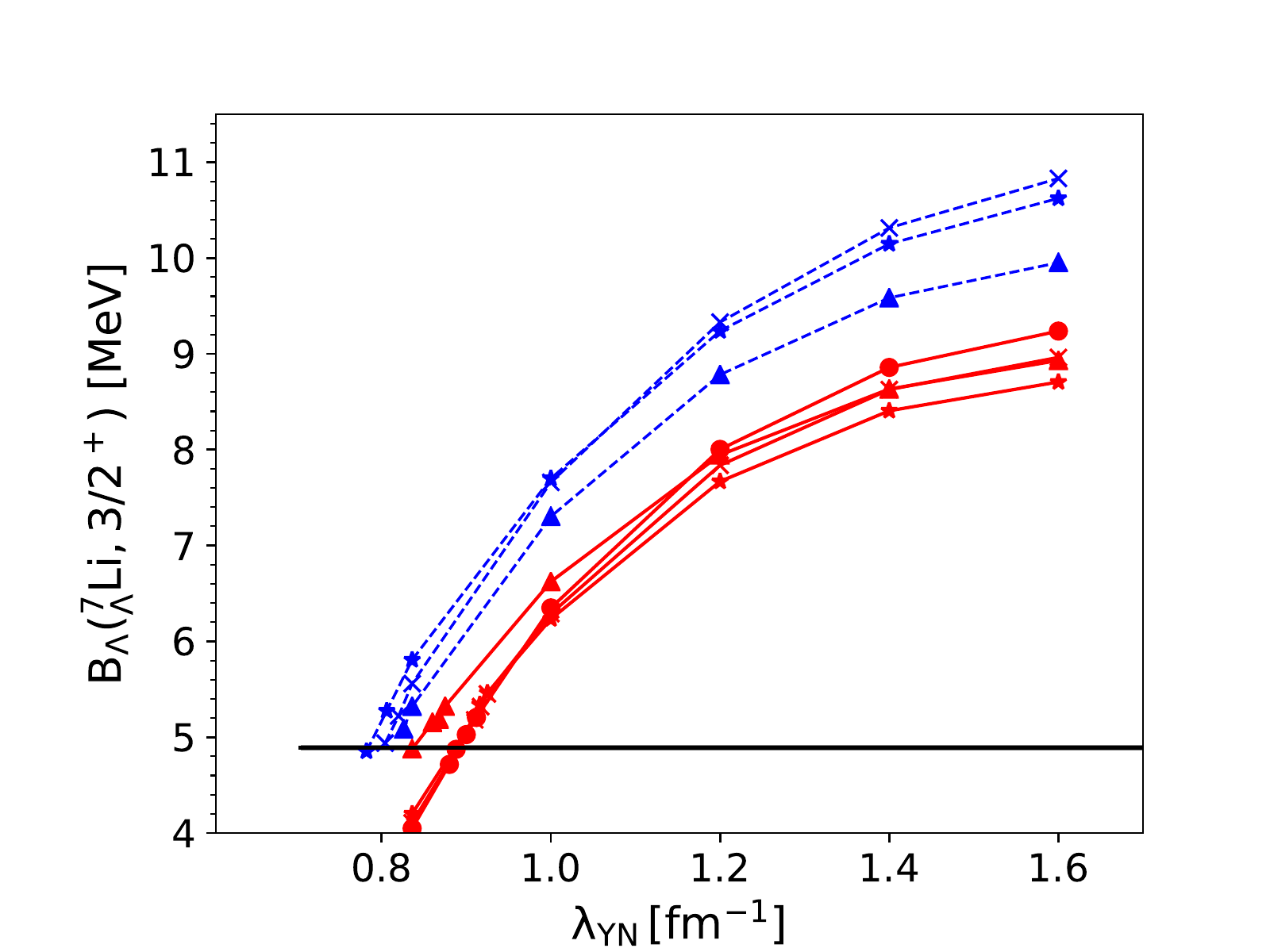}}  \phantom{\includegraphics[width=0.4\textwidth]{figures/7Lilambda-Jtotmax=3-13-19-NLO-Blambda.pdf}} \\
\end{center}
\caption{ $\Lambda$-separation energies of (a) $^4_{\Lambda}\text{He}(0^+)$, (b) $^4_{\Lambda}\text{He}(1^+)$,
  (c) $^5_{\Lambda}\text{He}(\frac{1}{2}^{+})$, (d) $^7_{\Lambda}\text{Li}(1/2^+)$, (e) $^7_{\Lambda}\text{Li}(3/2^+)$ 
  as a function of the SRG-YN flow parameter $\lambda_{YN}$.  
  Black lines with grey bands represent experimental value of $B_{\Lambda}$ and the uncertainties, respectively. 
  The calculations are based on the NN
  interaction SMS N\textsuperscript{4}LO{+}(450) with the SRG-NN evolution parameter of
  $\lambda_{NN}=1.6$~fm\textsuperscript{-1} in combination with the 
  NLO13 (red solid lines) and NLO19 (dashed blue lines) YN potentials
  for four regulators, $\Lambda_{Y} = $  500 (triangles), 550 (stars),
  600 (crosses) and  650 (circles) MeV.}
\label{fig:Blambda_YN13_19}
\end{figure*}
 
The  separation energies $B_{\Lambda}$ of the ground- and 
first-excited states of  the $A=4-7$ hypernuclei
evaluated for the two NLO YN potentials with various 
regulators  $\Lambda_{Y}=500 -650 $~MeV
are presented in Fig.~\ref{fig:Blambda_YN13_19}.
In that calculation, both YN interactions are evolved to the same range of the SRG-YN flow parameters,
$0.8 \leq \lambda_{YN} \leq 3.0$~fm\textsuperscript{-1}.  For  the two states of $^7_{\Lambda}$Li,  the calculations
have only been performed up to  $\lambda_{YN} \leq 1.6$~fm\textsuperscript{-1} in order to   save some
computational resources.  Overall,  the dependence of $B_{\Lambda}$ on the chiral regulator $\Lambda_{Y}$ is somewhat
stronger   for the NLO19  than for the  NLO13 potential.  This, however, does not relate to any physical reason
but simply reflects the fact that, in the NLO19 realization, one has less freedom to absorb regulator artifacts into the parameters of the chiral
interactions (low-energy constants, LECs) because some of the LECs are determined (and taken over) from fits to NN phase shifts 
in line with SU(3) flavor symmetry, see \cite{Haidenbauer:2019boi}.
 There are also noticeable differences between the $\Lambda$-separation energies obtained with the two
interactions, which apparently exceed  the  $\Lambda_{Y}$-dependence.  
For all states except $^4_{\Lambda}\text{He}(0^+)$, see panels (b-e), one observes a general tendency toward
larger $B_{\Lambda}$ values predicted by NLO19 than those calculated with NLO13. In other words, the interaction with a
weaker $\Lambda$-$\Sigma$ conversion potential generally leads  to larger 
$\Lambda$-separation energies.  That trend is, however, not clear for the ground state of $^4_{\Lambda}$He as
can be seen in panel (a).    We remark that a similar (chiral) regulator dependence and 
sensitivity to the YN potential has been  observed in the Faddeev-Yakubovsky results for $A=3$, $4$ hypernuclei,  
computed directly with the bare YN interactions \cite{Haidenbauer:2019boi}. 
There, it was already found that NLO19 
leads to somewhat stronger binding which 
might be a result of the weaker $\Lambda$-$\Sigma$ conversion of NLO19 compared to NLO13. The pronounced  variations of $B_{\Lambda}$
predicted by the two interactions  are a striking evidence for possible contributions of 3BFs to the 
$\Lambda$ separation energy. These discrepancies are expected to be  largely  removed once   proper  chiral  YNN
forces are taken into account explicitly \cite{Petschauer:2015elq}. 
 
Let us mention that the strong sensitivity  of   the $\Lambda$-separation energies of  $^4_{\Lambda}\text{He}(1^+)$
and $^5_{\Lambda}\text{He}$  to the \mbox{$\Lambda$-$\Sigma$} transition potential can be understood using a 
simple approximation for the  effective spin-dependent $\Lambda N$ potential in  $s$-shell hypernuclei, which
can be written as \cite{PhysRev.153.1091,Gibson:1994yp}
\begin{align} \label{eq:relation_in_effective_pot}
 \begin{split}
  ^3_{\Lambda}\text{H}: \quad  & \tilde{V}_{\Lambda N} \approx \frac{3}{4} V^{s}_{\Lambda N} + \frac{1}{4} V^{t}_{\Lambda N}\\[5pt]
 ^4_{\Lambda}\text{He}(0^+): \quad  & \tilde{V}_{\Lambda N} \approx \frac{1}{2} V^{s}_{\Lambda N} + \frac{1}{2} V^{t}_{\Lambda N}\\[5pt]
  ^4_{\Lambda}\text{He}(1^+): \quad  & \tilde{V}_{\Lambda N} \approx \frac{1}{6} V^{s}_{\Lambda N} + \frac{5}{6} V^{t}_{\Lambda N}\\[5pt]
   ^5_{\Lambda}\text{He}: \quad  & \tilde{V}_{\Lambda N} \approx \frac{1}{4} V^{s}_{\Lambda N} + \frac{3}{4} V^{t}_{\Lambda N},\\[5pt]
 \end{split}
\end{align}
where $V^{s}_{\Lambda N}$ and  $V^{t}_{\Lambda N}$  are the singlet- and triplet two-body potentials, respectively.
It  follows clearly from Eq.~(\ref{eq:relation_in_effective_pot}) that the  two states,  $^4_{\Lambda}\text{He}(1^+)$  and
$^5_{\Lambda}\text{He}$, are dominated by the spin-triplet $\Lambda N$ interaction, which  is, as already  mentioned,
strongly influenced by the $\Lambda$-$\Sigma$ conversion.  
Interestingly, as can be seen in Fig.~\ref{fig:Blambda_YN13_19}, the results for
$^4_\Lambda$He$\left( 1^+\right)$, $^5_\Lambda$He$\left( {1}/{2}^+\right)$ and $^7_\Lambda$Li$\left( {3}/{2}^+\right)$ in
panels (b), (c) and (e)  are clearly different for the NLO13 and NLO19 set of interactions. 
To a lesser extend this can also be seen for $^7_\Lambda$Li$\left( {1}/{2}^+\right)$ in panel (d). 
Since $^4_{\Lambda}\text{He}(1^+)$ and $^5_{\Lambda}\text{He}$ are  dominated by the $^3$S$_1$ interaction,\break 
cf. Eq.~(\ref{eq:relation_in_effective_pot}), this suggests that the $^3$S$_1$ contribution is also very important 
for $^7_\Lambda$Li, especially for the ${3}/{2}^+$ state. A future  more detailed study will be necessary
to validate this hypothesis. 

\begin{figure*}[htbp]  
        \begin{center}
\subfigure[]{\includegraphics[width=0.37\textwidth,trim={0.0cm 0.0cm 0.0cm 0.0cm},clip]{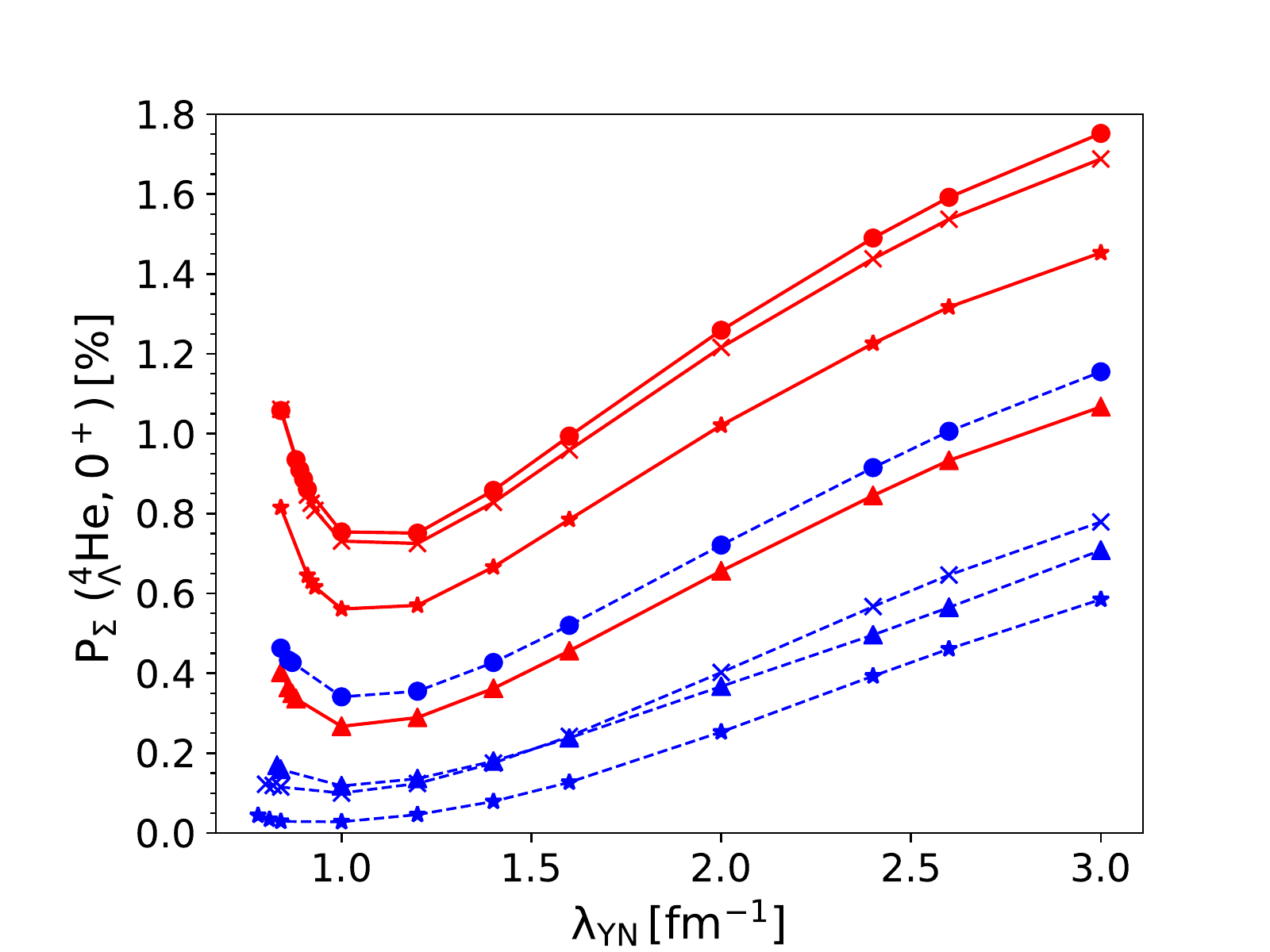}}
\subfigure[]{\includegraphics[width=0.37\textwidth,trim={0.0cm 0.0cm 0.0cm 0.0cm},clip]{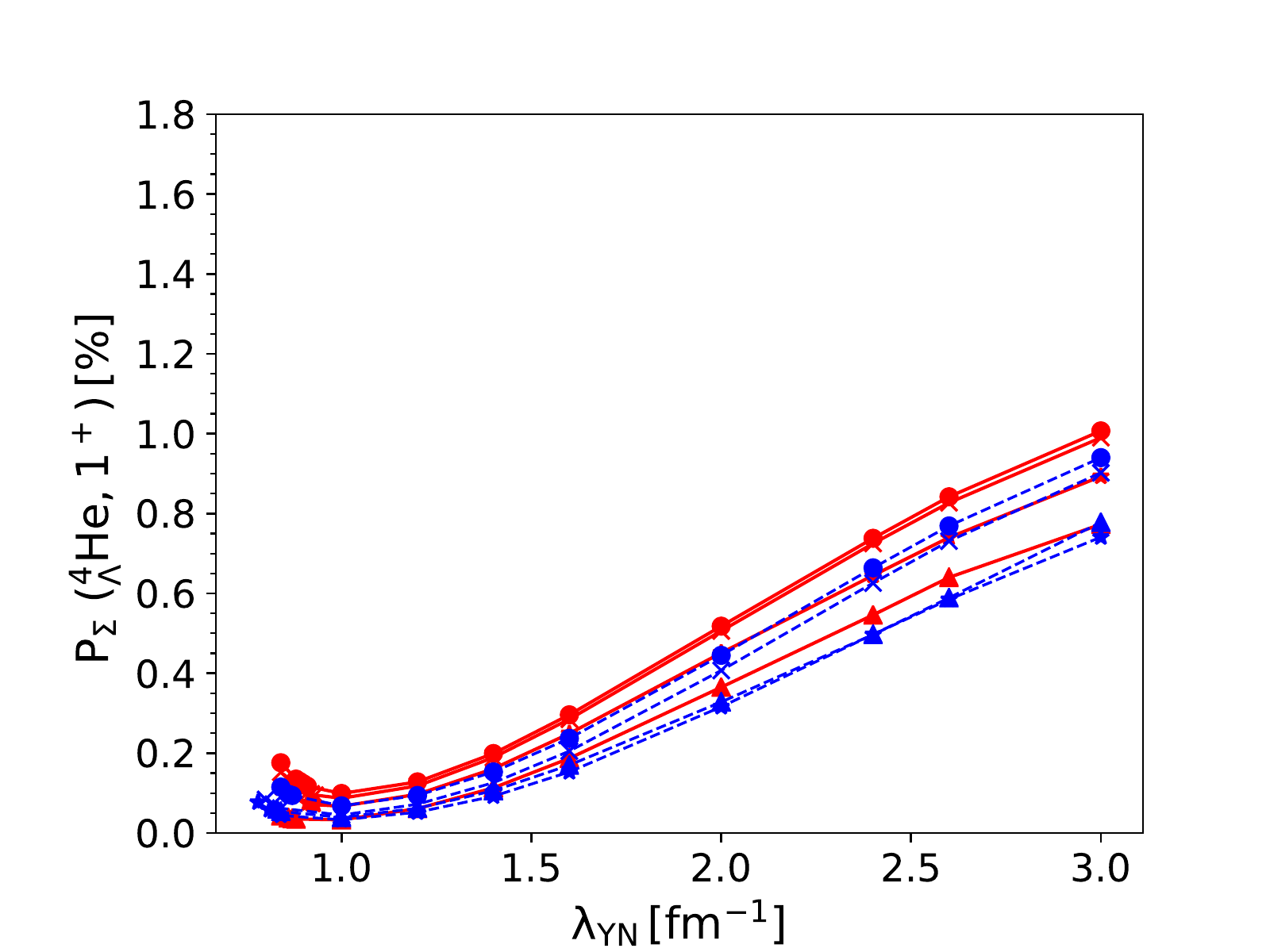}} \\
\subfigure[]{\includegraphics[width=0.37\textwidth,trim={0.0cm 0.0cm 0.0cm 0.0cm},clip]{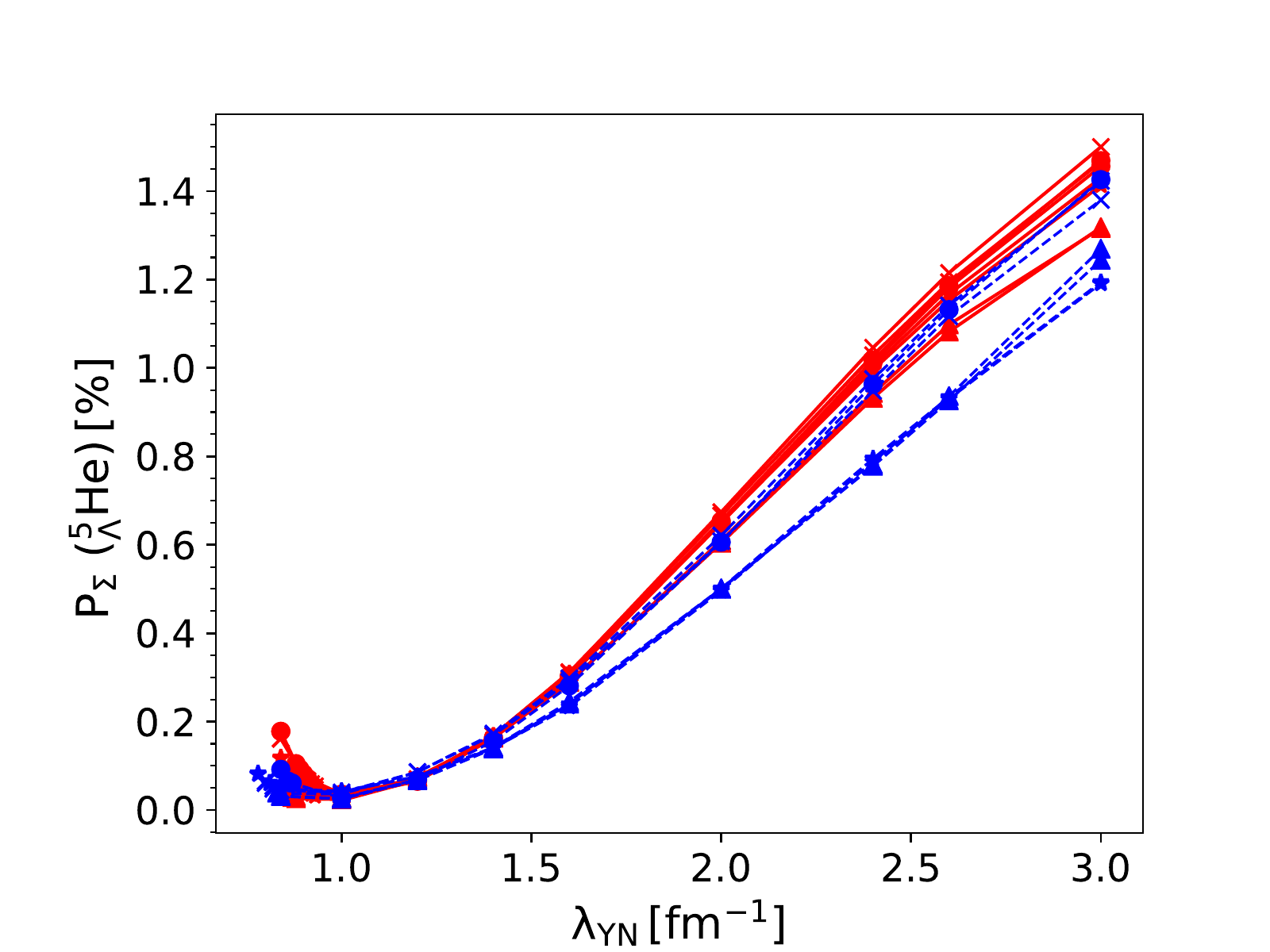}} 
\subfigure[]{\includegraphics[width=0.37\textwidth,trim={0.0cm 0.0cm 0.0cm 0.0cm},clip]{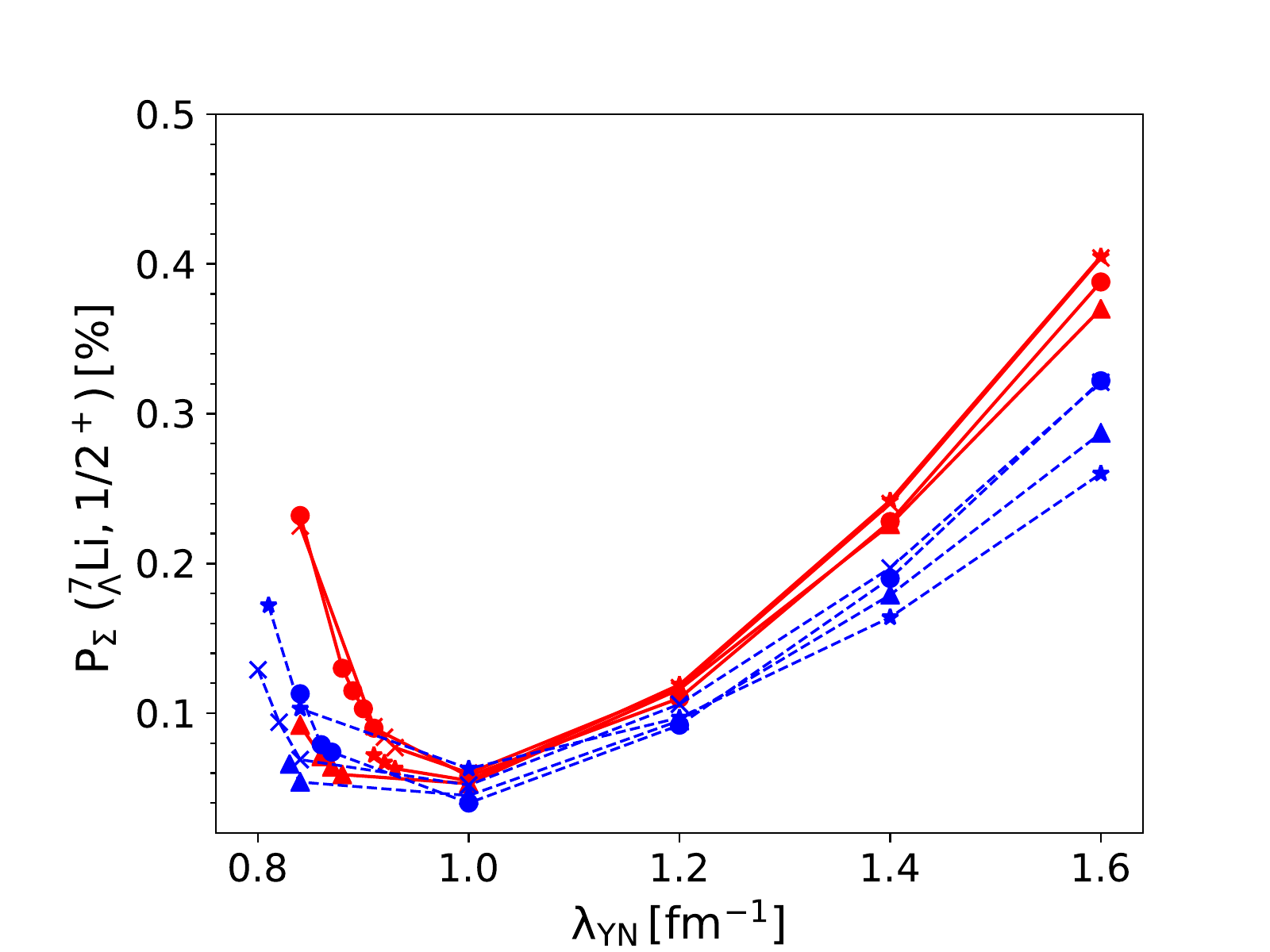}} \\
\subfigure[]{ \includegraphics[width=0.37\textwidth,trim={0.0cm 0.0cm 0.0cm 0.0cm},clip]{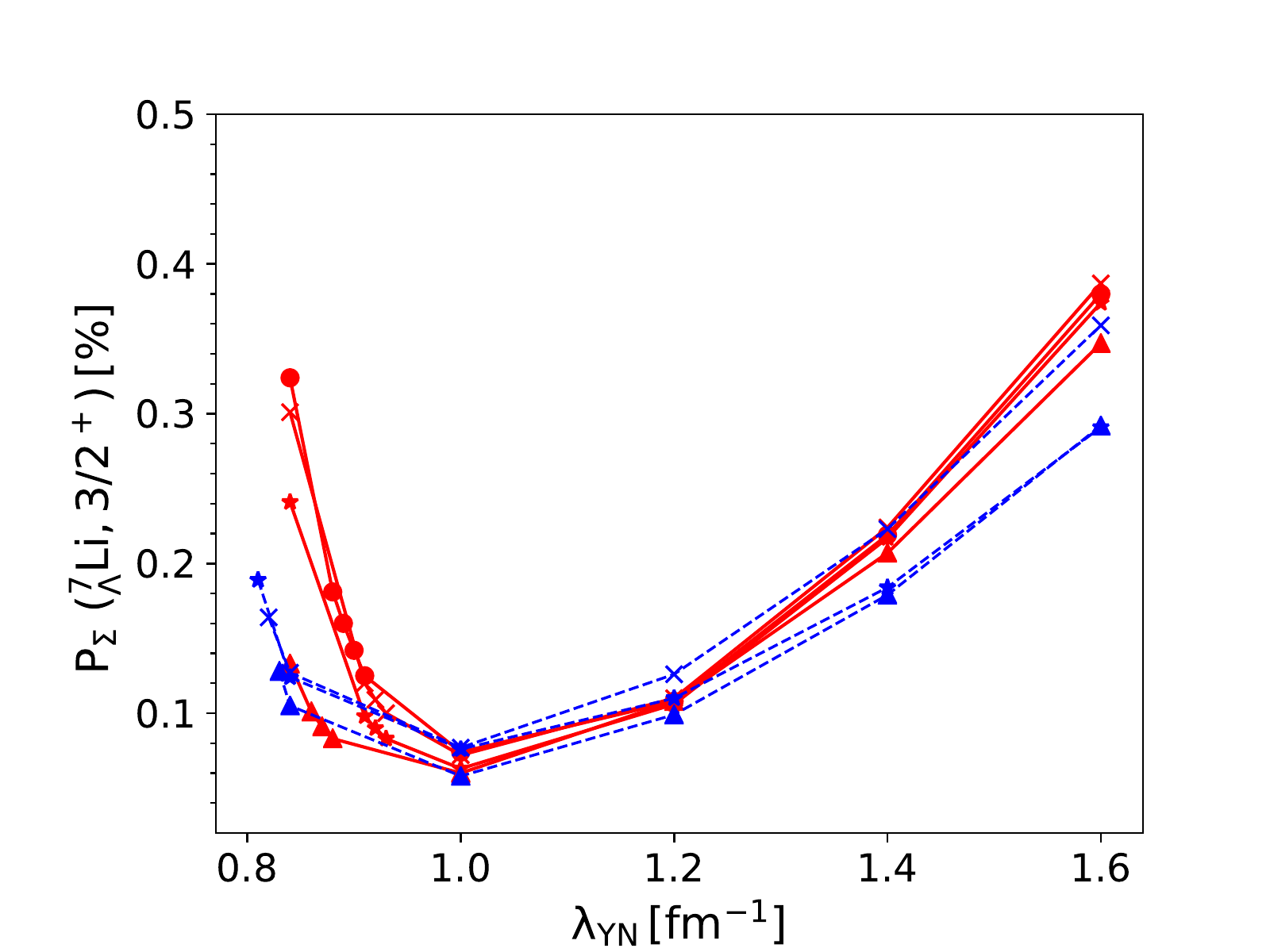}} 
\phantom{ \includegraphics[width=0.37\textwidth,trim={0.0cm 0.0cm 0.0cm 0.0cm},clip]{figures/7Lilambda-Jtotmax=3-Tmax=1-Psigma.pdf}} \\
\end{center}
              \caption{ Probabilities of finding the $\Sigma$ hyperon in 
                the wave functions of (a) $^4_{\Lambda}\text{He}(0^+)$, (b) $^4_{\Lambda}\text{He}(1^+)$,
                (c) $^5_{\Lambda}\text{He}({1}/{2}^+)$, (d) $^7_{\Lambda}\text{Li}(1/2^+)$, (e) $^7_{\Lambda}\text{Li}(3/2^+)$ 
                as a function of SRG-YN flow parameter $\lambda_{YN}$. Same NN potential, symbols and lines as
                in Fig.~\ref{fig:Blambda_YN13_19}.}
\label{fig:Psigma_YN13_19}
\end{figure*}

In this context, the probabilities of finding a $\Sigma$ particle 
in the hypernuclear wave functions ($P_{\Sigma}$) are of great interest, too.
Clearly, they are an indication for the strength 
of the $\Lambda$-$\Sigma$ conversion of the YN interaction. 
Moreover, it can be expected that there are some correlations to 
the charge-symmetry breaking (CSB) of $\Lambda$ separation energies 
of mirror hypernuclei as well \cite{Nogga:2001ef,Nogga:2013pwa}. 
Our calculated $\Sigma$-probabilities for    $A=4-7$ hypernuclei  obtained with the two NLO potentials
are shown in Fig.~\ref{fig:Psigma_YN13_19}.  
It is interesting that in all systems $P_\Sigma$ decreases with decreasing $\lambda_{YN}$ for $\lambda_{YN} \ge 1$~fm$^{-1}$ but increases
again for $\lambda_{YN} < 1$~fm$^{-1}$. Additionally,
The results displayed in panel (a) clearly  indicate a noticeable  dependence of  $P_{\Sigma}(^4_{\Lambda}\text{He}, 0^+)$
on the chiral cutoff $\Lambda_{Y}$. That regulator dependence, however, becomes somewhat less visible for  all
other states, see panels (b-e). Also, the variation of the $\Sigma$-probabilities caused by the two chiral
interactions is most pronounced for $^4_{\Lambda}\text{He}(0^+)$.   This is exactly opposite to the observations for
the $\Lambda$-separation energies as discussed above.   Moreover, there is an overall  tendency toward larger
$P_{\Sigma}$ predicted by the interaction with a  stronger 
$\Lambda$-$\Sigma$ transition (i.e. NLO13) although it
is  somewhat  blurred by  the regulator dependence.  
We further note that, while there is a visible difference between the $\Sigma$-probabilities  of the
$s$-shell  spin-doublet states  (in particular for  the predictions of NLO13), the $p$-shell  doublet
$P_{\Sigma}(^7_{\Lambda}\text{Li}, {1}/{2}^+)$  and  $P_{\Sigma}(^7_{\Lambda}\text{Li}, {3}/{2}^+)$  are quite similar
for both interactions.  Clearly,  one sees that the $\Lambda$-separation energies and $\Sigma$-probabilities
in $A=4-7$ hypernuclei are somewhat correlated. However, we do not observe a definite  one-to-one correlation
between the two quantities. 

\subsection{Correlation of \texorpdfstring{$\Lambda$}{Lambda}-separation energies }
\label{sec:resultscorrelation}

\begin{figure*}[htbp]  
        \begin{center}
\subfigure[]{\includegraphics[width=0.4\textwidth,trim={0.0cm 0.0cm 0.0cm 0.0cm},clip]{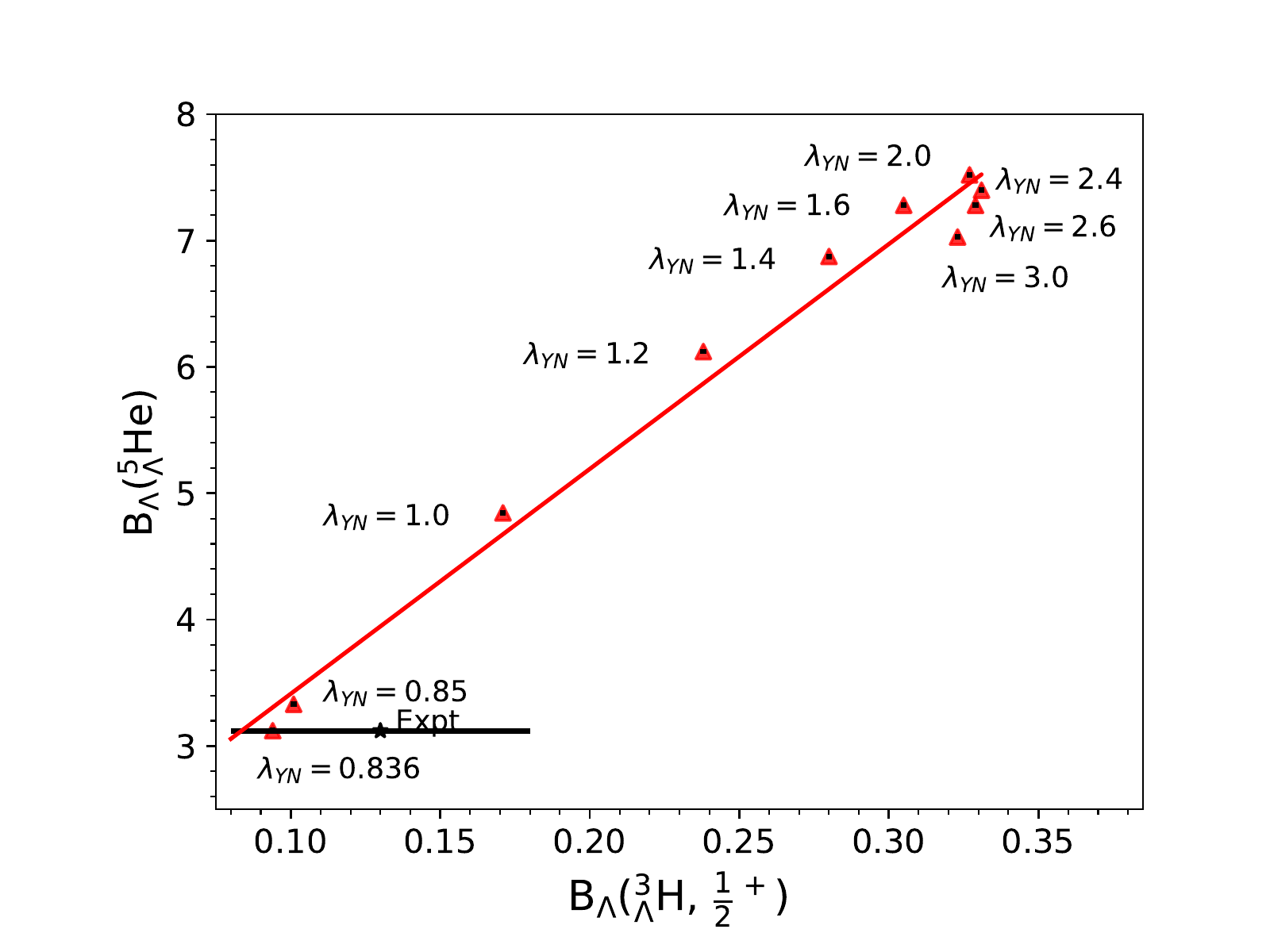}}        
\subfigure[]{\includegraphics[width=0.4\textwidth,trim={0.0cm 0.0cm 0.0cm 0.0cm},clip]{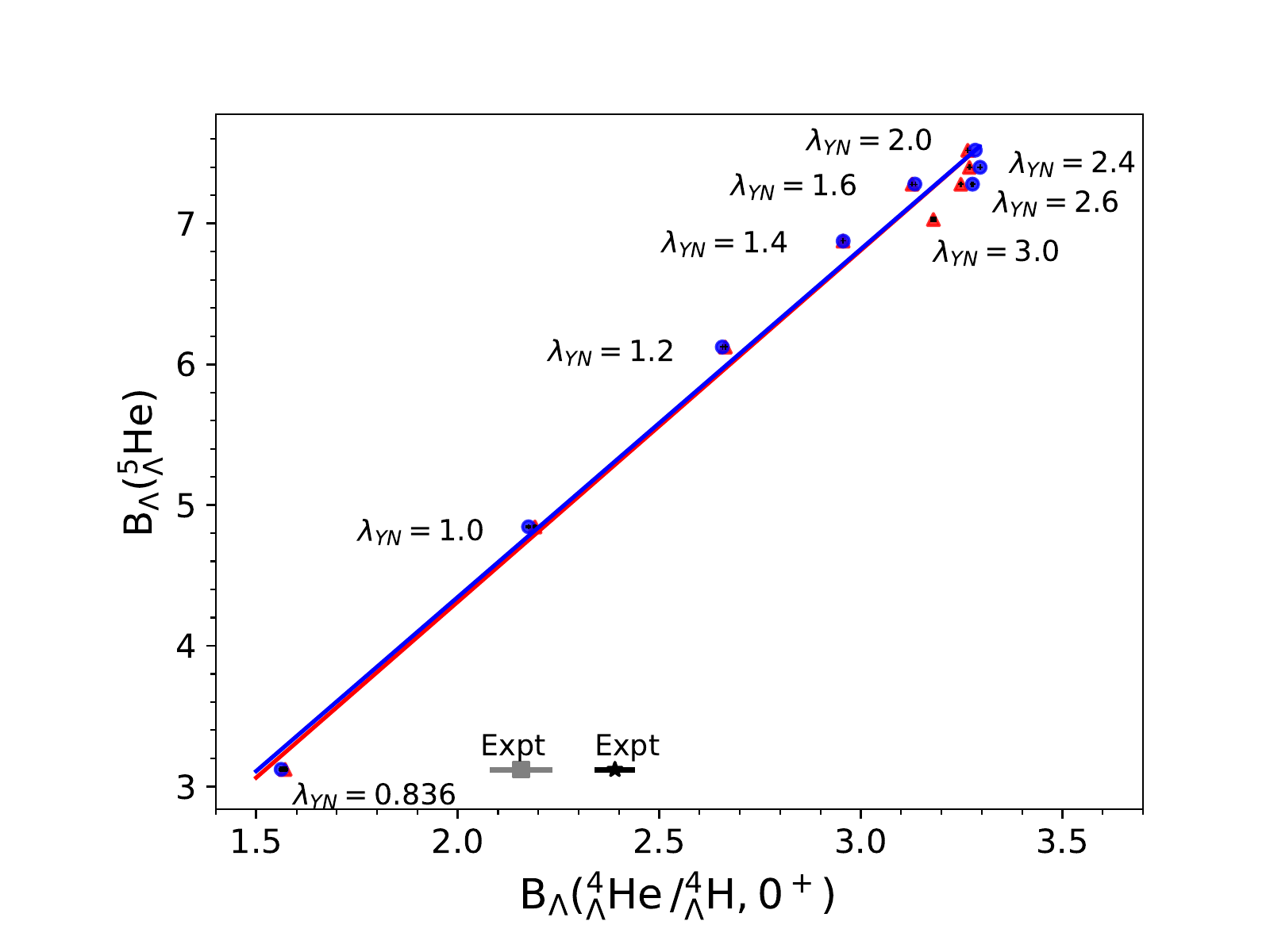}} \\        
\subfigure[]{ \includegraphics[width=0.4\textwidth,trim={0.0cm 0.0cm 0.0cm 0.0cm},clip]{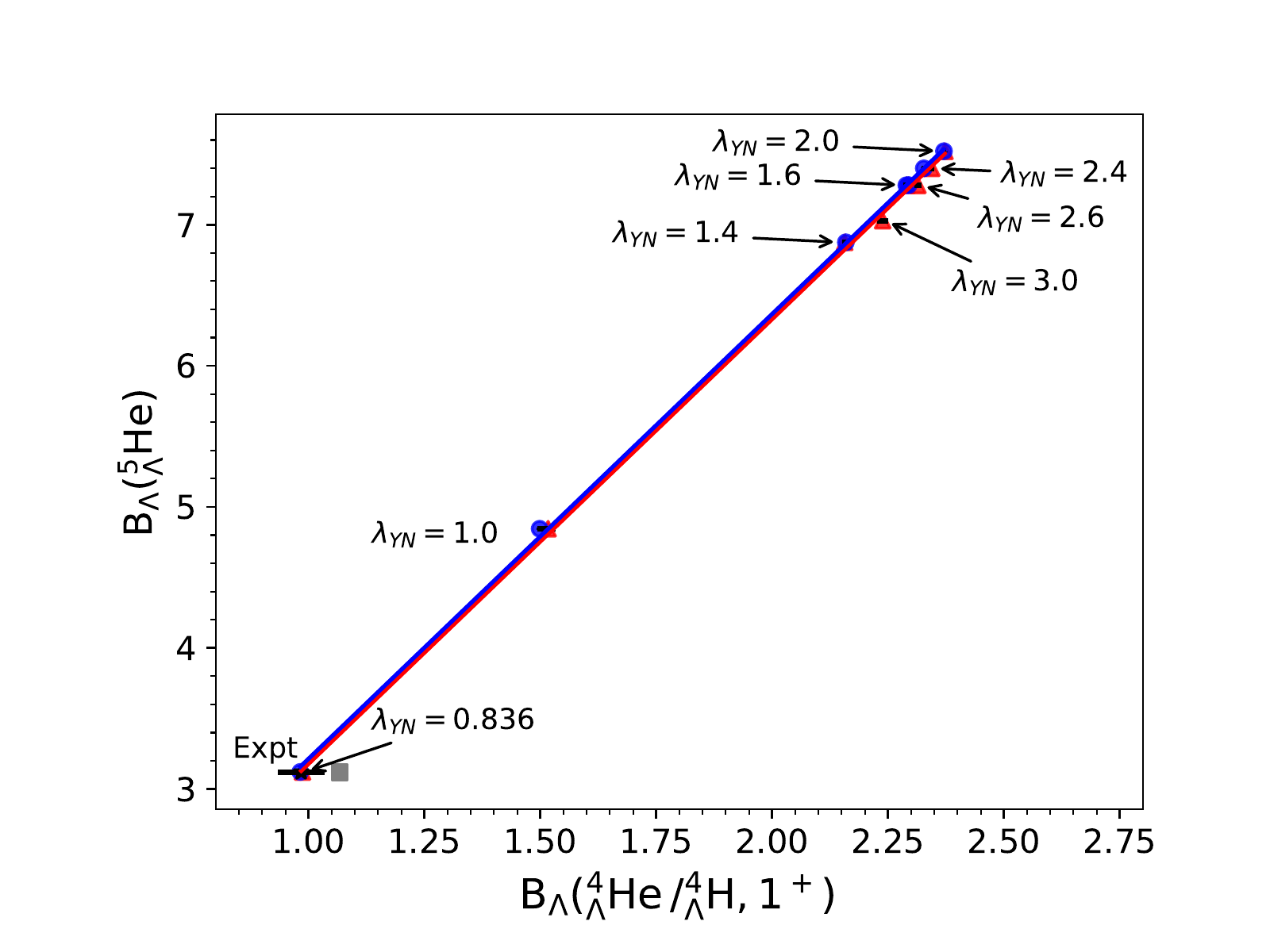}}        
\subfigure[]{\includegraphics[width=0.4\textwidth,trim={0.0cm 0.0cm 0.0cm 0.0cm},clip]{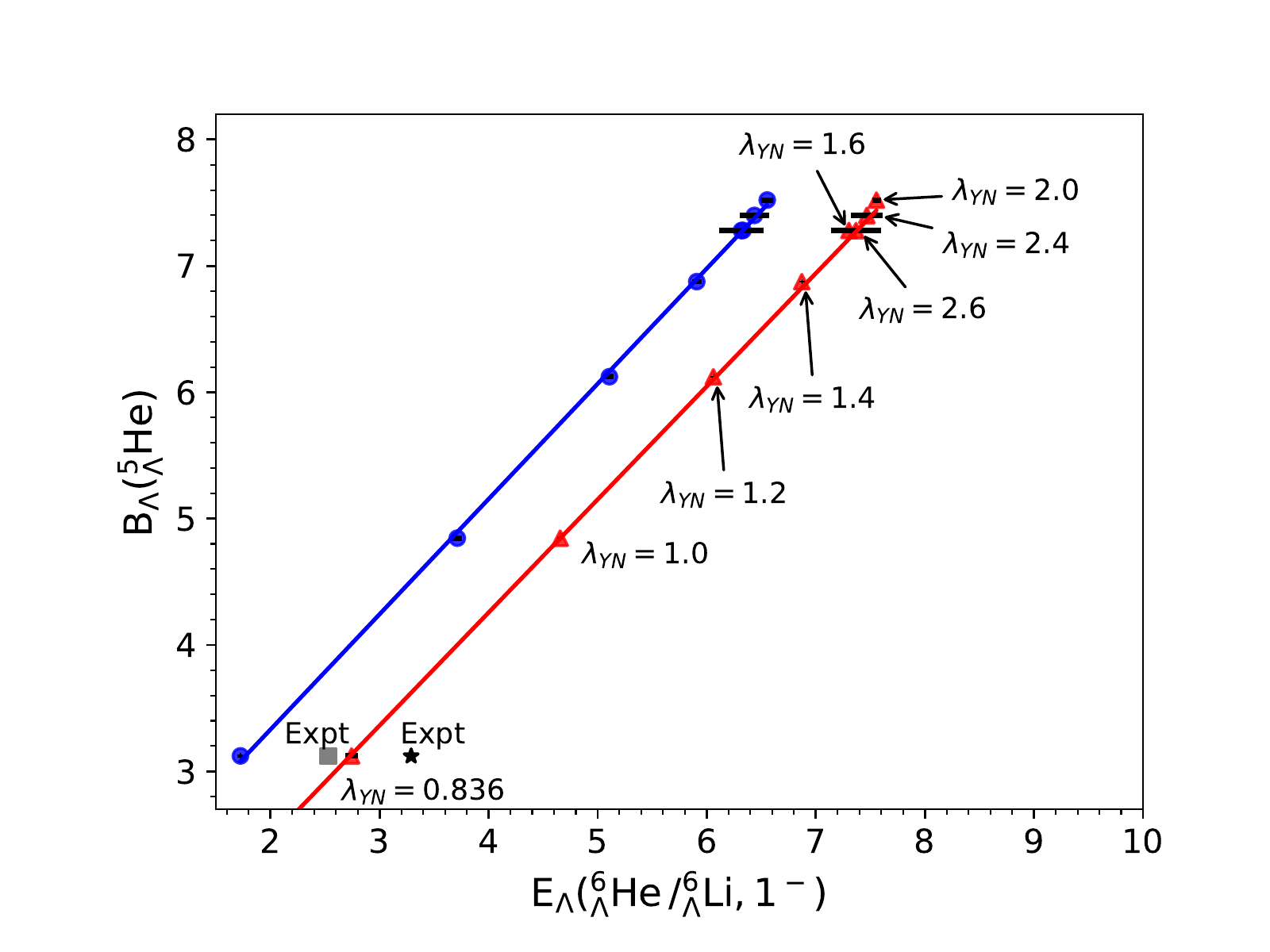}} \\        
\subfigure[]{ \includegraphics[width=0.4\textwidth,trim={0.0cm 0.0cm 0.0cm 0.0cm},clip]{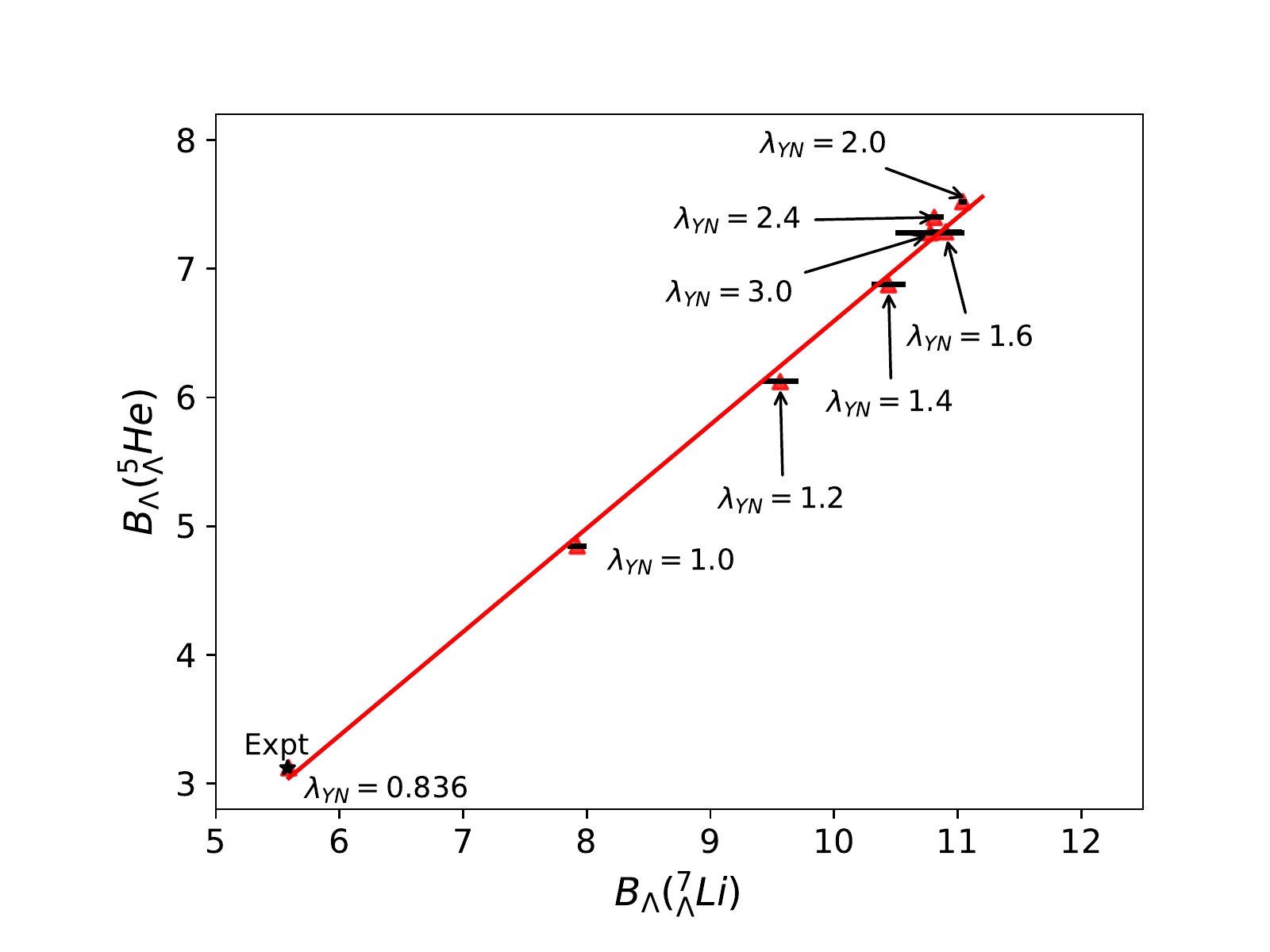}}        
\subfigure[]{\includegraphics[width=0.4\textwidth,trim={0.0cm 0.0cm 0.0cm 0.0cm},clip]{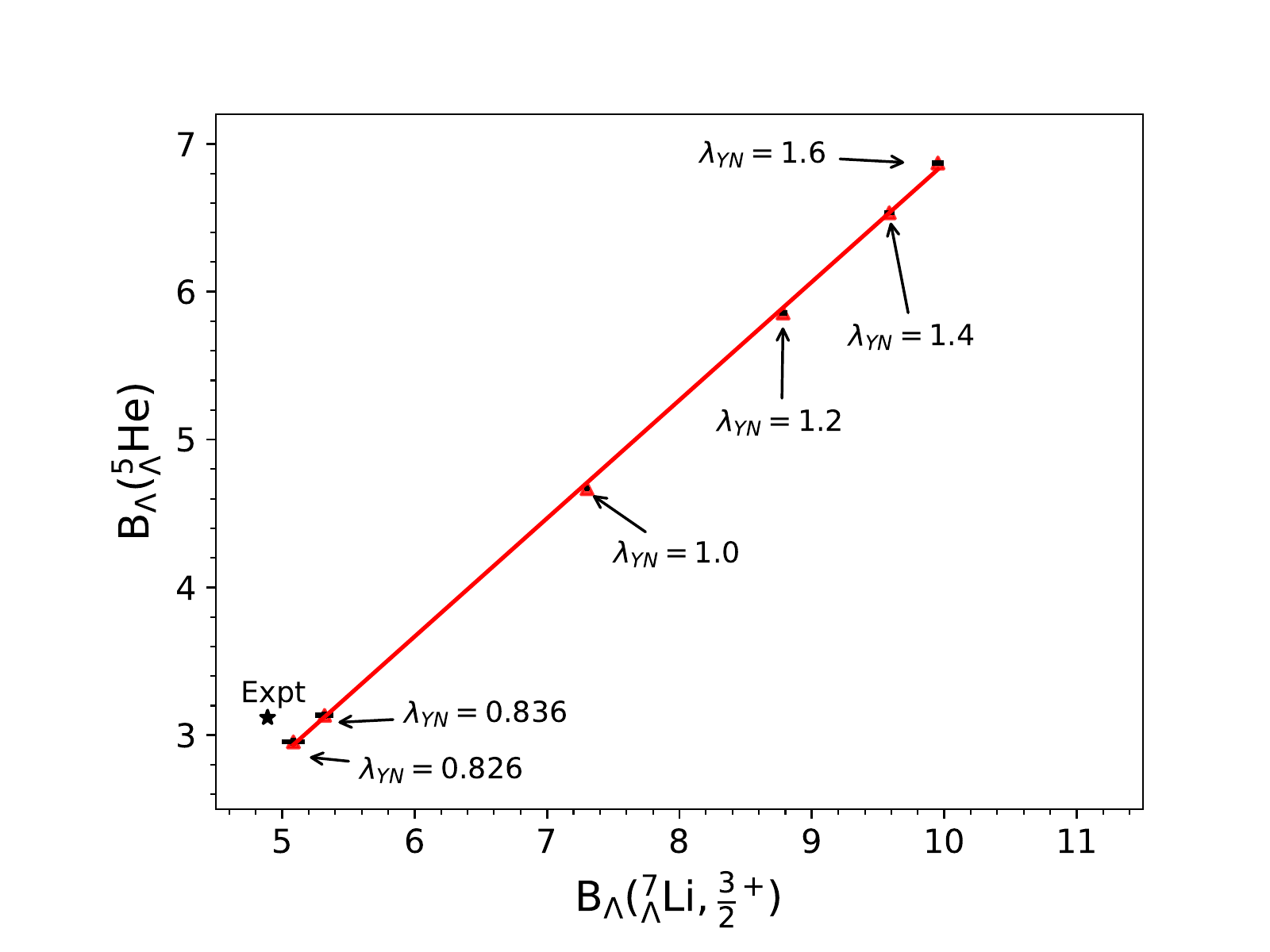}} \\
\end{center}
       \caption{Correlations of $\Lambda$-separation energies 
       between $^5_{\Lambda}$He and (a) $^3_{\Lambda}$H, (b) the $0^+$ state
       of $^4_{\Lambda}$He (red) and $^4_{\Lambda}$H (blue), (c)  the $1^+$ state
       of $^4_{\Lambda}$He (red)  and   $^4_{\Lambda}$H (blue), (d) $^6_{\Lambda}$He (red) and $^6_{\Lambda}$Li (blue),
       (e) $^7_{\Lambda}\text{Li}({1}/{2}^{+},0)$  and (f)
       $^7_{\Lambda}\text{Li}({3}/{2}^{+},0)$,
       for a wide range of  flow parameters $\lambda_{YN}$.  
       The error bars represent numerical uncertainties which are small in most of the cases.  The experimental
       $\Lambda$-separation energy   for $^5_{\Lambda}\text{He}$ is from \cite{Davis:2005mb}. The results  for
       other systems are taken from (a)  \cite{Davis:2005mb}, (b)-(c) \cite{Yamamoto:2015avw} for
       $^4_{\Lambda}\text{He}$ (black asterisk)   and   $^4_{\Lambda}\text{H}$ (grey square),
       (d)  \cite{Hiyama:1996gv} for $^6_{\Lambda}\text{He}$ (black asterisk) and $^6_{\Lambda}\text{Li}$
       (grey square), (e) \cite{Davis:2005mb}  and (f)  \cite{Tamura:2000ea}. The Idaho-N$^3$LO(500) evolved to 1.6~fm$^{-1}$ and NLO19(600) was used for the NN and YN interaction, respectively.}
\label{fig:Correlation_plots}
\end{figure*}

In Section~\ref{sec:YNeffect},  we have observed surprisingly similar trends
of the $\Lambda$-separation energies for all investigated hypernuclei  
with respect to  the running SRG-YN flow  parameter $\lambda_{YN}$. 
This  probably  hints at some intriguing correlations between the $\Lambda$-separation energies  
of these systems.  In order to quantitatively study these correlations, we compute
$B_{\Lambda}$ for all considered hypernuclei, for the same range of   $\lambda_{YN}$ evolution parameters, and compare the results with 
each other for selected values of $\lambda_{YN}$.
It is known that $^5_{\Lambda}$He is the experimentally best studied hypernucleus so far. Also, our J-NCSM
results for this hypernucleus  are well-converged. 
We therefore use $^5_{\Lambda}$He as a benchmark system 
and plot  $B_{\Lambda}(^5_{\Lambda}\text{He})$
against the separation energies  of  other hypernuclear  
systems ($A=3-7$), see Fig.~\ref{fig:Correlation_plots}. 
For that, we choose
Idaho-N\textsuperscript{3}LO(500) evolved to an SRG-NN flow variable  
of $\lambda_{NN}=1.6$~fm\textsuperscript{-1} for the NN interaction 
and NLO19 with a regulator of $\Lambda_{Y} = 600 $~MeV for the
YN interaction. However, we want to emphasize that similar trends 
are observed for SMS N\textsuperscript{4}LO{+}(450) and in 
combination with other YN interactions, see also \cite{LePhD:2020}.
Let us first look at the correlation between the $\Lambda$ removal energies of the $^5_{\Lambda}$He hypernucleus
and  of the hypertriton.   Here  $B_{\Lambda}(^3_{\Lambda}\text{H})$  are  computed  within the  Faddeev-Yakubovky
approach since NCSM calculations are very difficult for this weakly bound system. The correlation plot is presented
in panel~(a) of  Fig.~\ref{fig:Correlation_plots}. Here each symbol represents the numerical $B_{\Lambda}$ of
the two systems calculated at the same flow parameter $\lambda_{YN}$, and it also includes the estimated uncertainties
that  are small in most of the cases. 
The straight line is obtained from a linear fit to the results, 
reminding  one of the Tjon line between the binding energies of $^4$He 
and $^3$He \cite{Tjon:1975sme,Nogga:2000uu,Nogga:2001cz,Platter:2004he,Nogga:2004ab,Klein:2018lqz}. 
We observe a nearly perfect linear correlation between $B_{\Lambda}(^3_{\Lambda}\text{H})$ and 
$B_{\Lambda}(^5_{\Lambda}\text{He})$ for flow parameters up to 
 $\lambda_{YN}=2.0$~fm\textsuperscript{-1} and a 
slight deviation from the straight  line as $\lambda_{YN}$  further increases. The latter can be attributed
to the  possible contribution of 3BFs  \cite{LePhD:2020}. 
Interestingly, the correlation line goes through
the experimental $\Lambda$-separation energies  of the two systems at  $\lambda_{YN}=0.836$~fm\textsuperscript{-1}. 
The value of $\lambda_{YN}$, at which the $^5_{\Lambda}\text{He}$ hypernucleus is properly described, will be referred
to as the magic flow parameter $\lambda^{m}_{YN}$. For that value, 
the separation energy of $^3_\Lambda$H is 92~keV. Using the bare 
NLO19(600) and the same NN interaction, we found 119~keV
which is in reasonable agreement with the result at $\lambda^{m}_{YN}$. 
Obviously, the 
concrete value of $\lambda^{m}_{YN}$ will depend on the YN interactions 
as well as their regulators.

The correlation plots for the ground  and excited states of $^4_{\Lambda}\text{He}/^4_{\Lambda}\text{H}$ are
displayed in panels (b) and (c), respectively.  While  there is a strictly linear correlation between
the   separation energies  
$B_{\Lambda}(^4_{\Lambda}\text{He} / ^4_{\Lambda}\text{H},  1^+)$ and $B_{\Lambda}(^5_{\Lambda}\text{He})$,  the correlation
line for   $B_{\Lambda}(^4_{\Lambda}\text{He} / ^4_{\Lambda}\text{H},  0^+)$ and $B_{\Lambda}(^5_{\Lambda}\text{He})$  exhibits
a small loop to the right for large values of $\lambda_{YN}$,  $\lambda_{YN} \ge 2.4$~fm\textsuperscript{-1}
similar to the behavior of  the correlation line for $B_{\Lambda}(^3_{\Lambda}\text{H})$ and $B_{\Lambda}(^5_{\Lambda}\text{He})$.
Also, from panels (b) and (c),  one  easily notices   almost     identical results  for the isospin mirrors
$^4_{\Lambda}$He and $^4_{\Lambda}$H.
This is because there are no CSB terms in the employed version of the chiral YN potential.
The CSB effect arising from the point Coulomb interactions is included in the calculation, but its contribution is
minor \cite{Bodmer:1985km,Nogga:2019bwo}.   It is interesting that, 
at the magic flow parameter, $\lambda^{m}_{YN}=0.836$~fm\textsuperscript{-1},  the experimental  value of
$B_{\Lambda}(^4_{\Lambda}\text{He}, 1^+)$ is  exactly reproduced  while  the ground state is somewhat underbound.
Furthermore, at this $\lambda^{m}_{YN}$  our J-NCSM results for  the spin doublet of $^4_{\Lambda}\text{He}$, 
$B_{\Lambda}(0^+ (1^+)) = 1.57 (0.97)$~MeV, 
are surprisingly close to the those obtained within the  exact Faddeev-Yaku\-bovsky method using the  non-evolved bare YN  interactions, $B_{\Lambda}(0^+ (1^+)) = 1.61  (1.18)$~MeV. The slight deviation  between the two results is consistent with the size of 3BFs expected from 
the power counting of chiral EFT \cite{Haidenbauer:2019boi}.

Similarly,  almost perfectly   linear correlations are also
found between $B_{\Lambda}(^5_{\Lambda}\text{He})$  and the ground-state energies  $E_{\Lambda}$ of the
$p$-shell    $^6_{\Lambda}\text{He}$ and $^6_{\Lambda}\text{Li}$ 
hypernuclei,   panel~(d),  as well as  the $\Lambda$-separation 
energies $B_{\Lambda}$ of  the  ground and first excited states in  $^7_{\Lambda}\text{Li}$, panels (e) and (f),
respectively.  Note that the resonance energies  $E_{\Lambda}(^6_{\Lambda}\text{Li}/^6_{\Lambda}\text{He})$  
are computed 
as the difference between the  hypernuclear binding energies  $E(^6_{\Lambda}\text{Li}/^6_{\Lambda}\text{He})$   and
the binding energy  $E(^4\text{He})$. This removes most of the  
NN-interaction dependence. In
panel~(d), one notices a pronounced difference in the binding energies  $E_{\Lambda}$ of $^6_{\Lambda}\text{He}$ and
$^6_{\Lambda}\text{Li}$ (about $1.08$~MeV), which simply results from different contributions of the  Coulomb interactions
of the two nuclear cores $^5{\text{He}}$ and $^5{\text{Li}}$.   
We remark that the NLO19(600) YN potential with the magic flow 
parameter $\lambda^{m}_{YN}=0.836$~fm\textsuperscript{-1}  
underbinds the  $^6_{\Lambda}\text{He}/^6_{\Lambda}\text{Li}$  systems  while it slightly overbinds  the
first excited state in $^7_{\Lambda}$Li. The obtained $\Lambda$-separation energy for the ground state, \break
$B_{\Lambda}(^7_{\Lambda}\text{Li},\,{1}/{2}^+) =5.59 \pm 0.01$~MeV, is, however, in very good agreement with
the result from emulsion experiments,
$B_{\Lambda}(^7_{\Lambda}\text{Li}, {1}/{2}^+) = 5.58 \pm 0.03$~MeV \cite{Davis:2005mb}. 
It should be noted that counter experiments reported a somewhat
larger value for $^7_{\Lambda}\text{Li}({1}/{2}^{+},0)$, namely \break $B_{\Lambda}(^7_{\Lambda}\text{Li}, {1}/{2}^+) = 5.85 \pm
0.13 \pm 0.1$~MeV \cite{Agnello:2009kt}.

The observed linear correlations between the separation energies of different hypernucler systems is rather striking and interesting. 
It will be important to examine those correlations using different 
YN bare interactions in order to check whether this useful  
property is a universal feature or just a signature of the 
chiral interactions. Nevertheless, our 
finding for the chiral forces with SRG evolution suggests that the missing SRG-induced
three-body forces might be parameterized by only one adjustable parameter (effects of SRG-induced higher-body
forces on $B_{\Lambda}$ are expected  to be insignificant \cite{Wirth:2016iwn}). If this is the case, one 
is able to minimize the effects of the omitted three-body forces 
by tuning the
SRG-YN flow parameters $\lambda_{YN}$  to  the magic value for which a particular hypernucleus, for
example $^5_{\Lambda}\text{He}$, is properly described. This magic flow parameter $\lambda^{m}_{YN}$  then can
serve as a good starting point for hypernuclear calculations 
requiring a SRG-YN evolution -- which, in turn,
may provide a good opportunity to study hypernuclear structure 
as well as the YN forces in a less
expensive but  realistic approach. A possible application of this 
finding has been considered in \cite{Le:2019gjp,LePhD:2020}.
In this context let us mention that similar linear correlations have
been also observed in Ref.~\cite{Contessi:2019csf} for the double-$\Lambda$ hypernuclei
$^{\ \, 5}_{\Lambda\Lambda}$H and $^{\ \, 6}_{\Lambda\Lambda}$He. 

As discussed in Ref.~\cite{Haidenbauer:2019boi}, 
the contribution of chiral 3BFs is comparable to the uncertainty 
at NLO of approximately 200-300~keV 
for $A=4$. The full $\lambda_{YN}$ dependence of the result 
is an order of magnitude  larger than what is expected for 3BFs by 
chiral power counting. This situation is very different from 
that for ordinary nuclei where SRG-induced and chiral 3BFs 
are of comparable size. Wirth and Roth have pointed 
out that the size of the SRG-induced 3BFs 
is probably enhanced because the $\Sigma$ contribution 
is significantly weakened when $\lambda_{YN}$ is lowered
\cite{Wirth:2016iwn}. Our observation here is that, for 
extreme values of $\lambda_{YN}$ below 1~fm$^{-1}$, the 
$P_\Sigma$ value increases again and the overbinding 
disappears. For such $\lambda_{YN}$, the contribution 
of 3BFs is again in line with the expectation from 
the chiral power counting. Especially, it seems to be 
neglible for $^3_\Lambda$H.

\section{Conclusions}
\label{sec:concl}
In this work, we have extended the nuclear J-NCSM to describe baryonic systems with strangeness $S=-1$. 
The inclusion of the strangeness degree of freedom significantly complicates the implementation of the approach in part  
because the  particle conversion $\Lambda$-$\Sigma$ is explicitly  taken into account. Accordingly, the Jacobi
basis now  consists  of  two orthogonal subsets, characterized by the $\Lambda$ and $\Sigma$ 
hyperons. For the applications of the two-body NN 
and YN forces, we introduced two auxiliary  bases that explicitly single out
the involved NN and YN pairs, respectively. Like the coefficients of fractional parentage, 
the  expansion coefficients can  also be computed  in a preparatory step  separately from any
binding-energy calculations. Once they are known, 
the evaluation of the many-body Hamiltonian matrix elements in the 
Jacobi basis (and therefore the energy calculations) are straightforward. 

As a first application of the Jacobi NCSM, we utilized the approach 
to investigate hypernuclear systems with $A=4-7$. Here, the   $\Lambda$-separation (binding) energies are
extracted systematically via a two-step procedure that enables an effective  removal of the HO-$\omega$
sensitivity  of the final results as well as a reliable estimation of the numerical uncertainties. 
We performed the energy calculations based on various SRG-evolved chiral interactions. In particular, we
considered the Idaho N\textsuperscript{3}LO and SMS N\textsuperscript{4}LO NN potentials in combination with 
the next-to-leading order YN interactions,  NLO13 and NLO19. 
We found  that at low values of the SRG YN flow parameter,  $\lambda_{YN}  \leq 1.4$~fm\textsuperscript{-1},
the  separation energies are not very sensitive to the NN potentials.   
The dependence somewhat increases for higher $\lambda_{YN}$, however, the relative variations remain quite
similar for all systems.  
  
It turned out that, for some of the considered hypernuclei, there are large differences between the predictions 
of the two practically phase-equivalent YN potentials NLO13 and NLO19. 
Those can be attributed to possible (but so far neglected) contributions of chiral three-body (YNN) forces \cite{Petschauer:2015elq}. 
We also observed that there are almost perfect linear 
correlations between the $\Lambda$ separation energies of the $A=4-7$ hypernuclei calculated for a wide range
of the SRG-YN flow parameter. Interestingly, at the magic  value   $\lambda_{YN}^{m}$ that yields the
empirical $B_{\Lambda}(^5_{\Lambda}\text{He})$, the separation energies of 
$^3_{\Lambda}\text{H}$ and $^4_{\Lambda}\text{He}(0^+,1^+)$ are in good agreement with the results for the
non-evolved YN interactions  (at least within the expected contributions of the chiral 3BF), 
while the one for  $^7_{\Lambda}\text{Li}$ is  surprisingly close to the experiment. This may  suggest that
by tuning the SRG parameter such that the $^5_{\Lambda}$He  hypernucleus is correctly reproduced, one
can effectively minimize the effects of  the missing  SRG-induced 3BF.  Therefore,  the special flow
parameter  $\lambda_{YN}^{m}$   can be a good starting point for 
hypernuclear calculations  that   require an SRG evolution.
Such calculations will be useful to develop improved YN interactions. Eventually, taking SRG-induced and chiral 3BF into account will be necessary. Work in this direction is in progress.

\vspace{0.5cm}
\noindent
{\bf Acknowledgements:}
We thank Susanna Liebig and  Marcel Graus for collaboration during
an early stage of this investigation.
This work is supported in part by DFG and NSFC through funds provided to the Sino-German CRC 110 ``Symmetries and
the Emergence of Structure in QCD'' (DFG Grant No. TRR~110). We also acknowledge support of the THEIA net-working activity 
of the Strong 2020 Project. The numerical calculations have been performed on JURECA and the JURECA booster 
of the JSC, J\"ulich, Germany. The work of UGM was supported in part by the
Chinese Academy of Sciences (CAS) President's  International Fellowship Initiative (PIFI) (Grant No. 2018DM0034)
and by VolkswagenStiftung (Grant No. 93562).
\appendix

\section{Transition \texorpdfstring{$\big\langle \big(\alpha^{*(1)}\big)^{*(Y)}| \alpha^{*(YN)} \big\rangle$}{A-N-Y to A-YN} }\label{Appsec:transition}
\begin{figure} [t]
\centering
\begin{tikzpicture}[scale=1.6]
                \filldraw[color=black, ultra thick, fill=gray ]  (0.,0.) circle(.18cm)  node[anchor=base, right=1.2em ] {$ \tilde{\alpha}_{A-2}: \mathcal{\tilde{N}}_{A-2} \tilde{J}_{A-2} \tilde{T}_{A-2} \,\tilde{\zeta}_{A-2}$}
                  node[anchor=base,above=1.4 em,right =-0.3 em, black] {\small{3}};
                \filldraw[black]  (-0.9,0.) circle (0.5mm) node[anchor=base, above=0.4em] {$n_N l_{N} s_N t_N$} 
                node[anchor=base,below=0.3em,left =0.3 em, black] {\small{1}};
                \draw[baseline,thick, ->]  (-.18,0.) -- (-0.85,0.);
                \filldraw[red]  (-0.4,-0.55) circle (0.7mm) node[anchor=base,right=0.6em,black] {$n_Y l_{Y}s_Y t_Y$}  
                  node[anchor=base,below=0.3em,left =0.3 em, black] {\small{2}};
                \draw[baseline,thick, <-]  (-0.4,-0.48) -- (-0.4,0.);
                \end{tikzpicture}  
\caption{$\big|\big(\alpha^{*(1)}\big)^{*(Y)}\big\rangle$ state  with directions of momenta }
 \label{fig:psi_n_y}
\end{figure}
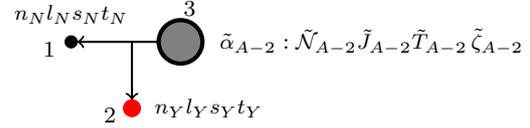

The states  $| \alpha^{*(YN)} \rangle$ and $\big|\big(\alpha^{*(1)} \big)^{*(Y)}\big\rangle$ with directions of momenta are illustrated in Figs. \ref{fig:psi_yn} and \ref{fig:psi_n_y}, respectively. The explicit quantum numbers of these states are
\begin{align} \label{eq:Bpsi_n_y}
\begin{split}
\big| \big (\alpha^{*(1)}\big )^{*(Y)} \big \rangle &= |\alpha^{*(1)}_{A-1}\rangle \otimes |Y\rangle\\[3pt]
 & = | \tilde{\mathcal{N}} {{J}}  {{T}}, \alpha^{*(1)}_{(A-1)} \, n_Y I_Y \tilde{t}_Y; \\[3pt]
 & \quad  (J^{*(1)}_{A-1} (l_Y s_Y) I_Y)  {J}, (T^{*(1)}_{A-1} \tilde{t}_Y) {T} \rangle 
  \equiv \big| \begin{tikzpicture}[baseline={([yshift=-.2ex]current bounding box.center)},scale=0.6]
                \filldraw[color=black, ultra thick, fill=gray ]  (0.,0.) circle(0.22cm);
                \filldraw[black]  (0.7,0.) circle (0.5mm) ;
                \draw[baseline,thick, -]  (0.22,0.) -- (0.65,0.);
                \filldraw[red]  (0.4,-0.4) circle (0.7mm) ;
                \draw[baseline,thick, -]  (0.4,-0.33) -- (0.4,0.);
                \end{tikzpicture} \big\rangle,\\[3pt]
\end{split}
\end{align}
with
\begin{align} \label{eq:Bpsi_n_y1} 
\begin{split}
|\alpha^{*(1)}_{(A-1)N}\rangle & =  | \mathcal{N}^{*(1)}_{(A-1)} {J}^{*(1)}_{A-1}{T}^{*(1)}_{A-1},
 \tilde{\alpha}_{(A-2)N}\, n_N  I_N t_N; \\[4pt]
&\qquad (\tilde{J}_{A-2}(l_N s_N)I_N)J^{*(1)}_{A-1}, (\tilde{T}_{A-2} t_N)T^{*(1)}_{A-1}\rangle\\
&   \equiv               
 \big|\begin{tikzpicture}[baseline={([yshift=-.5ex]current bounding box.center)},scale=0.65]
                \filldraw[color=black, ultra thick, fill=gray ]  (0.,0.) circle(0.22cm);
                \filldraw[black]  (0.7,0.) circle (0.5mm) ;
                \draw[baseline,thick, -]  (0.22,0.) -- (0.65,0.);
                \end{tikzpicture} \big\rangle,
\end{split}
\end{align} 
and 
\begin{align} \label{eq:YNsingleout2}
\begin{split}
|\alpha^{*(YN)} \rangle &= |\alpha_{YN} \rangle \otimes |\alpha_{A-2}\rangle\\[2pt]
& = | \mathcal{N} \mathcal{J} \mathcal{T}, \alpha_{YN}\, n_{\lambda} \lambda \,\alpha_{A-2};  ((l_{YN}(s_Y s_N)S_{YN})\\[2pt]
& \qquad J_{YN} (\lambda J_{A-2})I_{\lambda})\mathcal{J}, ( (t_Y t_N) T_{YN}T_{A-2}) \mathcal{T} \rangle \\
&\equiv \big| \,\begin{tikzpicture}[baseline={([yshift=-0.5ex]current bounding box.center)},scale=0.6]
                \filldraw[color=black, ultra thick, fill=gray ]  (0.,0.) circle(0.20cm);
                \filldraw[red] (-0.7,0.24)   circle(0.7mm); 
                \filldraw[black] (-0.7,-0.24)  circle(0.55mm); 
                \draw[baseline,thick, -]  (-0.7,-0.24) -- (-0.7,0.17);
                \draw[baseline,thick, -]  (-0.7,0.) -- (-0.2,0.0);
                \end{tikzpicture} \big\rangle.
\end{split}
\end{align}
Thereby, the total HO quantum number is given by $\tilde {\cal N} = \mathcal{N}^{*(1)}_{(A-1)} +2 n_Y +l_Y$ and ${\cal N} = {\cal N}_{YN} + {\cal N}_{A-2} + 2n_\lambda +\lambda $. 
The transition $\big\langle \big(\alpha^{*(1)}\big)^{*(Y)}| \alpha^{*(YN)} \big\rangle$  can be interpreted as 
a transformation between different Jacobi coordinates.   We  can therefore   make use of the 
general Jacobi-coordinate transformation formula Eq. (11) in  \cite{Liebig:2015kwa}. For that, we first  
 need  to specify the directions of the relative motions of particles (subclusters)  in the two states   $| \alpha^{*(YN)} \rangle$ and 
$\big|\big(\alpha^{*(1)} \big)^{*(Y)}\big\rangle$.   These directions are depicted in  Figs.~\ref{fig:psi_yn} and \ref{fig:psi_n_y}.
\begin{figure} [t]
\centering
\begin{tikzpicture}[scale=1.6]
                \filldraw[color=black, ultra thick, fill=gray ]  (0.,0.) circle(0.18cm) node[anchor=base,right=3em,below=1.2em] {$\alpha_{A-2}: \, \mathcal{N}_{A-2} J_{A-2} T_{A-2}\, \zeta_{A-2}$}
                node[anchor=base,above=1.4 em,right=-0.3em] {\small
{3}}  ;
                \filldraw[red] (-0.9,0.21)   circle(0.7mm) node[anchor=base,above=0.95em,right=-0.5 em, black] {\small{2}};
                \filldraw[black] (-0.9,-0.28)  circle(0.55mm) node[anchor=base,below=0.95em,right=-0.5 em, black] {\small{1}};
                \draw[baseline,thick, <-]  (-0.9,-0.22) -- (-0.9,0.15) node[anchor=base,below=0.5em,left=0em] {$\alpha_{YN}: \, \mathcal{N}_{YN} J_{YN} T_{YN}$};
                \draw[baseline,thick, ->]  (-0.9,0.) -- (-0.195,0.0) node[anchor=base,above=0.5em,left=0.2em] {$\lambda, n_{\lambda}$};
                \end{tikzpicture}  
\caption{$|\alpha^{*(YN)}\rangle$ state with directions of momenta }
\label{fig:psi_yn}
\end{figure}
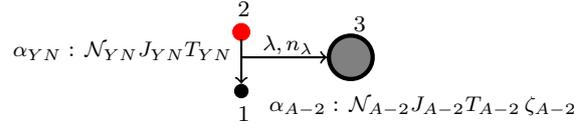 

Comparing the definitions of our two states  $| \alpha^{*(YN)} \rangle$ and $\big|\big(\alpha^{*(1)} \big)^{*(Y)}\big\rangle$  with the
corresponding ones  in Eq. (11) in \cite{Liebig:2015kwa},  one notices  that  the directions of  the relative  momenta are the same,
however, the  ordering of the coupling  of the angular momenta and isospins in  $\big|\big(\alpha^{*(1)} \big)^{*(Y)}\big\rangle$ and  $|\alpha\rangle_{(13)2}$  are different.
 The recoupling from
$
 (\tilde{T}_{A-2}\, t_N)T^{*(1)}_{A-1} \mbox{ to } ( t_N\, \tilde{T}_{A-2})T^{*(1)}_{A-1}    
$
requires a  simple  phase factor,
\begin{align} \label{eq:decoupleT}
\begin{split}
\big| \big(\tilde{T}_{A-2}\, t_N)T^{*(1)}_{A-1} \big\rangle    = (-1)^{\tilde{T}_{A-2} + t_N - T^{*(1)}_{A-1}}  \big|( t_N\, \tilde{T}_{A-2})T^{*(1)}_{A-1}\big\rangle.\\[3pt]
\end{split}
\end{align}
And,  changing 
 the coupling 
$
\big|(\tilde{J}_{A-2}(l_N s_N)I_N)J^{*(1)}_{A-1}  \big\rangle$      to   $\big |(l_N(s_N \tilde{J}_{A-2})S_{A-1})J^{*(1)}_{A-1} \big\rangle 
$
can be done with the help of    $6j$-symbols 
\begin{align} \label{eq:decouple}
\begin{split}
|\big(\tilde{J}_{A-2}( &l_N s_N)I_N)    J^{*(1)}_{A-1}\big\rangle     =
    (-1)^{I_N + 2\tilde{J}_{A-2} + l_N + s_N}\times\\
 &    \sum_{S_{A-1}=\tilde{J}_{A-2} + s_N} \hat{I}_N \hat{S}_{A-1}
                   \left\{ \begin{array}{ccc}
                  \tilde{J}_{A-2} & s_N &  S_{A-1}\\
                  l_N & J^{*(1)}_{A-1} & I_N 
                  \end{array} \right\} \\[6pt]
         &     \times  \big|(l_N(s_N \tilde{J}_{A-2})S_{A-1})J^{*(1)}_{A-1}\big\rangle, 
\end{split}
\end{align}
where   the abbreviation $\hat{I}_{N} = \sqrt{2I_N +1}$, etc.,  is introduced. 
Now taking into account Eqs.~(\ref{eq:decouple}) and (\ref{eq:decoupleT})  and then  making use of the Jacobi-coordinate transformation formula in \cite{Liebig:2015kwa}, 
 one obtains
\begin{align} \label{eq:Btransition}
\begin{split}
 \big\langle  &\big(  \alpha^{*(  1)}\big)^{*(Y)}  |   \alpha^{*(YN)} \big\rangle =\\[6pt]
 & \quad\,\,  \delta_{\mathcal{N} \tilde{\mathcal{N}}} \,
 \delta_{t_{Y} \tilde{t}_Y} \, \delta_{\tilde{T}_{A-2} T_{A-2}} \delta_{\tilde{J}_{A-2} J_{A-2}} 
 \delta_{\mathcal{\tilde{N}}_{A-2} \mathcal{N}_{A-2}}\delta_{\tilde{\zeta}_{A-2} \zeta_{A-2}}\\[10pt]
&\,\, \times  \hat{I}_{N}\,\hat{I}_Y \, \hat{J}_{YN} \, \hat{S}_{YN}\,\hat{I}_{A-2}
 \hat{J}^{*(1)}_{A-1} \, \hat{T}^{*(1)}_{A-1}\, \hat{T}_{YN}\\[11pt]
&\times (-1)^{3J_{A-2} + 2T_{A-2} + T_{YN} + S_{YN} +\lambda + t_Y + l_Y  + t_N + l_N + I_N+ 1}\\[7pt]
&\times \sum_{S_{A-1}=\tilde{J}_{A-2} + s_N} (-1)^{S_{A-1}} \hat{S}^{2}_{A-1} 
                   \left\{ \begin{array}{ccc}
                  {J}_{A-2} & s_N &  S_{A-1}\\
                  l_N & J^{*(1)}_{A-1} & I_N  
                  \end{array} \right\}\\[7pt]
&\times\sum_{L,S} \hat{L}^2 \hat{S}^2
           \left\{\begin{array}{ccc}
                   l_{N}  & S_{A-1} & J^{*(1)}_{A-1}\\
                   l_{Y} & s_{Y} & I_{Y}\\
                   L & S  &{J}
                  \end{array}\right\}    
                  \left\{\begin{array}{ccc}
                   l_{YN}  & S_{YN} & J_{YN}\\
                   \lambda & J_{A-2} & I_{A-2}\\
                   L & S  & {J}
                  \end{array}\right\}\\[6pt]
&\times  \langle n_N \,l_{N}\, n_Y\, l_{Y} : L \,|\,  n_{YN} \,l_{YN} \, n_{\lambda} \, \lambda: L \rangle_{d}\\[8pt]
                  &\times  \left\{ \begin{array}{ccc}
                  s_Y & s_N &  S_{YN}\\
                  J_{A-2} & S & S_{A-1}  
                  \end{array} \right\}
                   \left\{ \begin{array}{ccc}
                  t_Y & t_N &  T_{YN}\\
                  T_{A-2} & {T} & T^{*(1)}_{A-1}  
                  \end{array} \right\},
\end{split}
\end{align}
where the HO bracket
 $\langle n_N \,l_{N}\, n_Y\, l_{Y} : L \,|\,  n_{YN} \,l_{YN} \, n_{\lambda} \, \lambda: L \rangle_{d} $  follows  
 the same convention as in   \cite{Kamuntavicius:2001pf}  with the mass ratio given by 
\begin{align}
 d=\frac{(A-2) \, m(t_Y)}{(A-1)\,m_N + m(t_Y)}.
\end{align}

\section{Faddeev-Yakubovsky equations for hypernuclei}
\label{sec:FYequations}
Faddeev-Yakubovsky equations in momentum space  are a well established tool to solve the Schrödinger equations for light hypernuclei with $A=3$ or 4 \cite{Miyagawa:1993rd,Nogga:2001ef}. We use $A=4$ results to benchmark the NCSM and to provide results for bare interactions. For $A=3$, momentum space 
is much more efficient for the representation of wave functions and, therefore, for the solution of the 
Schrödinger equation than HO wave functions. The weak binding of $^3_\Lambda$H leads to an extremely slow convergence of the energy with respect to $\cal N$. 

Our solution follows Ref.~\cite{NoggaPhD:2001}. For $A=3$, we need to solve a set of coupled Faddeev equations 
\begin{eqnarray}
  | \psi_A \rangle & = & G_0 t_{12} (1-P_{12}) | \psi_B\rangle \cr  
  | \psi_B \rangle & = & G_0 t_{31} \left( | \psi_A\rangle -P_{12} | \psi_B \rangle \right)
\end{eqnarray}
for two Faddeev amplitudes $| \psi_A \rangle$ and $| \psi_B \rangle$. Here, we assume that particles 1 and 2 are nucleons and particle 3 is the hyperon. The permutation operator $P_{12}$ exchanges all coordinates and 
quantum numbers of the two nucleons. $G_0$ is the free 
three-baryon propagator. The two off-shell $t$-matrices $t_{12}$ and $t_{31}$ are solutions of the Lippmann-Schwinger equation of subsystem (12) or (31), respectively. 
For the solution, we use two momentum Jacobi bases: 
\begin{equation}
    | p_{12} p_3 \alpha \rangle = \left| p_{12} p_3 \!\left[ (l_{12}s_{12})j_{12}\!  \left(l_3 \frac{1}{2} \right)\! I_3 \right]\! J (t_{12} t_Y) T M_T  \right \rangle 
\end{equation}
and 
\begin{equation}
    | p_{31} p_2 \beta \rangle = \left| p_{31} p_2 \!\left[ (l_{31}s_{31})j_{31}\!  \left(l_2 \frac{1}{2} \right) \!I_2 \right]\! J (t_{31} \frac{1}{2}) T M_T \right \rangle .
\end{equation}
Here, $p_{ij}$ denotes the magnitude of the relative momentum in subsystem $(ij)$ 
and $p_k$ the relative momentum of particle $k$ relative to the other two particles. 
The angular dependence is expanded in corresponding orbital angular $l_{ij}$ and $l_k$. These are coupled to the spin of the two-baryon subsystem $s_{ij}$ to the total 
angular momentum of the subsystem $j_{ij}$. $l_k$ and the spin ${1}/{2}$ of 
the third baryon couple to the spectator angular momentum $I_k$. Finally, 
the total angular momentum $J$ of the three-body system is obtained by coupling $j_{ij}$ and $I_k$. The isospin of the pair $t_{ij}$ is coupled either 
with the isospin of the hyperon $t_Y=0,1$ or with the isospin of the 
nucleon ${1}/{2}$ to the total isospin $T$ and its third component $M_T$. 
For the hypertriton $M_T=0$ and $T=0$ is the by far dominant component of the 
wave function. 
For the solution, the two Faddeev  amplitudes $| \psi_A \rangle$ and $| \psi_B \rangle$ are expanded in terms of their natural set of basis states  $| p_{12} p_3 \alpha \rangle$ and $| p_{31} p_2 \beta \rangle $, respectivley. Because of the short-ranged hypernuclear interactions, the $t$-matrices converge quickly with respect to partial waves which is then also true for the Faddeev components if expressed in their natural set of basis states. Note that transitions 
between these states are required in order to solve 
the equations.  In this work, we restrict the partial wave so that $j_{ij}\le6$. This ensures that energies are converged to better than 1~keV. We also calculated the wave function 
\begin{equation}
| \Psi \rangle = | \psi_A \rangle + (1-P_{12})|\psi_B\rangle
\end{equation}
and checked explicitly that the expectation value of the 
Hamiltonian operator agrees with the energy obtained by solving the Faddeev equations. For this check, we need to include the $T=1$ and $T=2$ contributions 
to reach an accuracy of the order of 1 keV. 

For $A=4$ hypernuclei, we need to solve a set of five coupled Yakubovsky equations
\allowdisplaybreaks
\begin{eqnarray}
  | \psi_{1A} \rangle & =  & G_0 t_{12} (P_{13}P_{23}+P_{12}P_{23}) \nonumber \\ 
  & &\qquad\qquad         \left[  | \psi_{1A}\rangle + | \psi_{1B}\rangle + | \psi_{2A}\rangle \right] \nonumber \\[5pt]  
    | \psi_{1B} \rangle & = & G_0 t_{12}  
           \left[ (1-P_{12}) (1-P_{23}) | \psi_{1C}\rangle \right. \nonumber \\[5pt] 
    && \qquad \qquad \left. + (P_{13}P_{23}+P_{12}P_{23}) | \psi_{2B}\rangle \right] \nonumber \\[5pt]  
    | \psi_{1C} \rangle & = & G_0 t_{14}  
           \left[ | \psi_{1A}\rangle + | \psi_{1B}\rangle + | \psi_{2A}\rangle              -P_{12} | \psi_{1C}\rangle \right. \nonumber \\[5pt] 
 && \qquad \qquad  \left. +P_{13}P_{23} | \psi_{1C}\rangle 
                   +P_{12}P_{23} | \psi_{2B}\rangle \right] \nonumber \\[5pt]  
  | \psi_{2A} \rangle & = & G_0 t_{12} 
           \left[ (P_{12}-1)P_{13}  | \psi_{1C}\rangle + | \psi_{2B}\rangle \right] \nonumber \\[5pt]    
  | \psi_{2B} \rangle & = & G_0 t_{34} \left[  | \psi_{1A}\rangle + | \psi_{1B}\rangle + | \psi_{2A}\rangle \right] 
\end{eqnarray}
for the five Yakobovsky components $| \psi_{1A}\rangle $, $| \psi_{1B}\rangle $, $| \psi_{1C}\rangle $, $| \psi_{2A}\rangle $ and $| \psi_{2B}\rangle $. Each of these 
components is expanded in terms of its natural Jacobi coordinate, respectively,  as defined below \hfill
\begin{widetext}
\begin{eqnarray}
    | p_{12} p_3 q_4 \alpha_A \rangle 
    & = & \left| p_{12} p_3 q_4  \left[ \left[ (l_{12}s_{12})j_{12} \, \left(l_3 \frac{1}{2} \right) I_3 \right] j_{123} \ \left(l_4 \frac{1}{2} \right) I_4 \right] J \quad \left[ (t_{12} \frac{1}{2}) \tau_{123} t_Y \right] T M_T \right\rangle \ \nonumber \\[5pt]
    | p_{12} p_4 q_3 \alpha_B \rangle 
    & = & \left| p_{12} p_4 q_3  \left[ \left[ (l_{12}s_{12})j_{12} \, \left(l_4 \frac{1}{2} \right) I_4 \right] j_{124} \ \left(l_3 \frac{1}{2} \right) I_3 \right] J \quad \left[ (t_{12} t_Y ) \tau_{124} \frac{1}{2} \right] T M_T \right\rangle \ \nonumber \\[5pt]
    | p_{14} p_2 q_3 \alpha_C \rangle 
    & = & \left| p_{14} p_2 q_3  \left[ \left[ (l_{14}s_{14})j_{14} \, \left(l_2 \frac{1}{2} \right) I_2 \right] j_{124} \ \left(l_3 \frac{1}{2} \right) I_3 \right] J \quad \left[ (t_{14} \frac{1}{2}) \tau_{124} \frac{1}{2} \right] T M_T \right\rangle \ \nonumber \\[5pt]
    | p_{12} p_{34} q \beta_A \rangle 
    & = & \left| p_{12} p_{34} q  \left[ \left[ (l_{12}s_{12})j_{12} \, \lambda \right] I \ (l_{34}s_{34})j_{34} \right] J \quad  (t_{12} t_{34}) T M_T 
        \right\rangle \ \nonumber \\[5pt]
    | p_{34} p_{12} q  \beta_B \rangle 
    & = & \left| p_{34} p_{12} q  \left[ \left[ (l_{34}s_{34})j_{34} \, \lambda \right] I \ (l_{12}s_{12})j_{12} \right] J  \quad (t_{34} t_{12}) T M_T 
        \right\rangle  \ .
\end{eqnarray}
\end{widetext}
Here, the free propagator $G_0$ and the $t$-matrices $t_{ij}$ 
are of course embedded into the four-baryon system.  
The coupling scheme is much more complicated than in the three-baryon system.
Now there are two types of Jacobi coordinates required. The 
first three basis sets are of the ``3+1'' type. Here, three momenta $p_{ij}$, $p_k$ 
and $q_l$ are required that are relative momenta within the pair $ij$, of particle 
$k$ with respect to pair $ij$ and of particle $l$ with respect to the 
three-body subsytem $ijk$. Additionally to the quantum numbers of the 
three-body system, we have now introduced $j_{ijk}$ and $\tau_{ijk}$ for the total 
angular momentum and isospin of the three-body subsystem. $J$, $T$  and $M_T$ are the 
total angular momentum, isospin and third component of isospin of the 
four-baryon system. We have again omitted the spins and isospins of the 
two baryons in the inner most subsystem since only $t_4=t_Y$ differs from 
${1}/{2}$. The last two basis sets are of the ``2+2'' type. 
Here, relative momenta of two two-body subsytems $p_{ij}$ and $p_{kl}$ 
are introduced together with angular momenta and isospins for these 
subsystems. Additionally, the relative momentum of the two pairs $q$ 
and its angular momentum $\lambda$ is required. In order to finally define 
the total four-body angular momentum, an additional intermediate angular momentum 
$I$ needs to be introduced as seen in the definition of the states. 

For four-baryon states, it is not sufficient to constrain the 
two-body angular momenta in order to get a finite number of partial waves. 
Additional constraints on other angular momenta are necessary. For the calculations of this work, we chose $j_{ij}\le 5$, $l_i\le 6$, $\lambda \le 6 $,  
$l_{ij}+l_k+l_l \le 10$  and $l_{ij}+l_{kl}+\lambda \le 10$. In order 
to save computational resources, we restrict ourselves to the by far most 
important isospin component $T={1}/{2}$, although the contribution 
to the energy of the Yakubovsky equations induces an uncertainty of 10~keV. 
Interestingly, the other isospin components are more important when 
calculating the expectation value for which they contribute approximately 20~keV. 
Note that also our J-NCSM results are based on the dominant isospin 
components only. We therefore need to take an uncertainty of approximately 
20~keV in the four-baryon systems into account due to missing isospin components. 

Once the Yakubovsky components are found, we obtain the wave function by 
\begin{eqnarray}
  | \Psi \rangle & = & (1+P_{13}P_{23}+P_{12}P_{23}) | \psi_{1A} \rangle \nonumber \\[5pt]
        & & + (1+P_{13}P_{23}+P_{12}P_{23}) | \psi_{1B} \rangle \rangle \nonumber \\[5pt]
        & & + (1-P_{12}) (1+ P_{13}P_{23}+P_{12}P_{23}) | \psi_{1C} \rangle \nonumber \\[5pt]
        & & +(1+P_{13}P_{23}+P_{12}P_{23}) | \psi_{2A} \rangle\rangle \nonumber \\[5pt]
        & & +(1+P_{13}P_{23}+P_{12}P_{23}) | \psi_{2B} \rangle \ . 
\end{eqnarray}

This briefly summarizes the Faddeev-Yakubovsky approach as 
we used it for benchmarking our J-NCSM results.

\bibliographystyle{unsrturl}

\bibliography{hyp-literatur.bib,ncsm.bib,hypernuclei.bib,nn-interactions.bib,yn-interactions.bib,faddeev-yakubovsky.bib,srg.bib}

\end{document}